%% file: main.tex
\documentclass[12pt]{article}
\usepackage{amsmath}
\usepackage{graphicx,psfrag,epsf}
\usepackage{enumerate}
\usepackage{natbib}
\usepackage{url} % not crucial - just used below for the URL 

\usepackage{float}
\usepackage{multirow}
\usepackage{bm}
\usepackage{enumitem}
\usepackage{makecell}
\usepackage{amsthm}
\usepackage{siunitx}
\usepackage[dvipsnames]{xcolor}
\usepackage{amssymb}
\usepackage{amsfonts}
\usepackage{newverbs}
\usepackage{marvosym}
\usepackage[hidelinks]{hyperref}
\usepackage{sectsty}
\usepackage{booktabs,caption}
\usepackage[flushleft]{threeparttable}

\newtheorem{lemma}{Lemma}[section]
\newtheorem{definition}{Definition}[section]

\newlist{steps}{enumerate}{1}
\setlist[steps, 1]{label = Step \arabic*:}

\setcitestyle{aysep={}}
\subsectionfont{\bfseries\itshape}
\subsubsectionfont{\normalfont\itshape}
\newcommand{\blind}{0}

\begin{document}

\def\spacingset#1{\renewcommand{\baselinestretch}%
{#1}\small\normalsize} \spacingset{1}

%%%%%%%%%%%%%%%%%%%%%%%%%%%%%%%%%%%%%%%%%%%%%%%%%%%%%%%%%%%%%%%%%%%%%%%%%%%%%%

\if0\blind
{
  \title{\bf Vulnerability-CoVaR: Investigating the Crypto-market}
  \author{Martin Waltz\\
    Institute of Transportation Economics, Technische Universität Dresden,\\
    Dresden, Saxony 01062, Germany (martin.waltz@tu-dresden.de)\\
    and \\
    Abhay Kumar Singh\\
    Department of Applied Finance, Macquarie University,\\
    Sydney, NSW 2109, Australia (abhay.singh@mq.edu.au)\\
    and \\
    Ostap Okhrin\\ 
    Institute of Transportation Economics, Technische Universität Dresden,\\
    Dresden, Saxony 01062, Germany (ostap.okhrin@tu-dresden.de)}
    
  \maketitle
} \fi

\if1\blind
{
  \bigskip
  \bigskip
  \bigskip
  \begin{center}
    {\LARGE\bf Vulnerability-CoVaR: Investigating the Crypto-market}
\end{center}
  \medskip
} \fi

\begin{abstract}
This paper proposes an important extension to Conditional Value-at-Risk (CoVaR), the popular systemic risk measure, and investigates its properties on the cryptocurrency market. The proposed Vulnerability-CoVaR (VCoVaR) is defined as the Value-at-Risk (VaR) of a financial system or institution, given that at least one other institution is equal or below its VaR. The VCoVaR relaxes normality assumptions and is estimated via copula. While important theoretical findings of the measure are detailed, the empirical study analyzes how different distressing events of the cryptocurrencies impact the risk level of each other. The results show that Litecoin displays the largest impact on Bitcoin and that each cryptocurrency is significantly affected if an event of joint distress among the remaining market participants occurs. The VCoVaR is shown to capture domino effects better than other CoVaR extensions.
\end{abstract}

\noindent%
{\it Keywords:} copula, Conditional Value-at-Risk, cryptocurrency, systemic risk
\vfill

\newpage
\spacingset{1.45} % DON'T change the spacing!

%MY CHANGES START
\section{Introduction}
Various developments and crises over the last two decades, such as the financial crisis of 2009, have demonstrated how volatile, fragile, and interconnected the financial system and its institutions can be. This gives rise to  \emph{systemic risk}, which can be described as ‘the risk of the financial system as a whole’ \citep[p. 2]{cao2013multi}. The regulatory methodology focuses highly on protecting the financial system against systemic risk events by identifying globally systemically important financial institutions based on cross-jurisdictional activities, size, interconnectedness, substitutability, and complexity. These higher risk institutions are subject to higher loss absorbency requirements, which are imposed next to general liquidity and risk-based capital requirements \citep{basel2013}. However, the question of correctly quantifying systemic risk via appropriate measures remains a crucial task and has developed into a highly researched area. Classical univariate risk measures such as the VaR or the Expected Shortfall are constructed to quantify the risk of an isolated institution or asset class. Consequently, these univariate measures are unable to quantify the impact of an institution's distress on another institution or the whole financial system. As a result, alternative multivariate measures which overcome these limitations and are able to quantify the impact of the risk of a financial institution on other institutions in a system need to be defined.

The last decade has seen a rise of a completely new, highly volatile, and risky financial product known as Cryptocurrency (CC). The CC has recently received increased attention in academia \citep{vidal2021investigation, petukhina2021investing}. However, the economics of those financial assets are yet not well understood, and the risks hidden in this system require thorough investigation. The potential threats from CC have recently also been recognized by regulating authorities, see \cite{basel2019}. The following systemic risk discussion focuses solely on CC as financial assets, but the methods are general and are applicable to other inter-connected financial asset classes.

 \cite{bisias2012survey} and \cite{benoit2017risks} provide extensive surveys of current methodologies to quantify systemic risk. Among these several methods, the most widely-applied market-based measure is the Conditional Value-at-Risk (CoVaR) by \cite{tobias2016covar}, which expands the approach of the VaR to a conditional setting. The $CoVaR^{j|i}$ can be defined as a quantile of the conditional return distribution of CC (or a system) $j$ given that the CC $i$ is under distress, which means that if the usually stable Litecoin (LTC) becomes risky, will this risk be transferred to Bitcoin (BTC)? Based on that concept, \cite{tobias2016covar} define a measure called Delta-CoVaR by taking the difference between $CoVaR^{j|i}$ with $i$ being exactly at its VaR and with $i$ being in its median state, therefore highlighting the strength of the effect. A list of further systemic risk measures has been developed and analyzed by \cite{girardi2013systemic}, \cite{mainik2012dependence}, \cite{acharya2017measuring}, and \cite{brownlees2017srisk}. \cite{zhou2009banks} considers, among other measures, the Vulnerability Index (VI) that represents the probability that the CC of interest violates its VaR under the condition of at least one other CC violating its VaR. Several studies have expanded the CoVaR measure to a multiple case by incorporating more than one variable in the conditional event. \cite{cao2013multi} introduces the Multi-CoVaR (MCoVaR) with the condition of several CCs being simultaneously in distress. \cite{bernardi2019allocation} propose the System-CoVaR (SCoVaR), in which the conditional variables are aggregated via their sum. Further extensions are detailed in \cite{bernardi2018conditional}, \cite{di2015multivariate}, \cite{bernardi2017multiple}, and \cite{bonaccolto2021breakup}.

The main goal of this paper is to formalise a flexible approach that allows to capture a variety of distress events without having to specify a pre-specified distressing situation of the given system, e.g., distress of a specific element or group of elements. Therefore, complementary to SCoVaR and MCoVaR, this empirical study proposes the Vulnerability-CoVaR (VCoVaR), which translates the idea of the VI to the conditional quantile setting. The VCoVaR is defined as the VaR of a CC (or the CC system) given there exists at least one other CC being below or equal to its VaR. Copula-based estimation strategies and characteristics for CoVaR and all investigated CoVaR extensions (SCoVaR, MCoVaR, VCoVaR) are detailed and validated in a thorough simulation study. CoVaR, MCoVaR, and VCoVaR are found to be equal in certain dependence scenarios. Simulation-based analysis of the measures depending on the dependence structure and intensity reveal the desirable property of the VCoVaR of being a monotonically decreasing function of the dependence parameter for a selected list of Archimedean copulae (AC). As an important by-product of this research, a semi-automated univariate model selection procedure based on the minimization of an information criterion while fulfilling the requirements on the respective time series residuals is proposed, see Appendix B.

The paper is structured as follows: Section \ref{sec:SysRisk_crypto} illustrates why the VCoVaR is particularly appropriate for the CC market and further motivates the use of copula for estimation. Section \ref{sec:covar_theory} formally defines the measures and derives the copula-based estimation. Section \ref{sec:properties_rm} investigates the properties of the risk measures. Section \ref{sec:SimStudy} includes the simulation study, while Section \ref{sec:crypto} contains the application study of CCs. Section \ref{sec:conclusion} concludes. The \verb+R+ code to reproduce the results from this paper will be published in a GitHub repository as soon as the paper is accepted.

\section{Systemic Risk in the Cryptocurrency Market}\label{sec:SysRisk_crypto}
The literature identifies two highly relevant characteristic properties of the CC market: the existence of significant spillover effects and the occurrence of herding behaviour among CC market participants. The latter relates to the phenomenon that investors tend to imitate each others transaction behaviour instead of following their own information and belief basis \citep{hwang2004market}. The existence of spillover effects is displayed in \cite{borri2019conditional}, who applies the CoVaR of \cite{tobias2016covar} based on quantile regression to discover that CCs are highly exposed to tail-risk from other CCs. \cite{ji2019dynamic} use the methodology of \cite{diebold2015trans} to quantify return and volatility spillovers in the CC market. Pursuing a similar methodological approach, \cite{RiskConnectedness2020} find that risk spillovers are stronger in the direction from CCs with small market capitalization to those with larger capitalization. \cite{xu2020tail} run the TENET approach originally developed in \cite{hardle2016tenet} to conclude that the market of CCs is coined by significant effects of spillover risk and that the connectedness in the market increased steadily over the course of time. Further spillover analysis of the crypto-market can be found in \cite{koutmos2018return}, \cite{luu2019spillover}, and \cite{katsiampa2019volatility}, while the empirical findings are greatly summarized in the survey of \cite{kyriazis2019survey}. Along with spillover effects, CCs also show a strong behaviour of tail dependence, see \cite{tiwari2020empirical} and \cite{xu2020tail}, which can be modelled using the copula method.

Regarding the existence of herding behaviour, a relevant contribution is \cite{bouri2019herding}, who identify using the approach of \cite{stavroyiannis2017herding} significant herding effects whose intensity varies over time. \cite{vidal2019herding} give evidence for herding effects during downward market situations, based on the methodology of \cite{chang2000examination} and \cite{chiang2010empirical}. They notice that the behaviour of the main CCs is crucial for the investment decisions of traders. \cite{ballis2020testing} and \cite{kallinterakis2019investors} also follow the method of \cite{chang2000examination} and confirm the presence of herding effects, although detecting stronger effects during upward market situations. Finally, \cite{kyriazis2020herding} contains a survey about the empirical findings. 

These two properties - spillover effects and herding behaviour - of the CC market suggest that distress of a CC leads to subsequent distresses of other CCs, and consequently, a domino effect might take place, increasing the likelihood of a systemic risk event. Additionally, there is evidence that the CC market can be primarily influenced by one dominant CC, for example BTC, as stated in \cite{smales2020one}.

The VCoVaR is especially appropriate for the CC market because the measure is tailored for quantifying tail-dependence and domino effects. For example, in the case of extreme losses of Bitcoin (BTC) under the condition that at least one of Ethereum (ETH), Litecoin (LTC), Monero (XMR), and Ripple (XRP) is under distress, with the VCoVaR we capture all situations of such distress spreading processes in the system. It is not necessary to define which CC initially was under distress or how far the domino effect is already developed. The notion of \emph{at least one} includes all possible scenarios and is hence more appropriate in capturing domino effects than the existing alternatives CoVaR, MCoVaR, and SCoVaR, which focus only on one pre-specified distress situation. The use of copulae allows to model both tail dependencies and contagion risk, with the latter being especially pronounced in this market with one dominant CC. Consequently, the VCoVaR provides a flexible tool to depict the impact of such systemic risk scenarios due to its natural consideration of the special characteristics of the CC market.

\section{Conditional Multivariate Risk Measures}\label{sec:covar_theory}
\subsection{Definitions}\label{subsec:covar_def}
Before formally introducing the conditional measures, the univariate VaR measure is reviewed. Let $X_{i,t}$ be the return of CC $i$ at time $t$. The $VaR^{i}_{\alpha, t}$ at probability level $\alpha \in (0,1)$ is implicitly defined as:
\begin{equation}\label{def:var}
  P(X_{i,t} \leq VaR^{i}_{\alpha, t}) = \alpha.
\end{equation}
If $X_{i,t} \sim F_{i,t}$, one can alternatively write $VaR^{i}_{\alpha, t} = F_{i,t}^{-1}(\alpha)$, with $F_{i,t}^{-1}$ being the generalized inverse of $F_{i,t}$, defined as $F^{\leftarrow}(u) = \inf\{x: F(x) \geq u\}$.

Let $X_{j,t}$ be the return of CC (or the CC system) $j$ at time $t$. The original \cite{tobias2016covar} $CoVaR^{=, j|i}_{\alpha, \beta, t}$ with probability level $\beta$ for $j$ given $X_{i,t}$ equals its $VaR^{i}_{\alpha, t}$ is defined as:
\begin{equation}\label{eq:AB_covar_def}
    P(X_{j,t} \leq CoVaR^{=, j|i}_{\alpha, \beta, t} | X_{i,t} = VaR^{i}_{\alpha, t}) = \beta, \hspace{1cm} \text{for} \hspace{0.2cm} j \neq i.
\end{equation}
The $CoVaR^{=, j|i}_{\alpha, \beta, t}$ is the quantile of the conditional return distribution. Frequently applied probability levels in practice are $\alpha = \beta = 0.05$ or $\alpha = \beta = 0.01$. We consider general cases with $\alpha, \beta \in (0,1)$ for all measures. \cite{girardi2013systemic} modify (\ref{eq:AB_covar_def}) by adding inequality to the condition:
\begin{equation}\label{eq:GE_covar_def}
    P(X_{j,t} \leq CoVaR^{j|i}_{\alpha, \beta, t} | X_{i,t} \leq VaR^{i}_{\alpha, t}) = \beta.
\end{equation}
It is argued that this definition is reasonable as it considers more extreme distressing events of CC $i$ and gives the opportunity to apply standard backtesting procedures, e.g., \cite{kupiec1995techniques}. \cite{mainik2012dependence} showed for selected bivariate distributions that the CoVaR in (\ref{eq:AB_covar_def}) is not a monotonically increasing function of the dependence coefficient between $(X_{j,t}, X_{i,t})$, while the one in (\ref{eq:GE_covar_def}) is monotonically increasing. Note that this translates into monotonically decreasing functions in our case, as \cite{mainik2012dependence} considered loss variables. This characteristic is referred to as dependence consistency. More precisely, Theorem 3.6 in \cite{mainik2012dependence} guarantees the measure in (\ref{eq:GE_covar_def}) is dependence consistent if $(X_{j,t}, X_{i,t})$ follows a bivariate elliptical distribution or an elliptical copula. Similar properties have been found for the Gumbel copula.

However, the relationships in the CC world are unlikely to be fully captured with a bivariate distribution. It is necessary to find alternatives, including several variables for the conditional event, to capture more complex scenarios in which $p > 1$ CCs are in distress. In the following, let $X_{t} = (X_{1,t},\ldots, X_{p,t})^{\top}$ be the vector of returns of CCs, with indices collected in the vector $\mathbf{i} = 1, \ldots, p$ at time $t$ where $j$ is not part of these CCs. The first considered extension, the SCoVaR, aggregates the variables in the conditional event by taking their sum and was introduced in \cite{bernardi2019allocation}. Building on this idea, the SCoVaR in this paper is implicitly defined as follows:
\begin{definition}[System-CoVaR]\label{def:scovar}
Given the return $X_{j,t}$ of cryptocurrency/system $j$ and the returns $X_{t}$ of cryptocurrencies $\mathbf{i}$, the SCoVaR is defined as:
\begin{equation}\label{eq:s_covar_def}
    P\left\{X_{j,t} \leq SCoVaR^{j|\mathbf{i}}_{\alpha, \beta, t} \middle| \sum_{i = 1}^{p} X_{i,t} \leq VaR_{\alpha,t}\left(\sum_{i = 1}^{p} X_{i,t}\right)\right\} = \beta.
\end{equation}
\end{definition}
\noindent \cite{bernardi2019allocation} impose the additional restriction that every variable in the conditional event is below or equal its individual VaR, what leads to a different form of (\ref{eq:s_covar_def}), namely:
\begin{equation*}
    P\left\{X_{j,t} \leq SCoVaR^{j|\mathbf{i}}_{\alpha, \beta, t} \middle| \sum_{i = 1}^{p} X_{i,t} \leq VaR_{\alpha,t}\left(\sum_{i = 1}^{p} X_{i,t}\right), \forall i: X_{i,t} \leq VaR^{i}_{\alpha, t}\right\} = \beta.
\end{equation*}
Building on their formulation, the authors find a generalization of the Expected Shortfall measure, which is used to pursue a game theoretic approach of risk allocation. However, this paper separates these naturally different restrictions into the SCoVaR as in Definition \ref{def:scovar} and the MCoVaR, which is introduced in the following.

The MCoVaR is the second extension and was introduced in \cite{cao2013multi}. This measure covers cases when all $X_{i,t}$ are simultaneously equal or below their $VaR^{i}_{\alpha, t}$ level. Thus, using probability levels $\alpha$ and $\beta$, it is defined as:
\begin{definition}[Multi-CoVaR]
Given the return $X_{j,t}$ of cryptocurrency/system $j$ and the returns $X_{t}$ of cryptocurrencies $\mathbf{i}$, the MCoVaR is defined as:
\begin{equation}\label{eq:m_covar_def}
    P(X_{j,t} \leq MCoVaR^{j|\mathbf{i}}_{\alpha, \beta, t} | \forall i: X_{i,t} \leq VaR^{i}_{\alpha, t})  = \beta.
\end{equation}
\end{definition}
\noindent Although it is possible to consider different $\alpha$-levels for each $X_{i,t}$ to balance individual effects, for simplicity, it is assumed that all measures impose a common $\alpha$-level for the conditional variables. \cite{cao2013multi} defines a measure of systemic risk contribution by taking the difference of the MCoVaR as in (\ref{eq:m_covar_def}) and the MCoVaR when the $X_{i,t}$ are at a normal state. As for the SCoVaR, the aim of this paper is also to study the properties and estimation of the MCoVaR given in (\ref{eq:m_covar_def}).

Along these lines, we propose the VCoVaR, which is to the best of our knowledge not existent in the current literature, although allowing for a new perspective on systemic risk. It translates the idea of the VI of \cite{zhou2009banks} into a conditional quantile setting. The VI was originally defined on loss distributions and measures the probability of $X_{j,t}$ violating its VaR given there exists at least one other CC violating its VaR. Transferring this approach, the VCoVaR is implicitly defined as follows:
\begin{definition}[Vulnerability-CoVaR]
Given the return $X_{j,t}$ of cryptocurrency/system $j$ and the returns $X_{t}$ of cryptocurrencies $\mathbf{i}$, the VCoVaR is defined as:
\begin{equation}\label{eq:v_covar_def}
    P(X_{j,t} \leq VCoVaR^{j|\mathbf{i}}_{\alpha, \beta, t} | \exists i: X_{i,t} \leq VaR^{i}_{\alpha, t})  = \beta.
\end{equation}
\end{definition}
This approach allows to cover a variety of distress events and naturally generalizes the CoVaR of (\ref{eq:GE_covar_def}) and the MCoVaR of (\ref{eq:m_covar_def}). It is straightforward to see that the conditional event of the MCoVaR is a subset of the conditional events of the VCoVaR. In a setting of positive dependencies, the distressing event of the MCoVaR relates to the worst case covered in the VCoVaR, namely all $X_{i,t}$ are below or equal to their VaR. On the other side, the VCoVaR is able to cover situations that are less negative than the bivariate CoVaR. Having, e.g., the return of three CCs LTC, XMR, and XRP as conditional variables, the VCoVaR captures situations in which XMR violates its VaR while LTC and XRP do not. This crypto market situation can be assessed more positive than the one of the bivariate CoVaR with XMR in the conditional event as additional positive information about LTC and XRP exist.

\subsection{Estimation of Systemic Risk Measures}\label{subsec:covar_est}
\subsubsection{CoVaR Estimation}
The original CoVaR of \cite{tobias2016covar} given in (\ref{eq:AB_covar_def}) was estimated using a quantile regression approach \citep{koenker1978regression}. \cite{girardi2013systemic} point out that - although the resulting $CoVaR^{=, j|i}_{q, t}$ estimate is time-variant - the impact of $VaR^{i}_{q,t}$ on $CoVaR^{=, j|i}_{q, t}$ is constant, which is unlikely to be the case in practice. In contrast, they propose to estimate their CoVaR modification based on the bivariate distribution of $(X_{j,t},X_{i,t})$, thus rewrite (\ref{eq:GE_covar_def}) as:
\begin{equation*}
    \frac{P(X_{j,t} \leq CoVaR^{j|i}_{\alpha, \beta, t}, X_{i,t} \leq VaR^{i}_{\alpha, t})}{P(X_{i,t} \leq VaR^{i}_{\alpha, t})} = \beta,
\end{equation*}
which reduces to:
\begin{equation}\label{eq:GE_covar_transformed}
    P(X_{j,t} \leq CoVaR^{j|i}_{\alpha, \beta, t}, X_{i,t} \leq VaR^{i}_{\alpha, t})= \alpha\beta,
\end{equation}
as per definition $P(X_{i,t} \leq VaR^{i}_{\alpha, t}) = \alpha$, see (\ref{def:var}). On this basis, the following three-step procedure was proposed for the estimation:
\begin{steps}[align=left, leftmargin=*]
    \item Fit a suitable univariate time-series process (selected, e.g., through our newly proposed procedure, see Section \ref{subsubsec:univariate_models}) to $X_{i,t}$ and estimate $VaR^{i}_{\alpha, t}$.
    \item Estimate the bivariate conditional heteroscedasticity model (e.g.~the DCC-GARCH model of \citealp{engle2002dynamic}) to obtain an estimate of the time dependent bivariate density $\hat{f_t}(x_{j,t}, x_{i,t})$ with observations $x_{j,t}$, $x_{i,t}$ of $X_{j,t}$, $X_{i,t}$ with $i = 1,\ldots,p$.
    \item Solve for $CoVaR^{j|i}_{\alpha, \beta, t}$ the equation:
    \begin{equation}\label{eq:GE_int}
        \int_{-\infty}^{CoVaR^{j|i}_{\alpha,\beta, t}} \int_{-\infty}^{VaR^{i}_{\alpha, t}} \hat{f_t}(x_{j,t}, x_{i,t})dx_{j,t} dx_{i,t} = \alpha \beta.
    \end{equation}
\end{steps}
This procedure might be computationally demanding as it involves numerical evaluation of a double integral. Overcoming this issue, we base our estimation on copulae. Copulae are multivariate distribution functions with margins being $\mathcal{U}[0,1]$, see \cite{joe2014dependence}. Copulae give the opportunity to specify the dependence structure of random variables in a flexible way, allowing to go beyond the commonly applied multivariate Gaussian and $t$-distribution. This is also handy because CC returns are even less normal than fiat stocks, see, e.g. \cite{szczygielski2020one} for an extensive investigation of proper CC return distributions.

To estimate the CoVaR as given in (\ref{eq:GE_covar_transformed}), \cite{reboredo2015systemic} express the bivariate distribution function $F_{X_{j,t}, X_{i,t}}$ of $(X_{j,t}, X_{i,t})$ as:
\begin{align*}
 P(X_{j,t} \leq CoVaR^{j|i}_{\alpha, \beta, t}, X_{i,t} \leq VaR^{i}_{\alpha, t}) &= F_{X_{j,t}, X_{i,t}}(CoVaR^{j|i}_{\alpha, \beta, t}, VaR^{i}_{\alpha, t}) \\
 &= C_{X_{j,t},X_{i,t}}\{F_{X_{j,t}}(CoVaR^{j|i}_{\alpha, \beta, t}), F_{X_{i,t}}(VaR^{i}_{\alpha, t});\theta_t\},
\end{align*}
using the \cite{Sklar1959} theorem. $F_{X_{j,t}}$ and $F_{X_{i,t}}$ denote the marginal distributions of $X_{j,t}$ and $X_{i,t}$, respectively. $C_{X_{j,t},X_{i,t}}$ refers to the copula function with parameter $\theta_t$. The $CoVaR^{j|i}_{\alpha, \beta, t}$ is estimated by solving: 
\begin{equation}\label{eq:covar_est_cop}
 C_{X_{j,t},X_{i,t}}\{F_{X_{j,t}}(CoVaR^{j|i}_{\alpha, \beta, t}), \alpha;\theta_t\} = \alpha\beta,
\end{equation}
which uses $F_{X_{i,t}}(VaR^{i}_{\alpha, t}) = \alpha$. Note that in the case of AC (\ref{eq:covar_est_cop}) can be solved analytically for $CoVaR^{j|i}_{\alpha, \beta, t}$, see \cite{karimalis2018measuring}. Another crucial advantage is that it is not necessary to estimate the VaR of the conditional variable beforehand \citep{reboredo2015systemic}. To compute (\ref{eq:covar_est_cop}), it is sufficient to estimate the copula and the marginal distribution of $X_{j,t}$. This estimation strategy is transferred to the SCoVaR of (\ref{eq:s_covar_def}). Although it involves information of $p$ conditional variables, it can be estimated using (\ref{eq:covar_est_cop}) while replacing $X_{i,t}$ with $\sum_{i = 1}^{p} X_{i,t}$ for estimating the copula between $X_{j,t}$ and $\sum_{i = 1}^{p} X_{i,t}$.

\subsubsection{MCoVaR Estimation}
Set $VaR_{\alpha,t} = (VaR_{\alpha,t}^{1},\ldots, VaR_{\alpha,t}^{p})^{\top}$, where $X_{t} \leq VaR_{\alpha,t}$ holds componentwise. Furthermore, set $\bm{\alpha} = (\alpha, \ldots, \alpha)^{\top}$ and $F_{X_{t}}(VaR_{\alpha,t}) = \{F_{X_{1,t}}(VaR^{1}_{\alpha, t}), \ldots, F_{X_{p,t}}(VaR^{p}_{\alpha, t})\}^{\top}$. To estimate the MCoVaR, (\ref{eq:m_covar_def}) can be rewritten as:
\begin{equation}\label{eq:m_covar_transform}
    \frac{P(X_{j,t} \leq MCoVaR^{j|\mathbf{i}}_{\alpha, \beta, t}, X_{t} \leq VaR_{\alpha,t})}{P(X_{t} \leq VaR_{\alpha,t})}  = \beta.
\end{equation}
Similar to the procedure of \cite{girardi2013systemic}, \cite{cao2013multi} computes the individual VaR for each CC and assumes a parametric form of the $(p+1)$-dimensional distribution for all involved variables. On this basis, the denominator of (\ref{eq:m_covar_transform}) can be computed, leading to an expression with a multiple integral with $p+1$ variables and the MCoVaR as the only unknown. This is solved numerically, and \cite{cao2013multi} assumes a multivariate $t$-distribution driving the overall dependency in the application.

Furthermore, (\ref{eq:m_covar_transform}) is given in terms of copulae by:
\begin{equation*}
    \frac{C_{X_{j,t}, X_{t}}\{F_{X_{j,t}}(MCoVaR^{j|\mathbf{i}}_{\alpha, \beta, t}), F_{X_{t}}(VaR_{\alpha,t});\theta_{1,t}\}}{C_{X_{t}}\{F_{X_{t}}(VaR_{\alpha,t});\theta_{2,t}\}} = \beta,
\end{equation*}
leading to:
\begin{equation}\label{eq:m_covar_est}
    \frac{C_{X_{j,t},X_{t}}\{F_{X_{j,t}}(MCoVaR^{j|\mathbf{i}}_{\alpha, \beta, t}), \bm{\alpha};\theta_{1,t}\}}{C_{X_{t}}(\bm{\alpha};\theta_{2,t})} = \beta,
\end{equation}
where $\bm{\alpha} = (\alpha, \ldots, \alpha)^{\top} \in (0,1)^{p}$. $C_{X_{j,t}, X_{t}}$ denotes the $(p+1)$-dimensional copula of $(X_{j,t}, X_{t}) = (X_{j,t}, X_{1,t}, \ldots, X_{p,t})$ with the parameter $\theta_{1,t}$. $C_{X_{t}}$ refers to the $p$-dimensional copula of $X_{t}$ with parameter $\theta_{2,t}$. This expression can be solved, as the MCoVaR is the only unknown term. Consequently, it is sufficient to have an appropriate estimate of the copulae and the marginal distribution $F_{X_{j,t}}$. One can assume different structures for the copulae enabling different interpretations of the gained MCoVaR, which will be detailed in Section \ref{subsec:sim_analysis}. In practice, $C_{X_{j,t}, X_{t}}$ is estimated from the data, and the copula $C_{X_{t}}$ is gained through marginalization, setting $C_{X_{t}}(\bm{\alpha}; \theta_{2,t}) = C_{X_{j,t}, X_{t}}(1, \bm{\alpha}; \theta_{1,t})$, from the grounding property \citep{nelsen2007introduction}. Notice that (\ref{eq:m_covar_est}) yields an analytic solution for specific copula families. Let $\varphi_{\theta}$ be a generator function for an AC with parameter $\theta$ and $\varphi_{\theta}^{-1}$ the corresponding inverse. Let $C_{X_t}$ be some copula under the assumption, that $C_{X_{j,t}, X_t}(u_j, u_1, \ldots, u_p; \theta_{1,t}) = \varphi_{\theta_{1,t}}^{-1}\left[\varphi_{\theta_{1,t}}\left\{u_j\right\} + \varphi_{\theta_{1,t}}\left\{C_{X_t}(u_1, \ldots, u_p; \theta_{2,t})\right\}\right]$ with $(u_j, u_1, \ldots, u_p)^{\top} \in [0,1]^{p+1}$ is the proper copula function. Simplification of (\ref{eq:m_covar_est}) yields:
\begin{equation}\label{eq:mcovar_analytic}
    MCoVaR^{j|\mathbf{i}}_{\alpha, \beta, t} = F^{-1}_{X_{j,t}}\left(\varphi_{\theta_{1,t}}^{-1}\left[ \varphi_{\theta_{1,t}}\{C_{X_{t}}(\bm{\alpha};\theta_{2,t}) \beta\} - \varphi_{\theta_{1,t}}\{C_{X_t}(\bm{\alpha};\theta_{2,t})\}\right]\right).
\end{equation}
Special cases are where $C_{X_{j,t}, X_t}$ is an AC or a Hierarchical Archimedean copula (HAC), see \cite{okhrin2013structure}. In the former case, (\ref{eq:mcovar_analytic}) transforms to: $$ MCoVaR^{j|\mathbf{i}}_{\alpha, \beta, t} = F^{-1}_{X_{j,t}}\left\{\varphi_{\theta_{1,t}}^{-1}\left( \varphi_{\theta_{1,t}}\left[\varphi_{\theta_{1,t}}^{-1}\left\{p \varphi_{\theta_{1,t}}(\alpha)\right\} \beta\right] - p\varphi_{\theta_{1,t}}(\alpha)\right)\right\},$$ where the dependence intensity is expressed via the AC parameter $\theta_{1,t}$. Furthermore, from (\ref{eq:m_covar_est}) follows that $MCoVaR^{j|\mathbf{i}}_{\alpha, \beta, t} = CoVaR^{j|i}_{\alpha, \beta, t}$ for $p = 1$, as (\ref{eq:m_covar_est}) reduces to (\ref{eq:covar_est_cop}).

\subsubsection{VCoVaR Estimation}
In the following, the copula-based representation of the VCoVaR of (\ref{eq:v_covar_def}) is derived. Set $\bm{1-\alpha} = (1-\alpha, \ldots, 1-\alpha)^{\top}$ of length $p$.
\begin{lemma}\label{lemma:vcovar_est}
The VCoVaR defined in (\ref{eq:v_covar_def}) is equivalent to: 
\begin{equation}\label{eq:v_covar_est}
    \frac{F_{X_{j,t}}(VCoVaR^{j|\mathbf{i}}_{\alpha, \beta, t}) -
    \bar C_{X_{t}}(\bm{1-\alpha}; \theta_{2,t}) + \bar C_{X_{j,t}, X_{t}}\{1-F_{X_{j,t}}(VCoVaR^{j|\mathbf{i}}_{\alpha, \beta, t}), \mathbf{1-\bm{\alpha}}; \theta_{1,t}\}}{1 - \bar C_{X_{t}}(\bm{1-\alpha}; \theta_{2,t})} = \beta,
\end{equation}
where $\bar C_{X_{j,t}, X_{t}}$ denotes the $(p+1)$-dimensional survival copula associated with $(X_{j,t}, X_{t})$ and parameter $\theta_{1,t}$ and $\bar C_{X_{t}}$ denotes the $p$-dimensional survival copula of $X_{t}$, characterized by a parameter $\theta_{2,t}$.
\end{lemma}
Proofs of all lemmata are provided in Appendix A. The VCoVaR is estimated by solving (\ref{eq:v_covar_est}). The estimation approach includes survival functions, which can - analogously to the case of distribution functions - be decomposed using survival copulae, see \cite{georges2001multivariate}. A crucial characteristic is that the survival copula of a pair of random variables is the 180 degrees rotated version of its copula, similar holds for any dimension. In consequence, there exists a direct relationship between $C$ and $\bar C$. This allows estimation of the involved survival copulae in practice as follows: first, estimate the copula $C_{X_{j,t},X_{t}}$ from the data and rotate it to find $\bar C_{X_{j,t}, X_{t}}$. Second, extract $\bar C_{X_{t}}$ through marginalization, setting $\bar C_{X_{t}}(\bm{1-\alpha}; \theta_{2,t}) = \bar C_{X_{j,t}, X_{t}}(1, \bm{1-\alpha}; \theta_{1,t})$. The essential point is again the sufficiency of having an estimate of the (survival) copulae and the marginal distribution $F_{X_{j,t}}$. Additionally, the considered risk measures in (\ref{eq:covar_est_cop}), (\ref{eq:m_covar_est}), and (\ref{eq:v_covar_est}) are continuous transformations of these estimators. Therefore all asymptotic distributional properties of the risk measures are directly determined by application of the delta method, see \cite{oehlert1992note}.

The following Lemma shows the equivalence between the copula-based representations of CoVaR and VCoVaR if only one conditional variable is considered. 
\begin{lemma}\label{lemma:vcovar_p_1}
Given (\ref{eq:covar_est_cop}) and (\ref{eq:v_covar_est}), it holds that $VCoVaR^{j|\mathbf{i}}_{\alpha, \beta, t} = CoVaR^{j|i}_{\alpha, \beta, t}$ if $p = 1$.
\end{lemma}

\section{Properties of Systemic Risk Measures}\label{sec:properties_rm}
\subsection{Independence and Perfect Dependence}
\label{subsec:covar_certain_dep}
Let us establish a connection between the CoVaR measures introduced previously if all variables are independent or perfectly positive dependent. The case of perfect negative dependence is not considered, as countermonotonicity in higher dimensions is problematic.

\begin{lemma}\label{lemma:independence}
    Let $X_{j,t}, X_{1,t}, \ldots, X_{p,t}$ be independent, given expressions (\ref{eq:covar_est_cop}), (\ref{eq:m_covar_est}), and (\ref{eq:v_covar_est}), it holds that $CoVaR^{j|i}_{\alpha, \beta, t} = MCoVaR^{j|\mathbf{i}}_{\alpha, \beta, t} = VCoVaR^{j|\mathbf{i}}_{\alpha, \beta, t} = F_{X_{j,t}}^{-1}(\beta)$.
\end{lemma}

This observation is reasonable, as in the case of independence, the conditional probabilities of (\ref{eq:GE_covar_def}), (\ref{eq:m_covar_def}) and (\ref{eq:v_covar_def}) equal the respective unconditional probability, which results in the VaR at level $\beta$. Transferred to the market of CCs, in the case of independence, an extreme event for a crypto return under condition is completely irrelevant for the crypto return $X_{j,t}$.
\begin{lemma}\label{lemma:perfect_dependence}
    Let $X_{j,t}, X_{1,t}, \ldots, X_{p,t}$ be perfectly positive dependent, given expressions (\ref{eq:covar_est_cop}), (\ref{eq:m_covar_est}), and (\ref{eq:v_covar_est}), it holds that: $CoVaR^{j|i}_{\alpha, \beta, t} = MCoVaR^{j|\mathbf{i}}_{\alpha, \beta, t} = VCoVaR^{j|\mathbf{i}}_{\alpha, \beta, t} = F_{X_{j,t}}^{-1}(\alpha\beta)$.
\end{lemma}
All conditional measures equal the VaR at level $\alpha\beta$ in the given scenario, being directly influenced by the VaR level of the conditional variables, which means that in the case of perfect positive dependence, it is sufficient to consider one conditional crypto return, as additional CCs would not generate any additional information.

\subsection{General Positive Dependencies}\label{subsec:sim_analysis}
We want to gain further understanding of the measures based on (\ref{eq:covar_est_cop}), (\ref{eq:m_covar_est}), and (\ref{eq:v_covar_est}). Two major objectives are pursued. First, to detect the general behaviour of the measures as the function of the dependence parameter for a given copula. Second, to investigate differences between different copula families. This is realized by solving (\ref{eq:covar_est_cop}), (\ref{eq:m_covar_est}), and (\ref{eq:v_covar_est}) for a range of copula parameters. The marginal distribution $F_{X_{j,t}}$ is set to be standard normal.

For the bivariate CoVaR of (\ref{eq:covar_est_cop}), the Gaussian, $t$, Clayton, and Gumbel copulae are chosen, while we consider two approaches to analyze the MCoVaR of (\ref{eq:m_covar_est}) and the VCoVaR of (\ref{eq:v_covar_est}). Generally, we set $p = 2$. First, the copula $C_{X_{j,t}, X_{1,t}, X_{2,t}}(u_1, u_2, u_3)$ is assumed to be Clayton or Gumbel, thus belonging to the AC family. We excluded the Gaussian and $t$-copula as they require more correlation parameters in higher dimensions, and we want to start the analysis by varying one dependence parameter at a time. The copula of $(X_{1,t}, X_{2,t})$ is attained by marginalization: $C_{X_{1,t}, X_{2,t}}(u_2, u_3) = C_{X_{j,t}, X_{1,t}, X_{2,t}}(1, u_2, u_3)$. As a consequence, the measure depends only on one parameter for the Clayton and Gumbel copula and can be visualized comparable to the bivariate CoVaR case. This allows interpreting how the measure changes if the dependence of $(X_{j,t}, X_{1,t}, X_{2,t})$ as a whole changes. Second, the copula of $(X_{j,t}, X_{1,t}, X_{2,t})$ is assumed to be a HAC. The idea of a HAC is to nest AC in a hierarchical structure to allow for a more flexible specification of the dependence structure, as the property of AC of having one parameter materializes in practice often as a limitation. The following structure is imposed: $C_{X_{j,t}, X_{1,t}, X_{2,t}}(u_1, u_2, u_3) = C_1\{u_1, C_2(u_2,u_3)\}.$ The bivariate copula $C_2$ describes the dependency inside the conditional variables, while $C_1$ describes the dependency between the target variable and the conditional variables. Notice that for the VCoVaR each copula is rotated, as the measure is based on survival copulae.

For comparability, each copula parameter is converted into Kendall's $\tau$ \citep{joe2014dependence}. As in Section \ref{subsec:covar_certain_dep}, only positive dependencies with Kendall's $\tau \in [0,1]$ are considered, although $\tau$ does not range over the whole domain in some cases as numerical issues at the limits appeared. Note that $\tau < 0$ results in negative pair-dependencies, which is controversial in dimensions $d > 2$. The computations in the following are realized using the R-packages \verb+copula+ \citep{copulaPackage} and \verb+HAC+ \citep{HAC_JSS}.

\subsubsection{CoVaR Properties}
\begin{figure}[ht]
    \centering
    \includegraphics[width = 15cm]{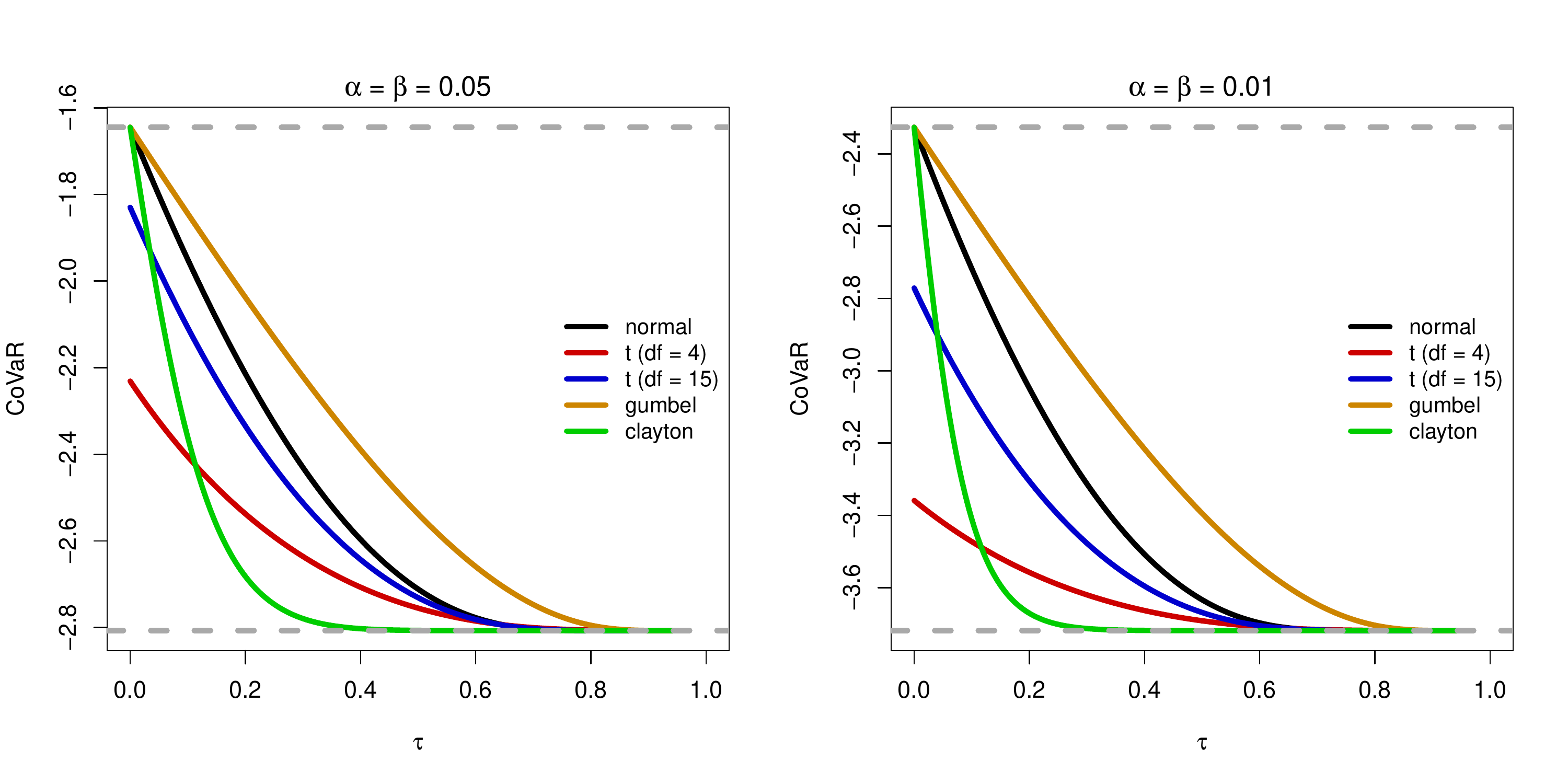}
    \caption[Bivariate CoVaR for different copulae and standard normal margins]{Bivariate CoVaR for different copulae and standard normal margins. The upper grey line corresponds to the independence case $F_{X_{j,t}}^{-1}(\beta)$ and the lower one to perfect positive dependence $F_{X_{j,t}}^{-1}(\alpha \beta)$, as derived in Section \ref{subsec:covar_certain_dep}.}
    \label{plot:sim_biv_covar}
\end{figure}

Starting with the bivariate CoVaR, Figure \ref{plot:sim_biv_covar} shows the measure depending on the selected copula and Kendall's $\tau$. The probability levels $\alpha = \beta = 0.05$ and $\alpha = \beta = 0.01$ are considered. In general, the CoVaR decreases monotonically for all copulae if Kendall's $\tau$ increases. This is consistent with the findings of \cite{mainik2012dependence}. For example, this implies the stronger LTC depends on XRP, the stronger a distressing event of XRP will impact LTC. Furthermore, the results of Section \ref{subsec:covar_certain_dep} are special cases for those copulae where independence ($\tau = 0$) and perfect positive dependence ($\tau = 1$) are attained. Given the standard normal distribution for the margins, these are $F_{X_{j,t}}^{-1}(\beta) \approx -1.645$ and $F_{X_{j,t}}^{-1}(\alpha\beta) \approx -2.807$, respectively, for $\alpha = \beta = 0.05$. The CoVaR converges for all copulae towards these theoretical limits, except for the $t$-copula if $\tau \rightarrow 0$. This is reasonable as $\tau = 0$ restricts the correlation of the $t$-copula to be 0, but does not affect the degrees of freedom $\nu$. However, the $t$-copula converges towards the Gaussian one if $\nu \rightarrow \infty$ \citep{eling2009modeling}. This is reflected in Figure \ref{plot:sim_biv_covar}, as with increasing $\nu$ the curve of the $t$-copula comes closer to the one of the Gaussian copula. The curve of the Clayton copula decreases very fast, while the curves of the Gumbel and Gaussian copulae decrease slowly. Thus, the Clayton copula leads to more conservative estimates of the CoVaR and should produce in an application fewer exceedances than the other copulae.

\subsubsection{MCoVaR and VCoVaR Properties}
\begin{figure}[ht]
    \centering
    \includegraphics[width = 15cm]{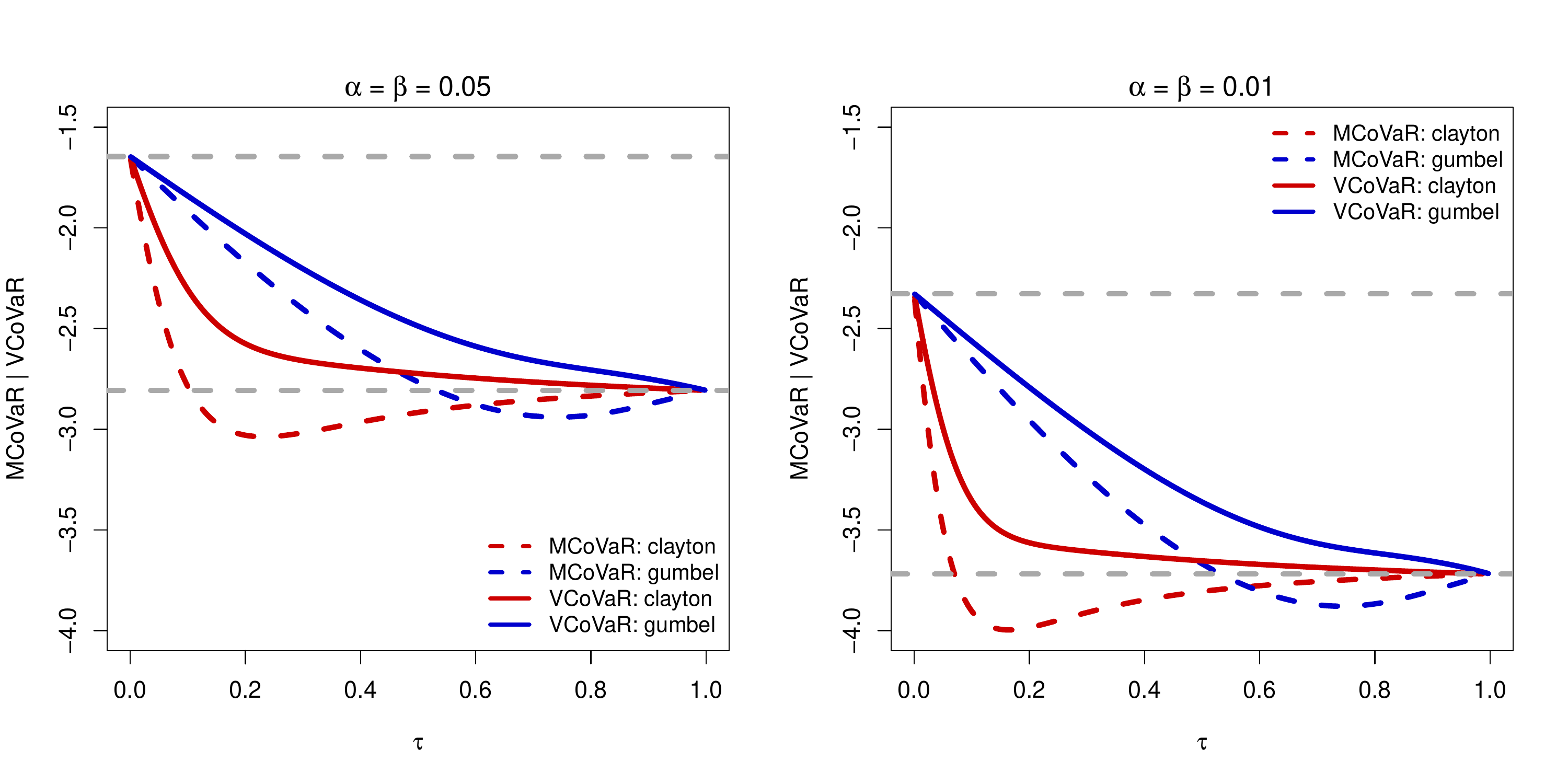}
    \caption{MCoVaR and VCoVaR assuming Archimedean copulae for $C_{X_{j,t}, X_{1,t}, X_{2,t}}$. The upper grey lines correspond to the independence case $F_{X_{j,t}}^{-1}(\beta)$ and the lower ones to perfect positive dependency $F_{X_{j,t}}^{-1}(\alpha \beta)$, as derived in Section \ref{subsec:covar_certain_dep}.}
    \label{plot:mcovar_vcovar_2d}
\end{figure}

Figure \ref{plot:mcovar_vcovar_2d} shows the results of the first approach with three-dimensional AC. The figure also contains the theoretical limits of Section \ref{subsec:covar_certain_dep}, and both the MCoVaR and the VCoVaR converge towards them if $\tau \rightarrow 0$ or $\tau \rightarrow 1$. This is the case for both considered probability levels. However, the MCoVaR is not a monotonically decreasing function of Kendall's $\tau$. For the Clayton copula, the MCoVaR achieves its minimum for $\tau \approx 0.2$, while for the Gumbel copula it is around $\tau \approx 0.75$. Thus, the MCoVaR measure of (\ref{eq:m_covar_est}) does not reflect the behaviour of the bivariate CoVaR for the given copula specification. Nevertheless, it is reasonable to detect lower values of the MCoVaR in comparison to the CoVaR, as the conditioning event describes a worse market situation. In contrast to the MCoVaR, the VCoVaR is a monotonically decreasing function of Kendall's $\tau$. Thus the measure decreases if the dependency of $(X_{j,t}, X_{1,t}, X_{2,t})$ as a whole intensifies. Furthermore, the curve of the Clayton copula again decreases faster than the one of the Gumbel copula. Transferring this to practice, we can expect the Clayton copula to lead to more conservative estimates for the VCoVaR in the empirical part.

\begin{figure}[ht]
\begin{tabular}{cc}
  \includegraphics[width=74mm]{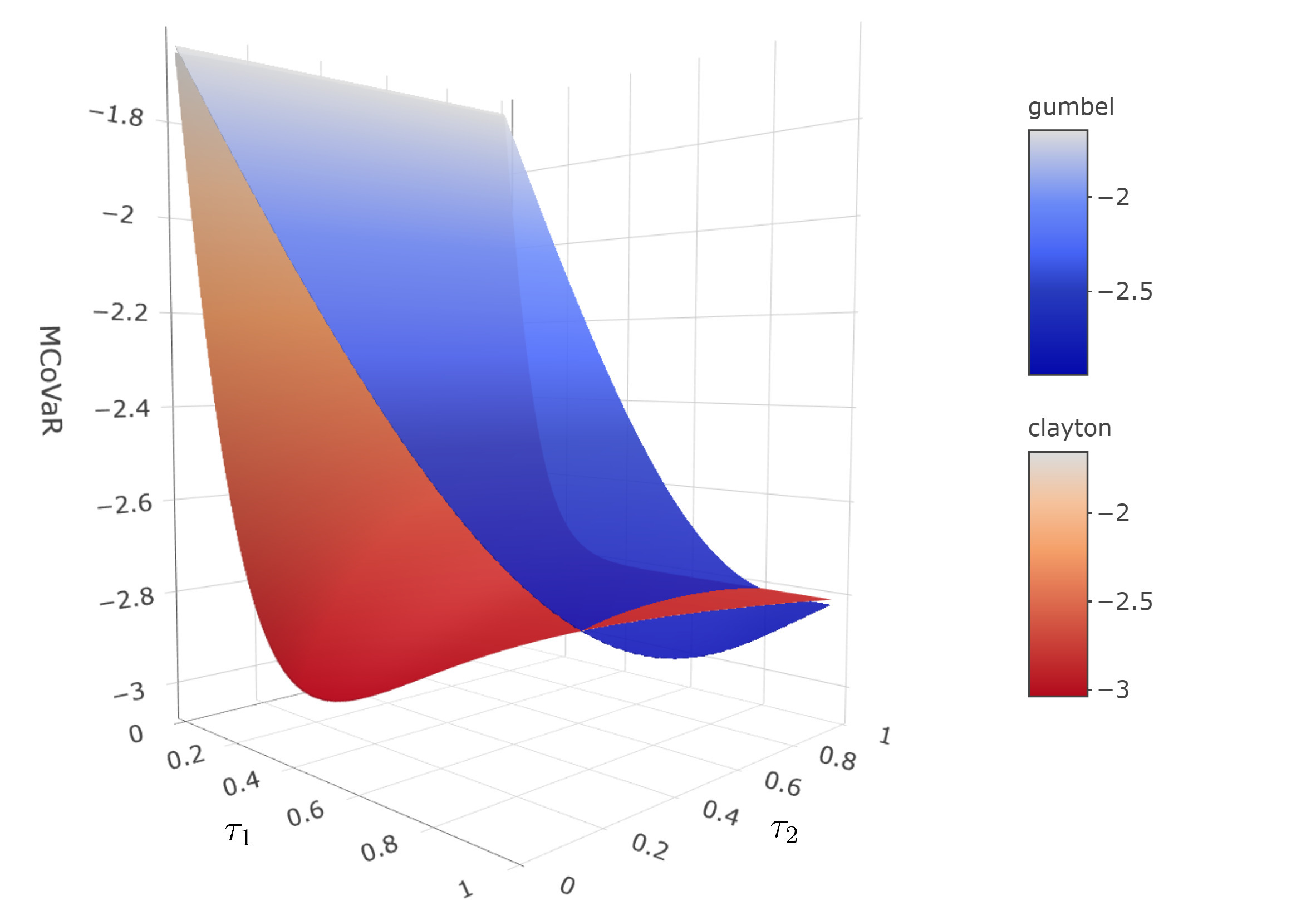} &   \includegraphics[width=74mm]{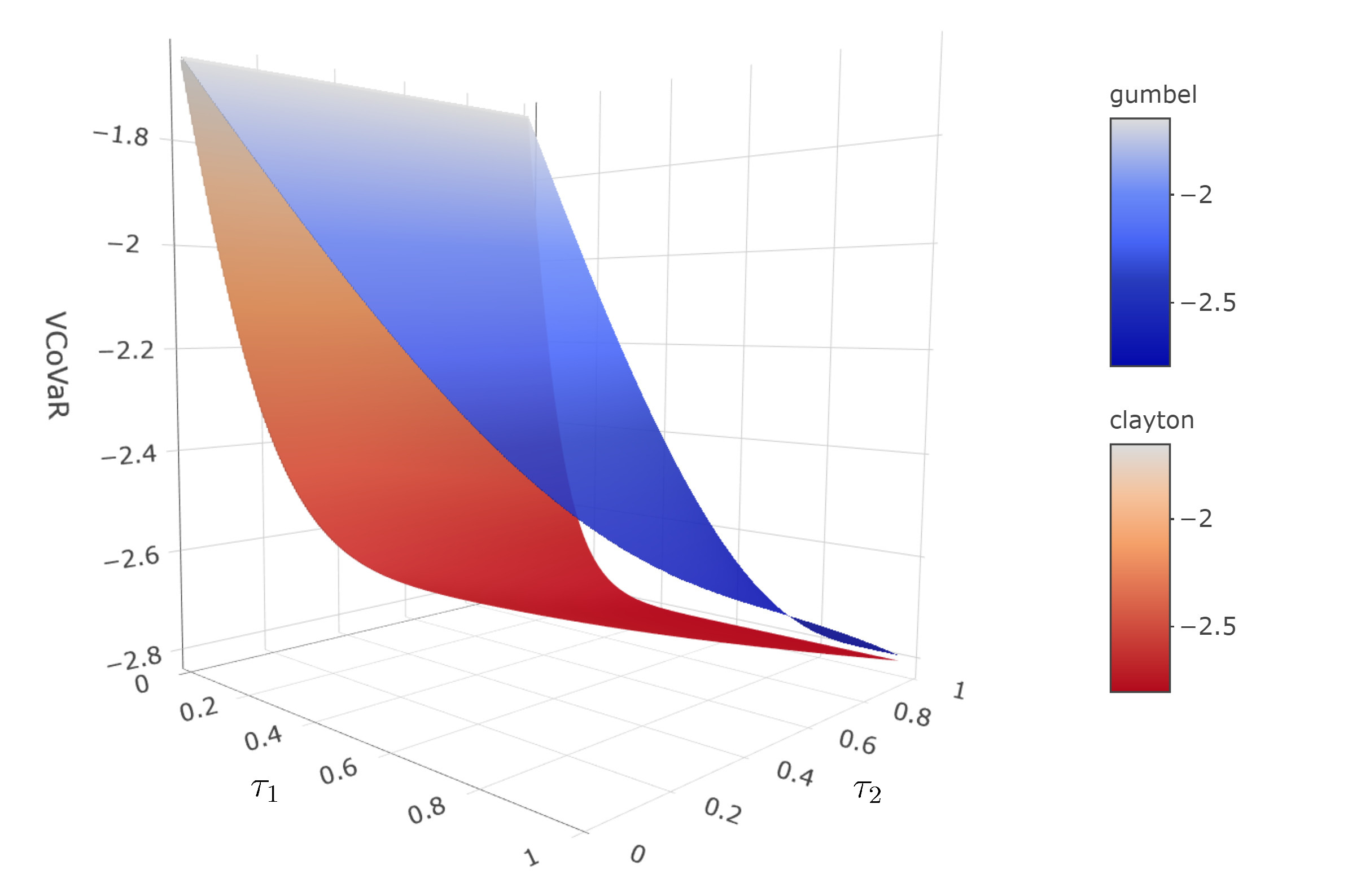} \\
(a) Multi-CoVaR & (b) Vulnerability-CoVaR \\[6pt]
\end{tabular}
\caption{MCoVaR and VCoVaR assuming Hierarchical Archimedean copulae with $C_{X_{j,t}, X_{1,t}, X_{2,t}}(u_1, u_2, u_3) = C_1\{u_1, C_2(u_2,u_3)\}$. $\tau_1$ refers to $C_1$ and $\tau_2$ to $C_2$. $\alpha = \beta = 0.05$.}
\label{plot:3D_surfaces}
\end{figure}

The second approach using the HAC allows decomposing these characteristics even further. Figure \ref{plot:3D_surfaces} shows the surfaces for Clayton and Gumbel generators if a HAC is assumed. Regarding notation, the figure contains $\tau_1$ and $\tau_2$, referring to $C_1$ and $C_2$, respectively. Note that a HAC is required to fulfil the sufficient nesting condition in order to be a proper copula \citep{okhrin2013structure}. This condition corresponds to $\tau_1 \leq \tau_2$ for considered copulae, the area in the surface plots violating this condition remains empty. The edges of the surfaces at $\tau_1 = \tau_2$ in Figure \ref{plot:3D_surfaces} equal the corresponding curves in Figure \ref{plot:mcovar_vcovar_2d}, as the HAC is the equivalent with the AC of the first approach. 

The MCoVaR decreases for both copulae if the dependency between the target variable and the conditional variables, represented through $\tau_1$, intensifies. For example, if BTC is the variable of interest and XRP and XMR are under condition, the impact of joint distresses of XRP and XMR on BTC increases if BTC has a larger dependency with XRP and XMR. This observation is reasonable and justifies the application of the MCoVaR. However, this mainly reflects the dependence consistency of the bivariate CoVaR. Given a fixed $\tau_2$, the value $C_2(u_2, u_3)$ can be calculated with $u_2, u_3$ set to $\alpha$, as seen from (\ref{eq:m_covar_est}). In consequence, the analysis with varying $\tau_1$ becomes analogous to a bivariate CoVaR analysis with an $\alpha$-level adjusted by the dependency between $X_{1,t}$ and $X_{2,t}$. If on the other side $\tau_1$ is fixed and $\tau_2$ increases, the MCoVaR is much less affected and tends to increase. Regarding our example with BTC, XMR, and XRP, this could potentially be interpreted as follows: if the conditional CCs XMR and XRP only have a weak dependence, the conditional event in (\ref{eq:m_covar_def}) will be unlikely. If such a joint event still happens, the market situation will be devastating and the MCoVaR needs to be very small. If in the other case, XRP and XMR depend highly on each other, distress of XRP suggests distress of XMR and vice versa. The fulfilment of the condition is a consequence of the dependency between XRP and XMR and not of the situation of the overall market. Thus, the values of the MCoVaR are slightly higher. However, the overall impact of $\tau_1$ on the MCoVaR is of primary importance and determines the general behaviour of the measure. Differentiating between the two considered copulae shows the Clayton copula decreases faster in $\tau_1$, although there is an overlapping of the two surfaces in Figure \ref{plot:3D_surfaces}(a) for high $\tau_1, \tau_2$.

The VCoVaR using the HAC structure in Figure \ref{plot:3D_surfaces}(b) decreases if $\tau_1$ increases. This observation confirms that the VCoVaR as calculated in (\ref{eq:v_covar_est}), is a reasonable extension of the bivariate CoVaR measure. In contrast to the MCoVaR, the VCoVaR decreases if the dependency between the conditional variables, expressed through $\tau_2$, increases. This could be explained via  different conditional events. Consider the example of BTC with XRP and XMR under condition again. If the dependency between XRP and XMR increases, the probability that the conditional event of the VCoVaR in (\ref{eq:v_covar_def}) yields a bad scenario, namely both CCs are in distress, increases. Consequently, the VCoVaR needs to decrease to capture these potentially worse situations. However, the impact of $\tau_2$ on the VCoVaR is less pronounced than the one of $\tau_1$. Finally, the Clayton copula produces again smaller VCoVaR values than the Gumbel copula.

\section{Simulation Study}\label{sec:SimStudy}
A simulation study is performed to test whether (\ref{eq:covar_est_cop}), (\ref{eq:m_covar_est}), and (\ref{eq:v_covar_est}) can reliably calculate the respective measure if the true copula is known, and no temporal dependency is present. The study is performed as follows: \emph{Step 1:} Assume $(X_{j,t}, X_{i,t})$ for the CoVaR or $(X_{j,t}, X_{1,t}, X_{2,t})$ for the M- and VCoVaR follow a certain copula with dependence parameter $\tau$. \emph{Step 2:} Sample $n = 10\,000$ iid observations from the copula with margins being $\mathcal{U}[0,1]$ and estimate the assumed copula via maximum likelihood (ML). \emph{Step 3:} Compute the respective conditional measure through (\ref{eq:covar_est_cop}), (\ref{eq:m_covar_est}), and (\ref{eq:v_covar_est}), and the VaR of the conditional variables as the empirical $\alpha$-quantile. \emph{Step 4:} Compute the violation rate of the respective measure using the sample equivalents of $\beta$ from the definitions in (\ref{eq:GE_covar_def}), (\ref{eq:m_covar_def}), and (\ref{eq:v_covar_def}), respectively. \emph{Step 5:} Repeat Steps 2 to 4 for $N = 100$ times and calculate the average violation rate. 

\input{tables/SimStudy}

The Gumbel and Clayton copulae are analyzed with Kendall's $\tau \in \{0.25, 0.50, 0.75\}$. The sample equivalents of $\beta$ in Step 4 are calculated by considering only the simulated observations which fulfill the conditional event. Of those observations, the number of violations of the respective CoVaR measure is computed. The equivalent of $\beta$ is then the ratio between the latter and the number of observations fulfilling the conditional event, which were considered in the first stage. This is equivalent to the procedure of \cite{mainik2012dependence}, in which the bivariate CoVaR was analyzed. Table \ref{table:sim_study} shows that for both selected probability levels, both copulae, and all values of Kendall's $\tau$, the violation rates are close to nominal level $\beta$. Concluding this simulation study, the copula-based (\ref{eq:covar_est_cop}), (\ref{eq:m_covar_est}), and (\ref{eq:v_covar_est}) are able to reliably compute the respective measure if the correct copula is assumed for the data-generating process.

\section{Empirical Study}\label{sec:crypto}
\subsection{In-sample Estimation}\label{subsec:crypto_in_sample}
\subsubsection{Proceeding and Data Investigation}
\begin{figure}[ht]
    \centering
    \includegraphics[width = 16cm]{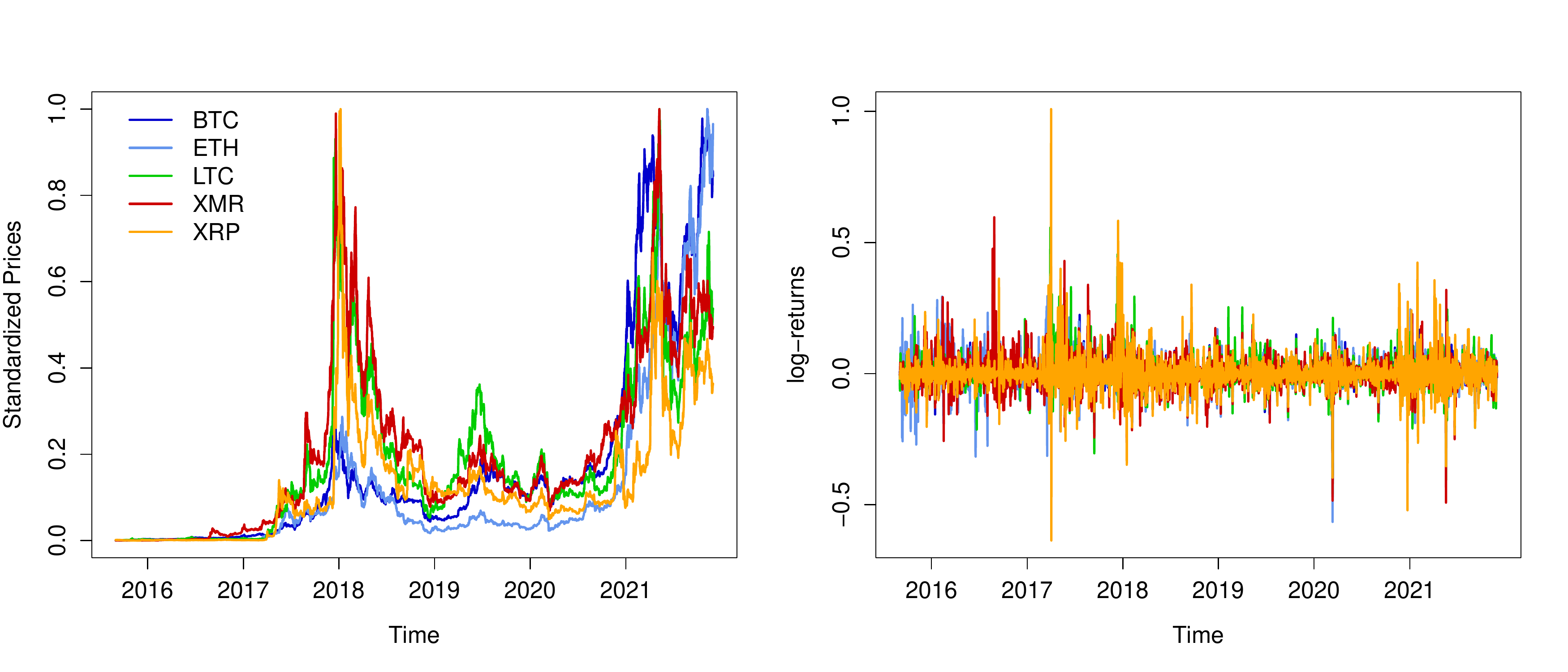}
    \caption[Standardized prices and log-returns of the cryptocurrencies]{Standardized prices and log-returns of the cryptocurrencies.}
    \label{plot:data_stock}
\end{figure}

The study uses five CCs BTC, ETH, LTC, XMR, and XRP. These five CCs constitute 64.54\% (\href{https://coinmarketcap.com/}{\url{https://coinmarketcap.com/}}, accessed: 07/12/2021) of the overall CC market capitalization and offer  relatively long time series compared to other CCs, thus providing a sufficient database. The data contains daily closing prices in USD stemming from the Community Network Data kindly provided by CoinMetrics (\href{https://coinmetrics.io/}{\url{https://coinmetrics.io/}}, accessed: 01/12/2021). The sample includes $n = 2\,283$ observations from 01/09/2015 to 30/11/2021 as CCs are traded every day, including weekends. For the analysis, the prices are transformed in log-returns. We calculate: (1) The bivariate CoVaR for all possible combinations of the five CC; (2) The SCoVaR, MCoVaR, and VCoVaR of each CC if the remaining four CCs are treated as conditional variables. We fix: $\alpha = \beta = 0.05$. Figure \ref{plot:data_stock} shows the prices (scaled to [0,1]) and the log-returns. Descriptive statistics, tests, and estimates of Kendall's $\tau$ for the log-returns are given in Appendix B.

To estimate the different measures of (\ref{eq:GE_covar_def}), (\ref{eq:s_covar_def}), (\ref{eq:m_covar_def}), and (\ref{eq:v_covar_def}), the following empirical procedure is used: \emph{Step 1:} Estimate the respective marginal model for each time series separately. \emph{Step 2:} Perform the parametric probability integral transformation on residuals to obtain iid $\mathcal{U}[0,1]$ data. \emph{Step 3:} Estimate the copula based on the resulting pseudo-sample observations. \emph{Step 4:} Solve (\ref{eq:covar_est_cop}), (\ref{eq:m_covar_est}), or (\ref{eq:v_covar_est}), respectively.

\subsubsection{Univariate Models}\label{subsubsec:univariate_models}
Empirical results from the literature indicate that CC returns exhibit characteristics as volatility clustering, fat tails, and leverage effects, see \cite{zhang2018some} and \cite{phillip2018new}. To account for these dynamics and building on the stationarity assumption, the margins are assumed to follow an autoregressive moving-average (ARMA) model for the conditional mean $\mu_t$ and a GJR-GARCH model \citep{glosten1993relation} for the conditional variance $\sigma_{t}^{2}$. For example, when $X_{j,t}$ denotes the log-return of a CC at time $t$, the full ARMA($p_l$, $q_l$)-GJR-GARCH(P,Q) model can be outlined as follows \citep{rugarch}:
\begingroup
\allowdisplaybreaks
\begin{align}
    &X_{j,t} = \mu_t + \varepsilon_t, \quad \mu_t = \mu + \sum_{i = 1}^{p_l} \phi_{i} (X_{j, t-i}-\mu) + \sum_{j = 1}^{q_l} \psi_j \varepsilon_{t-j}, \quad \varepsilon_t = \sigma_{t} z_t, \nonumber\\
    &\sigma_{t}^{2} = \omega + \sum_{i = 1}^{P} (\lambda_{i} + \gamma_i \mathbf{1}_{\{\varepsilon_{t-i} \leq 0\}} )\varepsilon_{t-i}^{2} + \sum_{j = 1}^{Q} \delta_j \sigma_{t-j}^{2},\label{eq:gjr_garch}
\end{align}
\endgroup

with iid  $z_t \sim F_{z}(0,1)$ and $\mathbf{1}_{\{\cdot\}}$ being an indicator function. This results in the ability of the GJR-GARCH specification to model positive and negative shocks, represented by $\varepsilon_t$, differently and accounts for the leverage effect \citep{rugarch}. For $F_{z}$, the skew-$t$ distribution with skewness $\zeta$ and shape $\nu$ of \cite{fernandez1998bayesian} is selected, as the time series exhibited a strong indication of non-normality and skewness. For parameter constraints and additional remarks on the model, see \cite{glosten1993relation}. In the context of this application, the ARMA-GJR-GARCH model acts as a filter for temporal dependencies inside the time series and the empirical counterparts of $z_t$ are extracted for further modeling. These standardized residuals $\hat z_t$ should be as serially independent as possible. In addition, it might be unnecessary to specify the model as presented above, and a more parsimonious version would be sufficient to capture the dynamics of the data. Considering these facts, we use a semi-automated process which selects the best fitting model according to an information criterion while fulfilling necessary requirements on the residuals, for details see Appendix B.

Building on the selected univariate models, the VaR for each time series at level $\alpha$ can be calculated parametrically as described in a forecasting context in \cite{kuester2006value}:
\begin{equation}\label{eq:VaR_GARCH}
    \widehat{VaR}_{\alpha, t} = \hat\mu_t + \hat\sigma_t F_{z}^{-1}(\alpha| \hat\zeta, \hat \nu),
\end{equation}
where $\hat \mu_t, \hat \sigma_t, \hat\zeta, \hat \nu$ are the estimates from the fitted univariate model. The generated in-sample VaR estimates are necessary for evaluation and comparison with the conditional measures. To validate the VaR estimates, Table \ref{table:VaR_rates} shows the realized equivalents of $\alpha$ in (\ref{def:var}) and the absolute number of observations, in which the respective log-return was equal or below the VaR estimate. The rates are close to $\alpha = 0.05$ in all cases, indicating the accuracy of the models.
\input{tables/VaR_rates}

\subsubsection{Copula Models}\label{subsubsec:copula_models}
With the univariate models being estimated, the standardized residuals $\hat z_t$ are parametrically transformed to $\mathcal{U}[0,1]$, denoted $u_{i,t} = F_{z}(\hat z_{i,t}| \hat \zeta_{i}, \hat v_{i})$ with $i \in \{1,\ldots, 5\}$ for BTC, ETH, LTC, XMR, and XRP. Based on these pseudo-observations, the copulae are estimated. The following time-invariant copula models are chosen for the application: Gaussian, $t$, Clayton, and Gumbel. Time-invariance relates to having the same copula parameter at every point of time $t$. In this case, the measures in (\ref{eq:covar_est_cop}), (\ref{eq:m_covar_est}), and (\ref{eq:v_covar_est}) become time-variant only through the dynamic nature of the univariate model of BTC, while the dependencies are assumed to be constant. However, it might be beneficial to investigate time-variant copulae to capture the potential dynamics of the dependencies, which is why we incorporate the dynamic model of \cite{patton2006modelling} and the DCC-copula approach of \cite{jin2009large}. Detailed descriptions of these models alongside resulting parameter estimates of all copulae can be found in Appendix B.

\subsubsection{Estimates of the Systemic Risk Measures}
\begin{figure}[htp]
    \centering
    \includegraphics[width=\textwidth]{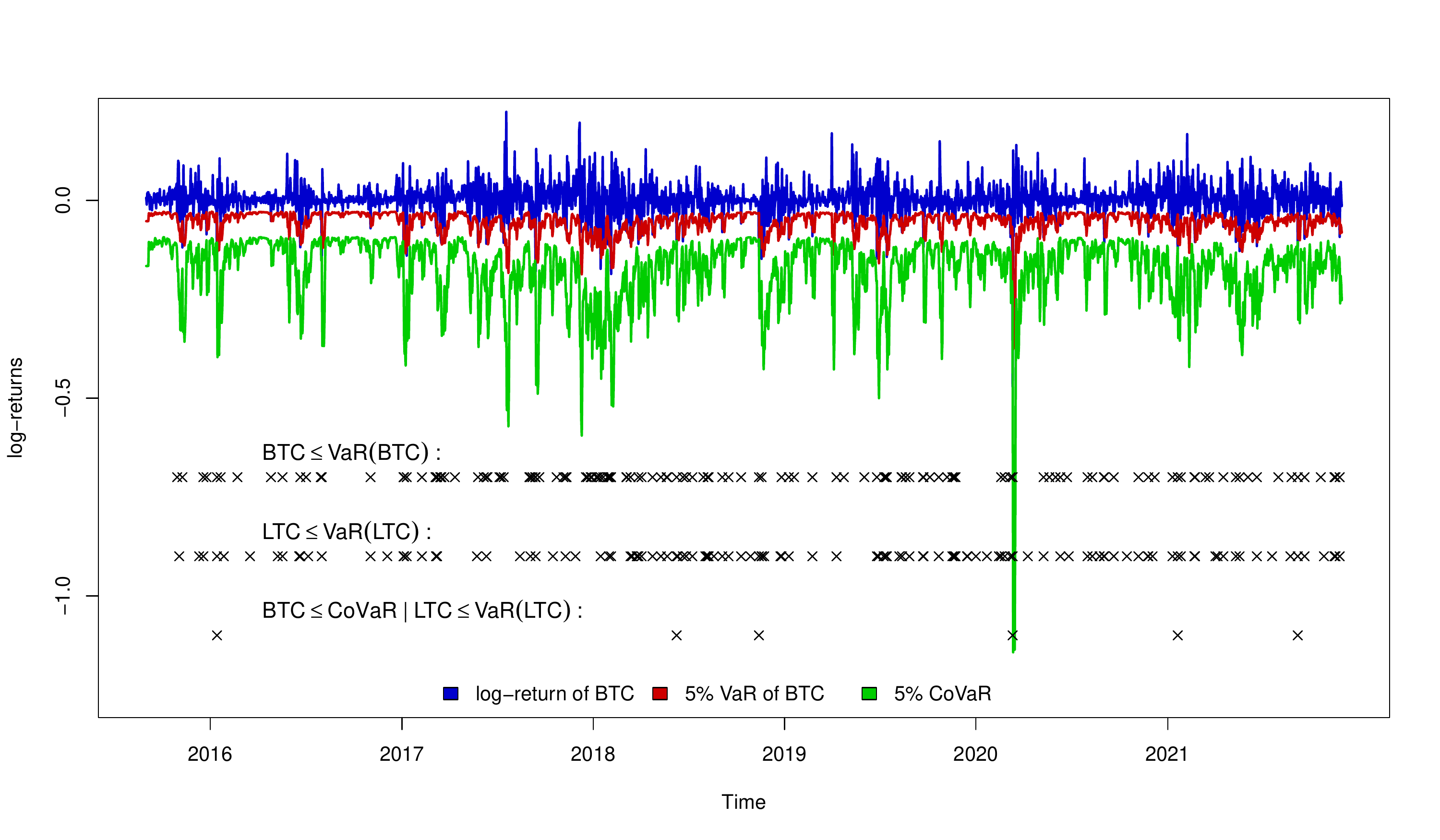}
    \caption{CoVaR of BTC with LTC under condition using a time-invariant $t$-copula. Realized violation rate: 0.0492.}
    \label{plot:crypto_BTC_LTC}
\end{figure}

\begin{figure}[htp]
    \centering
    \includegraphics[width=\textwidth]{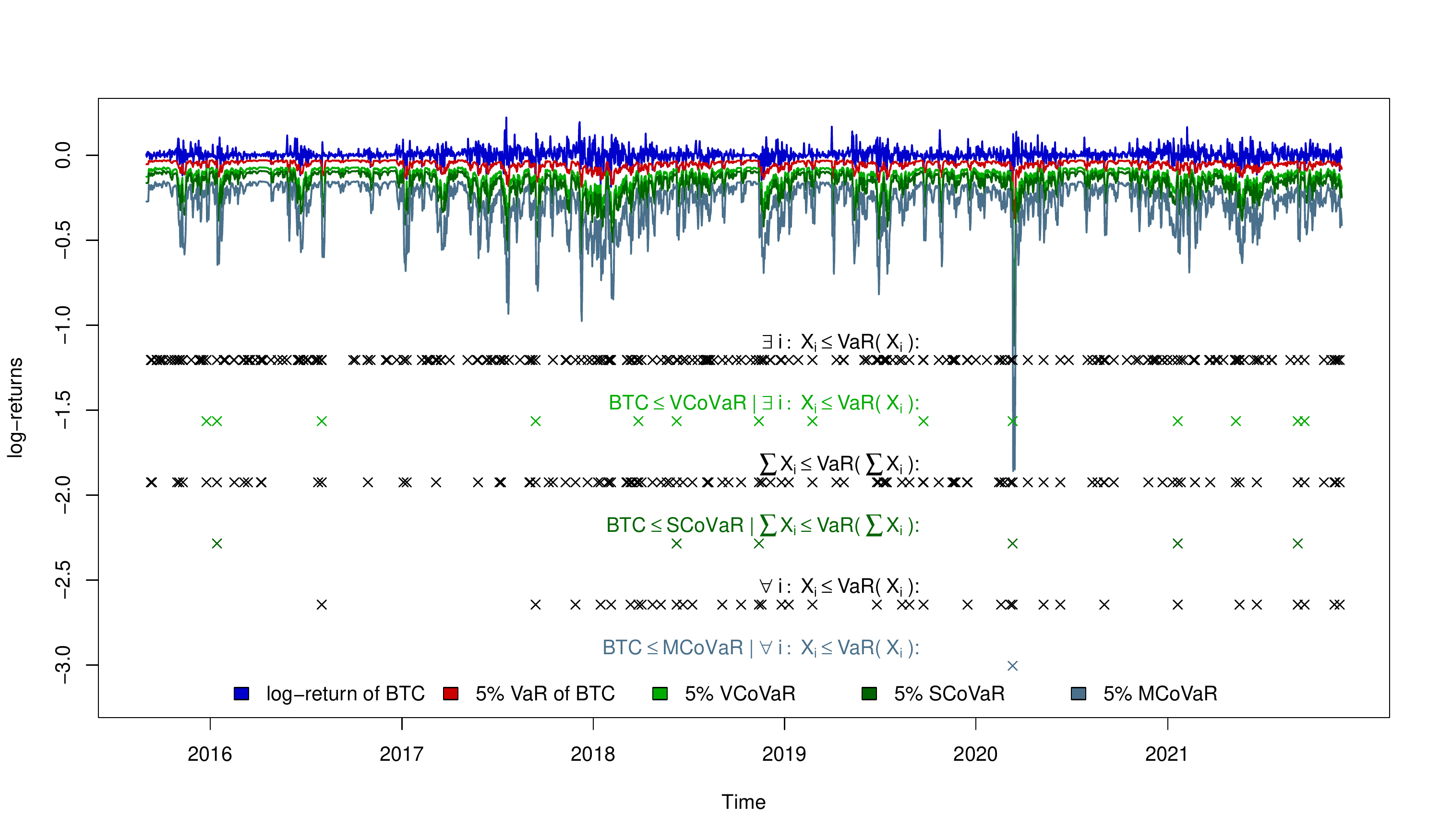}
    \caption{VCoVaR, SCoVaR, and MCoVaR of BTC with time-invariant $t$-copula.}
    \label{plot:crypto_multivariate_measures}
\end{figure}

Building on the estimated marginal distributions and copulae, (\ref{eq:covar_est_cop}), (\ref{eq:m_covar_est}), and (\ref{eq:v_covar_est}) are solved for the systemic risk measures. Table \ref{table:crypto_cop_rel_viol} shows descriptive statistics alongside violation rates, which are the sample equivalents of $\beta$ and computed similar to Section \ref{sec:SimStudy}. Figures \ref{plot:crypto_BTC_LTC} and \ref{plot:crypto_multivariate_measures} illustrate selected measures when the log-return of BTC is $X_{j,t}$. Appendix B contains further plots.

\input{tables/Violation_rates_all_copulae}

This leads to the following findings:
\begin{enumerate}
    \item On average, all conditional measures are below the respective univariate VaR, reflecting the positive dependencies in the crypto-market. Moreover, the figures show that the conditional measures are driven by similar dynamics as the VaR, which is a consequence of the chosen static $t$-copula and the inverse margin operation (\ref{eq:VaR_GARCH}).
    \item According to average bivariate CoVaR, BTC and LTC primarily affect each other. Consequently, Figure \ref{plot:crypto_BTC_LTC} displays the CoVaR of BTC drastically below its VaR. This finding agrees with \cite{luu2019spillover} and \cite{xu2020tail}, who noticed BTC as a risk recipient of other CCs. Investors might keep that in mind when driving towards BTC. ETH responds similarly to isolated distressing events of BTC, LTC, XMR, and XRP. However, a notable impact has LTC on XRP since it leads to the lowest average CoVaR estimates for this CC.
    \item Generally, the average SCoVaR estimate for each CC is below the bivariate CoVaR estimates. The only exception being the pair BTC and LTC. This implies that knowing LTC is in distress appears to be worse for BTC than knowing the summation of ETH, LTC, XMR, and XRP is in a critical state.
    \item The MCoVaR yields the lowest average estimates for each CC, illustrating how strongly a joint distressing event of other major CCs can impact a particular currency. In addition, the MCoVaR has the highest variance among all considered measures.
    \item Finally, the VCoVaR is estimated below the VaR, but larger than all conditional measures. This means that knowing that at least one out of the other four CC is in distress appears less critical than knowing that exactly one is in distress without information on the other three. This reflects the nature of the conditional event and can be interpreted in two ways. First, from a methodological point of view, the conditional event of the VCoVaR includes more scenarios than all other considered measures. Consequently, more observations will fulfil it, see Figure \ref{plot:crypto_multivariate_measures}. To achieve a violation rate at level $\beta$ for this amount of observations, the VCoVaR estimate has to lie closer to the log-return of BTC than the other conditional measures. Second, from an economic point of view, the cases of the conditional event of the VCoVaR when one CC is in distress have the additional information that the other CCs are not in distress, which the bivariate CoVaR does not include. For example, knowing LTC is in distress is worse than knowing LTC is in distress and ETH, XMR, and XRP are not. This information advantage is decisive, as it gives positive information about the market. This was discussed theoretically in Section \ref{subsec:covar_def} and perfectly materializes in this application. Furthermore, the VCoVaR exhibits the lowest variance among all other conditional measures.
    \item All measure can be reliably estimated via copulae. However, the selected dependence model is crucial for the estimation quality. In general, the best copulae are the $t$-copula and the Clayton one. The time-variant \cite{patton2006modelling} and DCC models can reliably be used as well, but offer only in distinct cases advantages over the static $t$ and Clayton models. The worst performance yields the Gumbel copula since it cannot the model the lower tail dependence adequately. These observations are in line with the results of the simulations in Section \ref{subsec:sim_analysis}. However, the violation rates should be treated carefully due to the evaluation approach, which only considers the observations fulfilling the respective conditional event.
\end{enumerate}

\subsection{Out-of-sample Estimation}\label{subsec:crypto_out_of_sample}
To validate the estimation performance in an out-of-sample setting, we analyze the CCs based on a moving window approach. As in Section \ref{subsec:crypto_in_sample} holds: $\alpha = \beta = 0.05$, and we calculate the same systemic risk measures. The window size covers $w = 500$ observations, and a one-day-ahead forecasting strategy is pursued. The univariate model is always assumed to be a GJR-GARCH(1,1) model with $\mu = 0$ and skew-$t$ innovations. The CoVaR, SCoVaR, MCoVaR, and VCoVaR forecasts for $t+1$ are evaluated based on calculating the realized equivalents of $\beta$ in (\ref{eq:GE_covar_def}), (\ref{eq:s_covar_def}), (\ref{eq:m_covar_def}), and (\ref{eq:v_covar_def}) while using the forecasted VaR and the log-return of $t+1$. This is motivated by \cite{girardi2013systemic}, although this work focuses directly on violation rates instead of tests like \cite{kupiec1995techniques}. To perform the one-day-ahead VaR forecasting, (\ref{eq:VaR_GARCH}) is calculated using the forecasts of the conditional mean and conditional variance. The procedure for the conditional measures for each window is as follows: \emph{Step 1:} Fit a GJR-GARCH(1,1) model to each time series and transform the resulting standardized residuals parametrically to $\mathcal{U}[0,1]$. \emph{Step 2:} Based on these pseudo-observations, estimate the respective copula via ML. \emph{Step 3:} Forecast the measure by solving (\ref{eq:covar_est_cop}), (\ref{eq:m_covar_est}), and (\ref{eq:v_covar_est}) using the one-day-ahead forecast of the univariate model (\ref{eq:gjr_garch}) of the $X_{j,t}$ variable:
\begin{equation*}
\hat\sigma_{j,t+1}^{2} = \hat\omega_j + (\hat\alpha_{j} + \hat\gamma_j \mathbf{1}_{\{\hat\varepsilon_{j,t} \leq 0\}} )\hat\varepsilon_{j,t}^{2} + \hat\beta_j \hat\sigma_{j,t}^{2}.
\end{equation*}

\input{tables/OOS_violations}

The resulting VaR violation rates for the CC are with 0.0572 for BTC, 0.0544 for ETH, 0.0578 for LTC, 0.0494 for XMR, and 0.0651 for XRP all close to $\alpha = 0.05$. Further, the rates for the systems are similarly accurate: 0.0511 for Sys:BTC, 0.0522 for Sys:ETH, 0.0483 for Sys:LTC, 0.0522 for Sys:XMR, and 0.0511 for Sys:XRP. Table \ref{table:crypto_oos} shows the rates for the conditional measures while using the time-invariant Gaussian, $t$, Clayton, and Gumbel copulae. Especially for the Clayton model, the CoVaR, SCoVaR, MCoVaR, and VCoVaR forecasts are close to $\beta = 0.05$, although the $t$-copula is also a valid choice in most cases. Furthermore, the performance of the copulae relative to each other is as in the in-sample scenario.

\section{Conclusion}\label{sec:conclusion}
Quantifying systemic risk in the financial system is a key focus for regulators and risk management practitioners. Especially the rapidly-growing and highly volatile market of CC has attracted the attention of regulating authorities and researchers due to its potential impact on the status of the global financial system. An important methodological contribution is the systemic risk measure CoVaR, introduced by \cite{tobias2016covar}. \cite{cao2013multi} proposed an extension called Multi-CoVaR, while \cite{bernardi2019allocation} introduced a measure named System-CoVaR. Complementing these extensions, a new measure, the Vulnerability-CoVaR is proposed. The VCoVaR offers the improved ability to capture domino effects and is advantageous in a system where there exists at least one other CC which is facing an extreme risk scenario. Due to evidence of spillover effects, tail-dependence, and herding behaviour, such domino effects are an existing threat and source of systemic risk in the CC market. 

 The simulation-based analysis of dependence consistency displays the property of the VCoVaR to be a monotonically decreasing function of the dependence parameter for selected AC. The empirical analysis on the CC market showed that LTC displays the largest impact on BTC out of the selected currencies, and LTC further affects XRP relatively strong. Generally, each CC is significantly affected if an event of joint distress of the remaining currencies occurs, which can be considered symptomatic for systemic risk events in the CC market. Interestingly, a situation of at least on CC being in distress appears less critical than a specification of one concrete CC in the conditional event. This observation reflects exactly the nature of the conditional event of the VCoVaR. Regarding the estimation quality, significant differences between the considered copulae are detected. However, for the $t$ and Clayton copulae, the conditional measures are reliably estimated, as the in-sample and out-of-sample violation rates are approximately in line with the selected probability level.

Future work on the topic of CoVaR extensions could extend the analysis of dependence consistency to broader families of copulae and higher dimensional scenarios. Especially in the case of the newly proposed VCoVaR, theoretical analysis and simulations apart from the considered Archimedean copula models could be conducted. In addition, considered CoVaR extensions could be applied to different asset markets with varied sample sizes, to validate the estimation quality using more observations. It could also be analysed whether the VCoVaR could be reasonably extended to a difference-based measure comparable to the Delta-CoVaR of \cite{tobias2016covar}. 
%MY CHANGES END

\medskip
\begin{center}
{\large\bf ACKNOWLEDGEMENTS}
\end{center}
The authors would like to thank Arsen Palestini for his fruitful answer regarding the System-CoVaR and Jianlin Zhang for writing an excellent Master's thesis, which inspired this work. The authors report there are no competing interests to declare.

\bigskip
\bibliographystyle{asa}

%MY CHANGES START
\bibliography{bibliography}

\newpage
\setcounter{equation}{0}
\renewcommand{\theequation}{A.\arabic{equation}}
\section*{Appendix A: Proofs}
\input{Proofs}
\input{Supp}

%MY CHANGES END
\end{document}

%% file: tables/SimStudy.tex
\begin{table}[ht]
\centering
\renewcommand{\arraystretch}{0.75}
\caption{Average violation rates for CoVaR, MCoVaR, and VCoVaR estimation.}
\begin{tabular}{lcccccc}
\toprule[0.05em]
\midrule
\multirow{2}{*}{Measure} & \multicolumn{2}{c}{$\tau = 0.25$} & \multicolumn{2}{c}{$\tau = 0.50$} & \multicolumn{2}{c}{$\tau = 0.75$} \\\cmidrule[0.05em]{2-7}
& Clayton & Gumbel & Clayton & Gumbel & Clayton & Gumbel \\ 
  \midrule
  \multicolumn{1}{c}{} & \multicolumn{6}{c}{$\alpha = \beta = 0.05$}\\
  \midrule
CoVaR & 0.0489 & 0.0493 & 0.0499 & 0.0507 & 0.0506 & 0.0506 \\ 
  MCoVaR & 0.0491 & 0.0516 & 0.0501 & 0.0501 & 0.0490 & 0.0480 \\ 
  VCoVaR & 0.0503 & 0.0505 & 0.0507 & 0.0495 & 0.0496 & 0.0502 \\ 
   \midrule
   \multicolumn{1}{c}{} & \multicolumn{6}{c}{$\alpha = \beta = 0.01$}\\
   \midrule
   CoVaR & 0.0102 & 0.0082 & 0.0102 & 0.0089 & 0.0093 & 0.0085 \\ 
  MCoVaR & 0.0080 & 0.0062 & 0.0082 & 0.0112 & 0.0101 & 0.0117 \\ 
  VCoVaR & 0.0110 & 0.0099 & 0.0115 & 0.0113 & 0.0091 & 0.0089 \\  
  \bottomrule[0.05em]
\end{tabular}
\label{table:sim_study}
\end{table}

%% file: tables/VaR_rates.tex
\begin{table}[htp]
\renewcommand{\arraystretch}{0.75}
\centering
\resizebox{\textwidth}{!}{
\begin{threeparttable}
\normalsize
\caption{\normalsize Violations for the VaR estimates.}
\begin{tabular}{lcccccccccc}
  \toprule[0.05em]
  \midrule
 & BTC & ETH & LTC & XMR & XRP & Sys:BTC & Sys:ETH & Sys:LTC & Sys:XMR & Sys:XRP \\
  \midrule
  Rate in $\%$ & 6.00 & 5.43 & 5.35 & 4.78 & 5.04 & 4.69 & 4.86 & 4.73 & 5.17 & 5.04 \\
  Number & 137 & 124 & 122 & 109 & 115 & 107 & 111 & 108 & 118 & 115 \\
  \bottomrule[0.05em]
\end{tabular}
\begin{tablenotes}
\setlength{\itemindent}{-2.49997pt}
\normalsize
      \item NOTE: Sys:BTC denotes the sum of the log-returns of ETH, LTC, XMR, and XRP. For the selected $\alpha = 0.05$ we should observe 114.1 violations.
\end{tablenotes}
\label{table:VaR_rates}
\end{threeparttable}}
\end{table}

%% file: tables/Violation_rates_all_copulae.tex
\begin{table}[htp]
\renewcommand{\arraystretch}{0.75}
\centering
\scriptsize
\resizebox{\textwidth}{!}{
\begin{threeparttable}
\caption{Statistics and in-sample rates of the risk measures.}
\begin{tabular}{lcccccccc}
\toprule[0.05em]
\midrule
\multirow{2}{*}{Measure} & \multirow{2}{*}{\makecell{Mean \\($t$-copula)}} & \multirow{2}{*}{\makecell{Sd \\($t$-copula)}} & \multicolumn{6}{c}{Violation rates for copula} \\ \cmidrule[0.05em]{4-9}
& & & Gaussian & $t$ & Clayton & Gumbel & Patton-$t$ & DCC-$t$ \\ 
  \midrule
  BTC-VaR & -0.0545 & 0.0271 & -& -& -& -& -& -\\
BTC-ETH & -0.1641 & 0.0791 & 0.0645 & \textbf{0.0484} & 0.0403 & 0.1774 & 0.0645 & 0.0645 \\ 
  BTC-LTC & -0.1720 & 0.0829 & 0.0574 & \textbf{0.0492} & \textbf{0.0492} & 0.1066 & \textbf{0.0492} & 0.0574 \\ 
  BTC-XMR & -0.1642 & 0.0792 & 0.0917 & 0.0642 & \textbf{0.0550} & 0.2477 & 0.0734 & 0.0826 \\ 
  BTC-XRP & -0.1609 & 0.0776 & 0.1130 & 0.0609 & \textbf{0.0522} & 0.2435 & 0.0696 & 0.0957 \\ 
  BTC-SCoVaR & -0.1694 & 0.0816 & 0.0654 & \textbf{0.0561} & \textbf{0.0561} & 0.1776 & 0.0654 & 0.0654 \\ 
  BTC-MCoVaR & -0.2819 & 0.1352 & \textbf{0.0263} & \textbf{0.0263} & \textbf{0.0263} & 0.1316 & - & \textbf{0.0263} \\ 
  BTC-VCoVaR & -0.1309 & 0.0633 & 0.1070 & \textbf{0.0576} & 0.0412 & 0.1358 & - & 0.0741 \\ 
  \midrule
  ETH-VaR & -0.0860 & 0.0350 & -& -& -& -& -& -\\
  ETH-BTC & -0.2442 & 0.0972 & 0.0657 & 0.0292 & 0.0219 & 0.1387 & \textbf{0.0511} & 0.0365 \\ 
  ETH-LTC & -0.2472 & 0.0984 & 0.0738 & 0.0328 & 0.0246 & 0.1475 & \textbf{0.0410} & \textbf{0.0410} \\ 
  ETH-XMR & -0.2455 & 0.0977 & 0.0826 & 0.0367 & 0.0275 & 0.1560 & \textbf{0.0459} & \textbf{0.0459} \\ 
  ETH-XRP & -0.2458 & 0.0979 & 0.0870 & 0.0348 & 0.0261 & 0.1652 & \textbf{0.0609} & \textbf{0.0609} \\ 
  ETH-SCoVaR & -0.2533 & 0.1008 & \textbf{0.0541} & 0.0270 & 0.0270 & 0.1171 & 0.0270 & 0.0360 \\ 
  ETH-MCoVaR & -0.4026 & 0.1596 & 0.0227 & 0.0227 & 0.0227 & \textbf{0.0682} & - & 0.0227 \\ 
  ETH-VCoVaR & -0.1950 & 0.0779 & 0.0753 & \textbf{0.0460} & 0.0418 & 0.1004 & - & 0.0711 \\ 
  \midrule
  LTC-VaR & -0.0726 & 0.0359 & -& -& -& -& -& -\\
  LTC-BTC & -0.1974 & 0.0976 & 0.0876 & 0.0657 & \textbf{0.0511} & 0.1533 & 0.0730 & 0.0730 \\ 
  LTC-ETH & -0.1916 & 0.0948 & 0.1290 & 0.0726 & \textbf{0.0484} & 0.1774 & 0.0806 & 0.0645 \\ 
  LTC-XMR & -0.1863 & 0.0921 & 0.1468 & 0.0917 & \textbf{0.0550} & 0.2202 & 0.0917 & 0.1009 \\ 
  LTC-XRP & -0.1916 & 0.0947 & 0.1304 & 0.0783 & \textbf{0.0522} & 0.1826 & 0.0783 & 0.0870 \\ 
  LTC-SCoVaR & -0.1968 & 0.0973 & 0.0926 & 0.0648 & \textbf{0.0556} & 0.1852 & 0.0741 & 0.0741 \\ 
  LTC-MCoVaR & -0.3122 & 0.1544 & 0.0256 & \textbf{0.0513} & \textbf{0.0513} & 0.1282 & - & 0.0256 \\ 
  LTC-VCoVaR & -0.1551 & 0.0767 & 0.1000 & 0.0885 & \textbf{0.0731} & 0.1231 & - & 0.0885 \\ 
  \midrule
  XMR-VaR & -0.0876 & 0.0361 & -& -& -& -& -& -\\
  XMR-BTC & -0.2276 & 0.0925 & 0.0730 & \textbf{0.0438} & \textbf{0.0438} & 0.1387 & 0.0584 & 0.0657 \\ 
  XMR-ETH & -0.2286 & 0.0929 & 0.0726 & \textbf{0.0484} & \textbf{0.0484} & 0.1290 & 0.0565 & 0.0565 \\ 
  XMR-LTC & -0.2239 & 0.0910 & 0.0820 & 0.0574 & \textbf{0.0492} & 0.1557 & 0.0656 & 0.0656 \\ 
  XMR-XRP & -0.2204 & 0.0896 & 0.0870 & 0.0783 & \textbf{0.0522} & 0.1826 & 0.0783 & 0.0870 \\ 
  XMR-SCoVaR & -0.2324 & 0.0944 & 0.0763 & \textbf{0.0508} & \textbf{0.0508} & 0.1186 & 0.0593 & 0.0678 \\ 
  XMR-MCoVaR & -0.3388 & 0.1373 & \textbf{0.0488} & \textbf{0.0488} & \textbf{0.0488} & 0.1463 & - & \textbf{0.0488} \\ 
  XMR-VCoVaR & -0.1861 & 0.0757 & 0.0675 & 0.0397 & 0.0397 & 0.0833 & - & \textbf{0.0476} \\ 
  \midrule
  XRP-VaR & -0.0817 & 0.0552 & -& -& -& -& -& -\\
  XRP-BTC & -0.2250 & 0.1522 & 0.1095 & 0.0730 & \textbf{0.0365} & 0.1971 & 0.0730 & 0.0657 \\ 
  XRP-ETH & -0.2308 & 0.1561 & 0.1129 & 0.0887 & \textbf{0.0484} & 0.1935 & 0.0887 & 0.0887 \\ 
  XRP-LTC & -0.2321 & 0.1570 & 0.0984 & 0.0738 & \textbf{0.0328} & 0.1885 & 0.0902 & 0.0820 \\ 
  XRP-XMR & -0.2211 & 0.1495 & 0.1560 & 0.0917 & \textbf{0.0459} & 0.2661 & 0.1193 & 0.1009 \\ 
  XRP-SCoVaR & -0.2345 & 0.1586 & 0.1043 & 0.0783 & \textbf{0.0435} & 0.1652 & 0.0870 & 0.0783 \\ 
  XRP-MCoVaR & -0.3644 & 0.2463 & \textbf{0.0488} & \textbf{0.0488} & 0.0244 & 0.0976 & - & \textbf{0.0488} \\ 
  XRP-VCoVaR & -0.1821 & 0.1231 & 0.1004 & 0.0763 & \textbf{0.0562} & 0.1285 & - & 0.0763 \\ 
  \bottomrule[0.05em]
\end{tabular}
\begin{tablenotes}
\setlength{\itemindent}{-2.49997pt}
      \footnotesize
      \item NOTE: The copula with the closest rate to $\beta = 0.05$ for each row is marked bold. BTC-LTC denotes the CoVaR of BTC with LTC under condition. Mean and sd of each measure (except VaR) are given for the static $t$-copula case.
\end{tablenotes}
\label{table:crypto_cop_rel_viol}
\end{threeparttable}}
\end{table}

%% file: tables/OOS_violations.tex
\begin{table}

\renewcommand{\arraystretch}{0.775}
\centering
\footnotesize
%\resizebox{\textwidth}{!}{
%\begin{threeparttable}
\caption{Out-of-sample rates of the risk measures.}
\begin{tabular}{lcccc}
\toprule[0.05em]
\midrule
\multirow{2}{*}{Measure} & \multicolumn{4}{c}{Violation rates for copula} \\ \cmidrule[0.05em]{2-5}
& Gaussian & $t$ & Clayton & Gumbel\\ 
  \midrule
  BTC-ETH & 0.1237 & \textbf{0.0825} & 0.0928 & 0.1959 \\ 
  BTC-LTC & 0.0583 & \textbf{0.0485} & \textbf{0.0485} & 0.1165 \\ 
  BTC-XMR & 0.1023 & \textbf{0.0568} & \textbf{0.0568} & 0.1932 \\ 
  BTC-XRP & 0.1466 & 0.0862 & \textbf{0.0345} & 0.2328 \\ 
  BTC-SCoVaR & 0.1099 & \textbf{0.0549} & \textbf{0.0549} & 0.1758 \\ 
  BTC-MCoVaR & \textbf{0.0513} & \textbf{0.0513} & \textbf{0.0513} & 0.1026 \\ 
  BTC-VCoVaR & 0.0909 & 0.0642 & \textbf{0.0481} & 0.1390 \\ 
  \midrule
  ETH-BTC & 0.1078 & 0.0784 & \textbf{0.0686} & 0.1667 \\ 
  ETH-LTC & 0.0583 & 0.0583 & \textbf{0.0485} & 0.1068 \\ 
  ETH-XMR & 0.1023 & 0.0795 & \textbf{0.0455} & 0.1705 \\ 
  ETH-XRP & 0.0948 & 0.0690 & \textbf{0.0431} & 0.1552 \\ 
  ETH-SCoVaR & 0.0968 & 0.0753 & \textbf{0.0430} & 0.1290 \\ 
  ETH-MCoVaR & 0.0233 & 0.0233 & 0.0233 & \textbf{0.0698} \\ 
  ETH-VCoVaR & 0.0952 & 0.0635 & \textbf{0.0423} & 0.1270 \\ 
  \midrule
  LTC-BTC & 0.0588 & \textbf{0.0490} & 0.0392 & 0.1176 \\ 
  LTC-ETH & 0.0722 & \textbf{0.0515} & \textbf{0.0515} & 0.1031 \\ 
  LTC-XMR & 0.1023 & 0.0682 & \textbf{0.0568} & 0.1477 \\ 
  LTC-XRP & 0.0776 & \textbf{0.0431} & 0.0345 & 0.1121 \\ 
  LTC-SCoVaR & 0.0814 & 0.0581 & \textbf{0.0465} & 0.1279 \\ 
  LTC-MCoVaR & \textbf{0.0250} & \textbf{0.0750} & \textbf{0.0750} & 0.1250 \\ 
  LTC-VCoVaR & 0.0681 & \textbf{0.0576} & 0.0419 & 0.1099 \\ 
  \midrule
  XMR-BTC & 0.1078 & 0.0784 & \textbf{0.0490} & 0.1667 \\ 
  XMR-ETH & 0.0825 & 0.0825 & \textbf{0.0412} & 0.1340 \\ 
  XMR-LTC & 0.0874 & 0.0777 & \textbf{0.0485} & 0.1359 \\ 
  XMR-XRP & 0.0948 & 0.0690 & \textbf{0.0431} & 0.1466 \\ 
  XMR-SCoVaR & 0.0968 & 0.0860 & \textbf{0.0430} & 0.1290 \\ 
  XMR-MCoVaR & \textbf{0.0732} & \textbf{0.0732} & \textbf{0.0732} & 0.1220 \\ 
  XMR-VCoVaR & 0.0737 & \textbf{0.0474} & \textbf{0.0474} & 0.1105 \\ 
  \midrule
  XRP-BTC & 0.1667 & 0.0980 & \textbf{0.0882} & 0.2941 \\ 
  XRP-ETH & 0.1134 & \textbf{0.0825} & \textbf{0.0825} & 0.1959 \\ 
  XRP-LTC & 0.1165 & 0.0680 & \textbf{0.0583} & 0.1845 \\ 
  XRP-XMR & 0.1250 & \textbf{0.0795} & \textbf{0.0795} & 0.2386 \\ 
  XRP-SCoVaR & \textbf{0.0769} & \textbf{0.0769} & \textbf{0.0769} & 0.1429 \\ 
  XRP-MCoVaR & \textbf{0.0732} & 0.0244 & 0.0244 & \textbf{0.0732} \\ 
  XRP-VCoVaR & 0.1561 & 0.1040 & \textbf{0.0636} & 0.1676 \\ 
  \bottomrule[0.05em]
\end{tabular}
\parbox{\textwidth}{\small%
\vspace{0.5cm}
\footnotesize
NOTE: The copula with the closest rate to $\beta = 0.05$ for each row is marked bold. BTC-LTC denotes the CoVaR of BTC with LTC under condition.}
%\begin{tablenotes}
%\setlength{\itemindent}{-2.49997pt}
%      \footnotesize
%      \item NOTE: The copula with the closest rate to $\beta = 0.05$ for each row is marked bold. BTC-LTC denotes the CoVaR of BTC with LTC under condition. Mean and sd of each measure (except VaR) are given for the static $t$-copula case.
%\end{tablenotes}
\label{table:crypto_oos}
%\end{threeparttable}%}
\end{table}

%% file: Proofs.tex
\begin{proof}[Proof of Lemma 3.1]
Transform (\ref{eq:v_covar_def}) to:
\begin{equation}\label{eq:v_covar_deriv_1}
    \frac{P\{X_{j,t} \leq VCoVaR^{j|\mathbf{i}}_{\alpha, \beta, t},  (\exists i: X_{i,t} \leq VaR^{i}_{\alpha, t})\}}{P(\exists i: X_{i,t} \leq VaR^{i}_{\alpha, t})}  = \beta.
\end{equation}
Based on: $P(\exists i: X_{i,t} \leq VaR^{i}_{\alpha, t}) = 1 - P(\forall i: X_{i,t} > VaR^{i}_{\alpha, t})$, and:
\begin{align*}
P(X_{j,t} \leq VCoVaR^{j|\mathbf{i}}_{\alpha, \beta, t}) &= P\{X_{j,t} \leq VCoVaR^{j|\mathbf{i}}_{\alpha, \beta, t},  (\exists i: X_{i,t} \leq VaR^{i}_{\alpha, t})\} \\ 
&+ P\{X_{j,t} \leq VCoVaR^{j|\mathbf{i}}_{\alpha, \beta, t},  (\forall i: X_{i,t} > VaR^{i}_{\alpha, t})\},
\end{align*}
(\ref{eq:v_covar_deriv_1}) is rewritten:
\begin{equation}\label{eq:v_covar_deriv_2}
    \frac{P(X_{j,t} \leq VCoVaR^{j|\mathbf{i}}_{\alpha, \beta, t}) - P\{X_{j,t} \leq VCoVaR^{j|\mathbf{i}}_{\alpha, \beta, t},  (\forall i: X_{i,t} > VaR^{i}_{\alpha, t})\}}{1 - P(\forall i: X_{i,t} > VaR^{i}_{\alpha, t})}  = \beta.
\end{equation}
Recognizing:
\begin{align*}
    P(\forall i: X_{i,t} > VaR^{i}_{\alpha, t}) &= P\{X_{j,t} \leq VCoVaR^{j|\mathbf{i}}_{\alpha, \beta, t},  (\forall i: X_{i,t} > VaR^{i}_{\alpha, t})\} \\ &+ P\{X_{j,t} > VCoVaR^{j|\mathbf{i}}_{\alpha, \beta, t},  (\forall i: X_{i,t} > VaR^{i}_{\alpha, t})\},
\end{align*}
allows to transform (\ref{eq:v_covar_deriv_2}) to:
\begin{equation*}
\resizebox{1.0 \textwidth}{!}{$\frac{P(X_{j,t} \leq VCoVaR^{j|\mathbf{i}}_{\alpha, \beta, t}) - P(\forall i: X_{i,t} > VaR^{i}_{\alpha, t}) + P\{X_{j,t} > VCoVaR^{j|\mathbf{i}}_{\alpha, \beta, t},  (\forall i: X_{i,t} > VaR^{i}_{\alpha, t})\}}
    {1 - P(\forall i: X_{i,t} > VaR^{i}_{\alpha, t})}  = \beta.$}
\end{equation*}
\noindent Noticing this expression consists of survival functions, Lemma 3.1 follows.
\end{proof}

\begin{proof}[Proof of Lemma 3.2]
(\ref{eq:v_covar_est}) becomes:
\begin{equation}\label{eq:v_covar_p_1}
    \frac{F_{X_{j,t}}(VCoVaR^{j|i = 1}_{\alpha, \beta, t}) - (1-\alpha)
    + \bar C_{X_{j,t}, X_{1,t}}\{1-F_{X_{j,t}}(VCoVaR^{j|i = 1}_{\alpha, \beta, t}), 1-\alpha; \theta_{1,t}\}}
    {1 - (1-\alpha)} = \beta.
\end{equation}
Noticing $\bar C(u_1, u_2) = u_1 + u_2 - 1 + C(1-u_1, 1-u_2)$, rewrite:
\begin{align*}
    \bar C_{X_{j,t}, X_{1,t}}\{1-F_{X_{j,t}}(VCoVaR^{j|i = 1}_{\alpha, \beta, t}), 1-\alpha; \theta_{1,t}\} &= 1-F_{X_{j,t}}(VCoVaR^{j|i = 1}_{\alpha, \beta, t}) + 1-\alpha - 1 \\
    &+ C_{X_{j,t}, X_{1,t}}\{F_{X_{j,t}}(VCoVaR^{j|i = 1}_{\alpha, \beta, t}), \alpha; \theta_{1,t}\}.
\end{align*}
Simplifying and inserting into (\ref{eq:v_covar_p_1}) yields: $C_{X_{j,t},X_{1,t}}\{F_{X_{j,t}}(VCoVaR^{j|i=1}_{\alpha, \beta, t}), \alpha;\theta_{1,t}\} = \alpha\beta$,
equaling the result of the bivariate CoVaR in (\ref{eq:covar_est_cop}). Thus, $VCoVaR^{j|i=1}_{\alpha, \beta, t} = CoVaR^{j|i}_{\alpha, \beta, t}$.
\end{proof}

\begin{proof}[Proof of Lemma 4.1]
The CoVaR of (\ref{eq:covar_est_cop}) becomes:
$F_{X_{j,t}}(CoVaR^{j|i}_{\alpha, \beta, t}) \alpha = \alpha\beta$, which reduces for $\alpha, \beta \in (0,1)$ to: $CoVaR^{j|i}_{\alpha, \beta, t} = F_{X_{j,t}}^{-1}(\beta)$. $F_{X_{j,t}}^{-1}$ denotes the inverse marginal distribution of $X_{j,t}$. The MCoVaR of (\ref{eq:m_covar_est}) becomes:
\begin{equation*}
    \frac{F_{X_{j,t}}(MCoVaR^{j|\mathbf{i}}_{\alpha, \beta, t})\alpha^p}{\alpha^p} = \beta,
\end{equation*}
which simplifies to: $MCoVaR^{j|\mathbf{i}}_{\alpha, \beta, t} = F_{X_{j,t}}^{-1}(\beta)$. Finally, the VCoVaR in (\ref{eq:v_covar_est}) becomes:
\begin{equation*}
\frac{F_{X_{j,t}}(VCoVaR^{j|\mathbf{i}}_{\alpha, \beta, t}) -(1-\alpha)^p
    + \{1-F_{X_{j,t}}(VCoVaR^{j|\mathbf{i}}_{\alpha, \beta, t})\}(1-\alpha)^p}
    {1 - (1-\alpha)^p} = \beta.
\end{equation*}
Simplification yields:
\begin{equation*}
\frac{F_{X_{j,t}}(VCoVaR^{j|\mathbf{i}}_{\alpha, \beta, t})\{1-(1-\alpha)^p\}}
    {1 - (1-\alpha)^p} = \beta,
\end{equation*}
leading to: $VCoVaR^{j|\mathbf{i}}_{\alpha, \beta, t} = F_{X_{j,t}}^{-1}(\beta)$.
\end{proof}

\begin{proof}[Proof of Lemma 4.2]
The CoVaR of (\ref{eq:covar_est_cop}) becomes: $\min\{F_{X_{j,t}}(CoVaR^{j|i}_{\alpha, \beta, t}), \alpha\} = \alpha\beta,$ as the copula equals the upper Fréchet-Hoeffding bound. Solving leads to: $CoVaR^{j|i}_{\alpha, \beta, t} = F_{X_{j,t}}^{-1}(\alpha\beta)$, based on $\alpha, \beta \in (0,1)$. Similarly for the MCoVaR, (\ref{eq:m_covar_est}) transforms in: \begin{equation*}
    \frac{\min\{F_{X_{j,t}}(MCoVaR^{j|\mathbf{i}}_{\alpha, \beta, t}), \bm{\alpha}\}}{\min(\bm{\alpha})} = \beta,
\end{equation*}
which is solved as: $MCoVaR^{j|\mathbf{i}}_{\alpha, \beta, t} = F_{X_{j,t}}^{-1}(\alpha\beta)$. For the VCoVaR, (\ref{eq:v_covar_est}) becomes:
\begin{equation}\label{eq:perf_dep_helper}
\frac{F_{X_{j,t}}(VCoVaR^{j|\mathbf{i}}_{\alpha, \beta, t}) -
    \min(\bm{1-\alpha})
    + \min\{1-F_{X_{j,t}}(VCoVaR^{j|\mathbf{i}}_{\alpha, \beta, t}), \bm{1-\alpha}\}}
    {1 - \min(\bm{1-\alpha})} = \beta,
\end{equation}
which applies the upper Fréchet-Hoeffding bound to survival copulae, see \cite{lux2017improved}. Notice that $F_{X_{j,t}}(VCoVaR^{j|\mathbf{i}}_{\alpha, \beta, t}) \geq \alpha$ implies $\beta = 1$. This follows directly as (\ref{eq:perf_dep_helper}) would reduce in this case to:
\begin{equation*}
F_{X_{j,t}}(VCoVaR^{j|\mathbf{i}}_{\alpha, \beta, t}) - (1-\alpha) + \{1-F_{X_{j,t}}(VCoVaR^{j|\mathbf{i}}_{\alpha, \beta, t})\} = \alpha\beta,
\end{equation*}
which is simplified to: $\beta = 1$. As $\alpha, \beta \in (0,1)$ is imposed, it must hold: $F_{X_{j,t}}(VCoVaR^{j|\mathbf{i}}_{\alpha, \beta, t}) < \alpha$. In this scenario, (\ref{eq:perf_dep_helper}) can be transformed to:
\begin{equation*}
\frac{F_{X_{j,t}}(VCoVaR^{j|\mathbf{i}}_{\alpha, \beta, t}) -
    (1-\alpha) + (1-\alpha)}
    {\alpha} = \beta,
\end{equation*}
which yields: $VCoVaR^{j|\mathbf{i}}_{\alpha, \beta, t} = F_{X_{j,t}}^{-1}(\alpha \beta)$.
\end{proof}

%% file: Supp.tex
\renewcommand\thetable{B.\arabic{table}}
\renewcommand\thefigure{B.\arabic{figure}}
\setcounter{table}{0}
\setcounter{figure}{0}
\renewcommand{\thesubsection}{B.\arabic{subsection}}

\section*{Appendix B: Supplements to the Empirical Study}
\subsection{Data Description}
Tables \ref{table:des_stat_tests}, \ref{table:des_stat_tests_systems} provides descriptive statistics and tests of the log-returns of the empirical study for the CCs and system variables. For example, 'Sys:BTC' denotes the system for BTC, and is the equally weighted sum of the log-returns of ETH, LTC, XMR, and XRP. ADF refers to the test of \cite{said1984testing}, PP to \cite{phillips1988testing}, and KPSS to \cite{kwiatkowski1992testing}. An asterisk (*) indicates rejection of the null hypothesis at level $\alpha^{*} = 0.05$.

\input{tables/stats_tests_logreturns}

Table \ref{table:tau_log_ret} shows estimates of Kendall's $\tau$ between the log-returns. BTC and LTC display the strongest dependency between the five CCs, while $\tau$ between the systems is by design relatively high.

\input{tables/tau}

\subsection{Univariate Model Selection}
\subsubsection{Procedure}
We propose to use the following procedure to select the univariate models, while we perform the ML estimation in the \verb+R+-package \verb+rugarch+ \citep{rugarch}.
\begin{steps}[align=left, leftmargin=*]
    \item Calculate all possible ARMA($p_l, q_l$) models with $p_l,q_l < 6$. Of those models, consider the ones non-rejecting 
    the \cite{ljung1978measure} (LB) $H_0$ hypothesis of having no serial correlation in the residuals.
    \item From the models left after Step 1, select the one with the lowest Akaike information criterion (AIC, \citealp{akaike1974new}). Test the model for ARCH effects in the residuals using the test of \cite{mcleod1983diagnostic}.
    \begin{sloppypar}
    \item If the $H_0$ hypothesis of having no ARCH effects of the model of Step 2 is non-rejected, return the model of Step 2. If $H_0$ is rejected, calculate all possible ARMA($p_l,q_l$)-GARCH(P,Q) models with $p_l,q_l,P < 6$ and $Q < 2$. Of those models, consider the ones non-rejecting the LB test and the weighted version of the test of \cite{li1994squared} proposed in \cite{fisher2012new} (WLM) for correct ARCH model specification, applied to the standardized residuals.
    \end{sloppypar}
    \item From the models left after Step 3, select the one with the lowest AIC. Test the model for asymmetries/leverage effect using the Sign Bias tests of \cite{engle1993measuring}.
    \item If the $H_0$ hypotheses of having no asymmetries are non-rejected, return the model of Step 4. If an $H_0$ is rejected, calculate all ARMA($p_l,q_l$)-GJR-GARCH(P,Q) models with $p_l,q_l,P < 6$ and $Q < 2$. Of those models, consider the ones non-rejecting the LB, the WLM and the Sign Bias tests on the standardized residuals.
    \item From the models of Step 5, return the one with the lowest AIC.
\end{steps}
Following \cite{tsay2005analysis}, the degrees of freedom of the chi-square distribution of the test statistic of the LB test is reduced by $(p_l+q_l)$ to account for the parametrization of the ARMA model. The same holds for the test of \cite{mcleod1983diagnostic} of Step 2 in the selection procedure, which tests the null hypothesis of having no ARCH effects by applying the LB statistic to the squared residuals \citep{tsay2005analysis}. Common use in practice is to pursue the same procedure for the squared standardized residuals of an ARMA-GARCH model, see, e.g., \cite{reboredo2015systemic}. However, \cite{li1994squared} proposed an improved test statistic to test the null hypothesis of having no ARCH effects in GARCH-type models. In the given application, the modification of \cite{fisher2012new} is used while pursuing a correction of the respective degrees of freedom following the implementation of the \verb+rugarch+ package. The Sign Bias tests of \cite{engle1993measuring} regress the squared standardized residuals $\hat{z}_{t}^{2}$ on the lagged estimated residuals $\hat{\varepsilon}_{t-1}$ as follows:
\begin{equation*}
    \hat{z}_{t}^{2} = c_0 + c_1 \mathbf{1}_{\{\hat{\varepsilon}_{t-1} < 0\}} +
    c_2 \mathbf{1}_{\{\hat{\varepsilon}_{t-1} < 0\}}\hat{\varepsilon}_{t-1} + c_3 \mathbf{1}_{\{\hat{\varepsilon}_{t-1} \geq 0\}}\hat{\varepsilon}_{t-1} + u_t,
\end{equation*}
where $\mathbf{1}_{\{\cdot\}}$ is the indicator function, to examine the hypotheses $H_0: c_i = 0 $ with $i \in \{1,2,3\}$ and for the \emph{Joint effect} $H_0: c_1 = c_2 = c_3 = 0$. The test of $c_1$ is referred to as \emph{Sign Bias}, the one of $c_2$ as \emph{Negative Sign Bias}, and the one of $c_3$ as \emph{Positive Sign Bias} \citep{rugarch}. In case of rejection of a hypothesis, there are significant asymmetries in the data, and the model is adjusted.

Please note that this procedure results in finding the best fitting model (in the sense of AIC) while maintaining parsimony and fulfilling the conditions on the standardized residuals. All models were estimated twice - with $\mu$ estimated from the data and with $\mu = 0$ - to capture possible improvements. We remark that this procedure is statistically thoroughly debatable, as a test is regarded as fulfilled if its $H_0$ hypothesis is not rejected. However, such approaches are heavily pursued in practical time series testing, which is the reason we specified this kind of selection procedure. The resulting models of this procedure have not been used in the case of the systems, which are necessary for the SCoVaR. Since each system is the sum of four CC, the dynamics become extremely complex and the found models yielded large VaR violations. More research has to be done to analyze the dynamics in such systems and we stick to a standard GARCH(1,1) specification for these cases.

\subsubsection{Results}
The results of the univariate model selection are given in Tables \ref{table:margin_model_res} and \ref{table:margin_model_res_systems}. Standard errors are given in parentheses, and an asterisk (*) denotes significance at level $\alpha^{*} = 0.05$. The AIC of a model with log-likelihood $LL$ and number of parameters $\Tilde{k}$ is given as: $AIC = 2\Tilde{k} - 2LL$ \citep{akaike1974new}. LB is the test of \cite{ljung1978measure}, while LB$^2$ performs the test on the squared standardized residuals. WLM is the weighted version of the test in \cite{li1994squared} proposed by \cite{fisher2012new}. The Sign Bias (SB) tests relate to \cite{engle1993measuring}. 

All time series exhibit relevant ARCH effects, and the pure ARMA($p_l,q_l$) specification was in all cases not sufficient to pass the considered tests. Moreover, a symmetric GARCH(P,Q) specification seems to be adequate in most cases. An exception is BTC, with significant asymmetric effects causing the necessity of considering GJR-GARCH(P,Q) models for the variance equation. However, in the final model, the $\gamma$-estimates were not significant. Regarding $F_{z}$, the estimates of $\zeta$ and $\nu$ are significant in all cases, confirming the relevance of a non-normal distribution. Tables \ref{table:margin_model_res} and \ref{table:margin_model_res_systems} further show the AIC and relevant tests of the standardized residuals of the selected models. For comparison, the LB test on the squared standardized residuals is included, which is in line with the WLM test in most cases. In the following, the plots of the autocorrelation functions (ACF) and partial autocorrelation functions (PACF) of the standardized residuals of the selected models are displayed.

\input{tables/margins}

\begin{figure}[H]
    \centering
    \includegraphics[width = \textwidth]{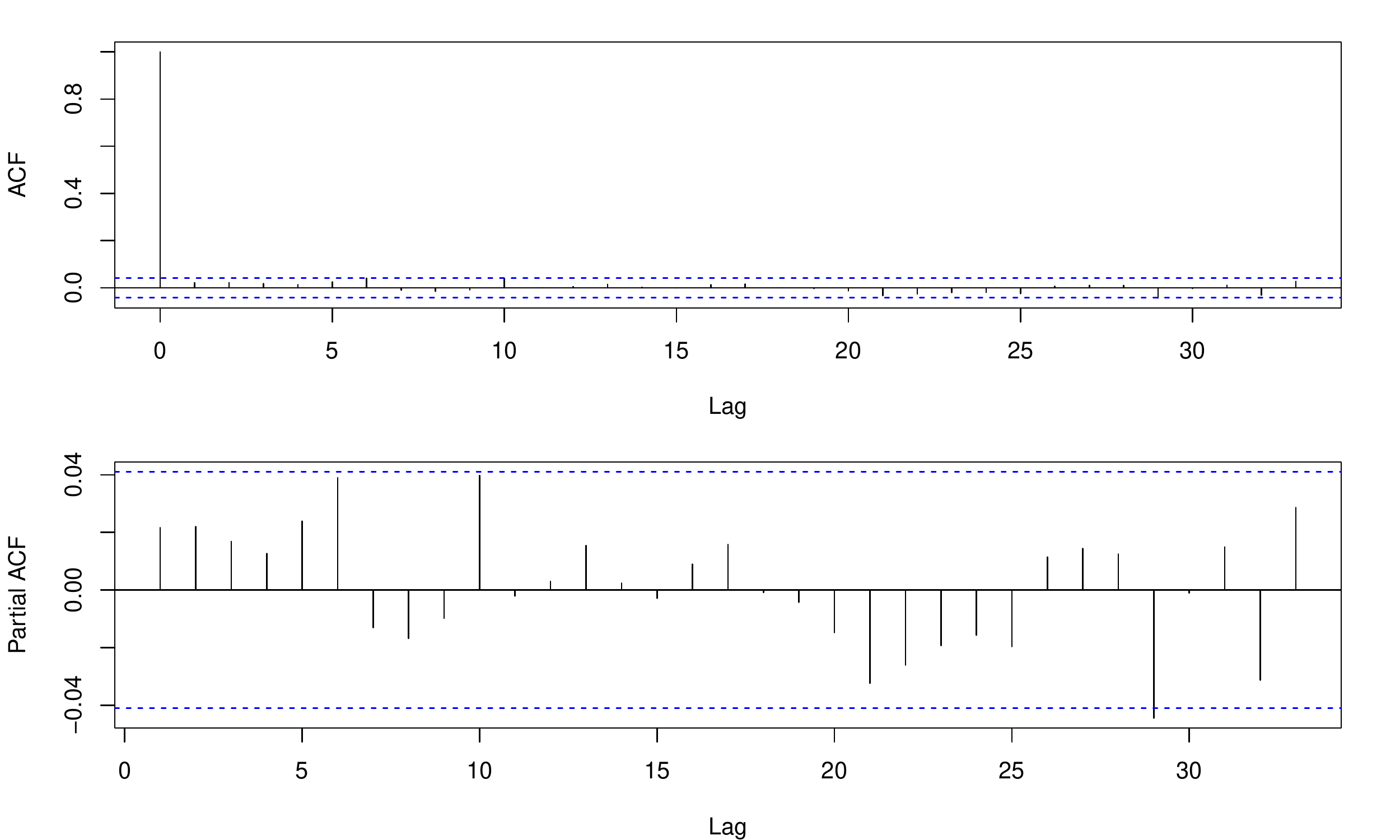}
    \caption{ACF/PACF plots of standardized residuals for BTC.}
\end{figure}

\begin{figure}[H]
    \centering
    \includegraphics[width = \textwidth]{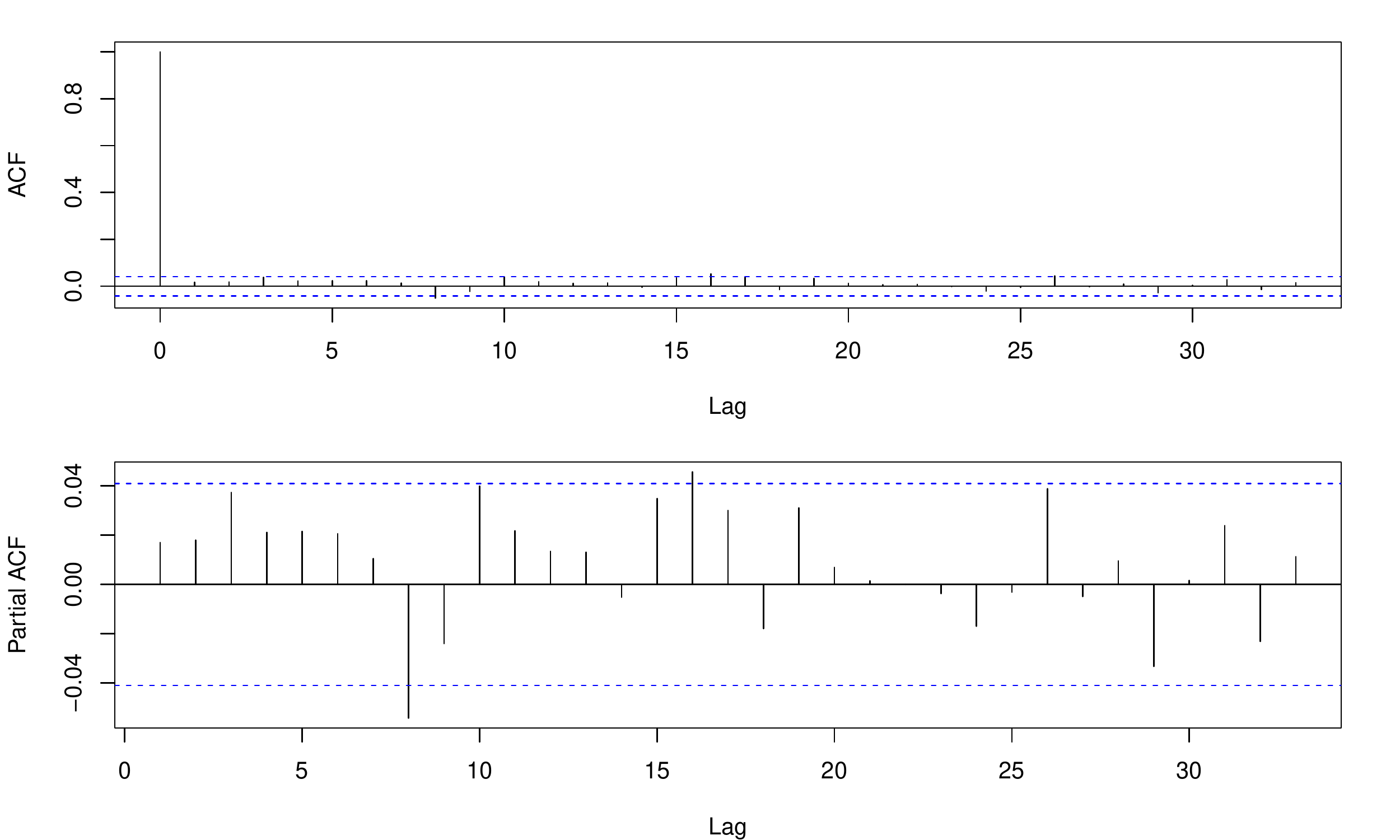}
    \caption{ACF/PACF plots of standardized residuals for ETH.}
\end{figure}

\begin{figure}[H]
    \centering
    \includegraphics[width = \textwidth]{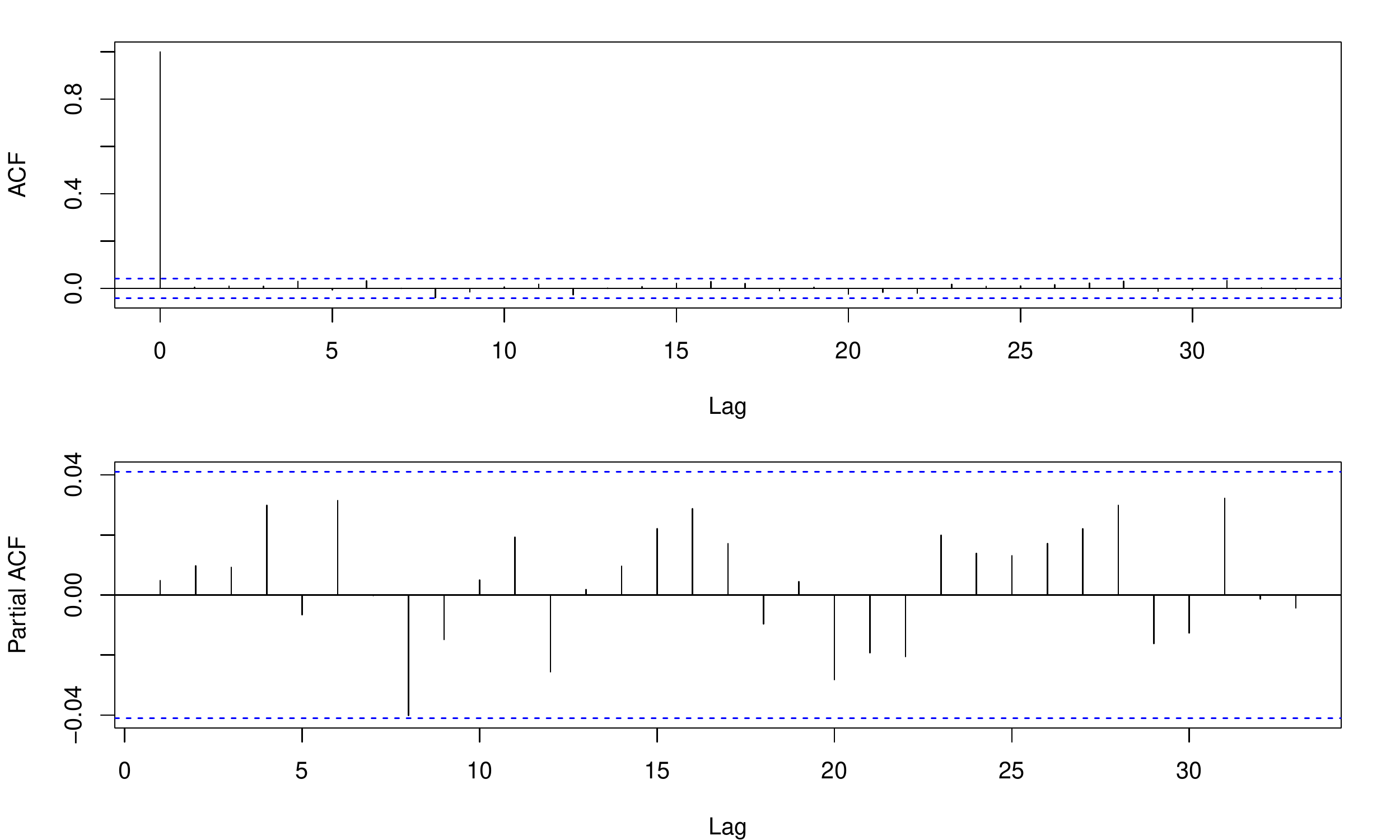}
    \caption{ACF/PACF plots of standardized residuals for LTC.}
\end{figure}

\begin{figure}[H]
    \centering
    \includegraphics[width = \textwidth]{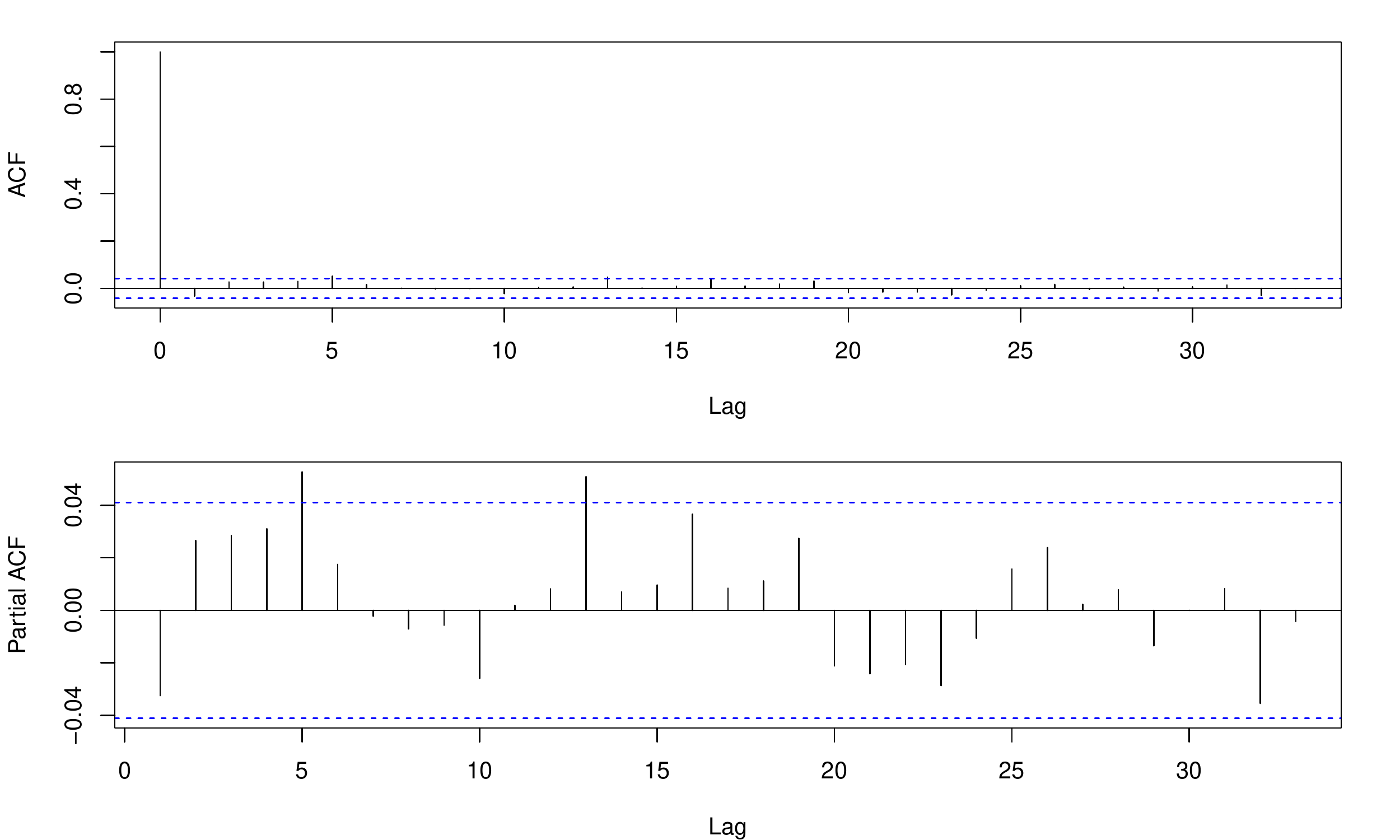}
    \caption{ACF/PACF plots of standardized residuals for XMR.}
\end{figure}

\begin{figure}[H]
    \centering
    \includegraphics[width = \textwidth]{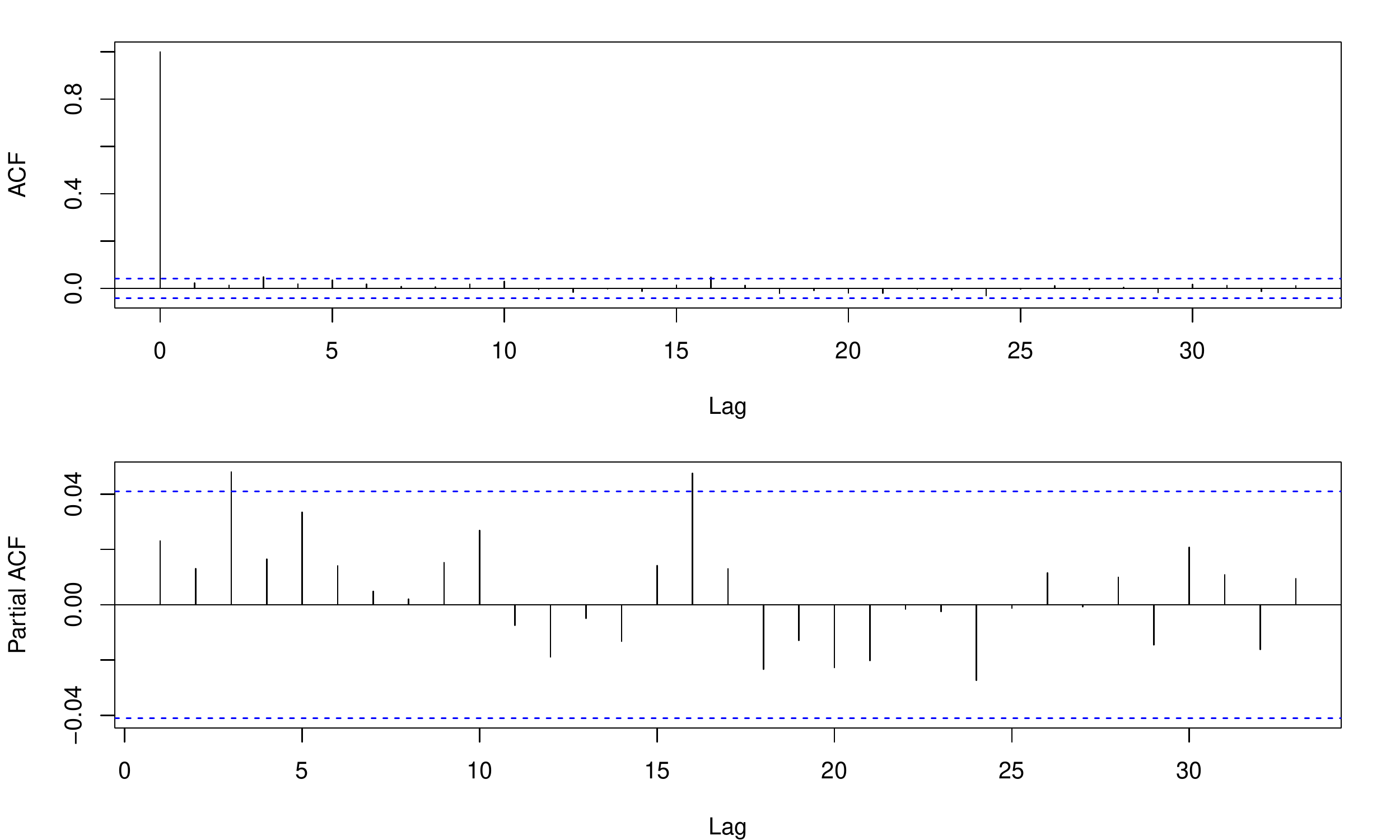}
    \caption{ACF/PACF plots of standardized residuals for XRP.}
\end{figure}

\begin{figure}[H]
    \centering
    \includegraphics[width = \textwidth]{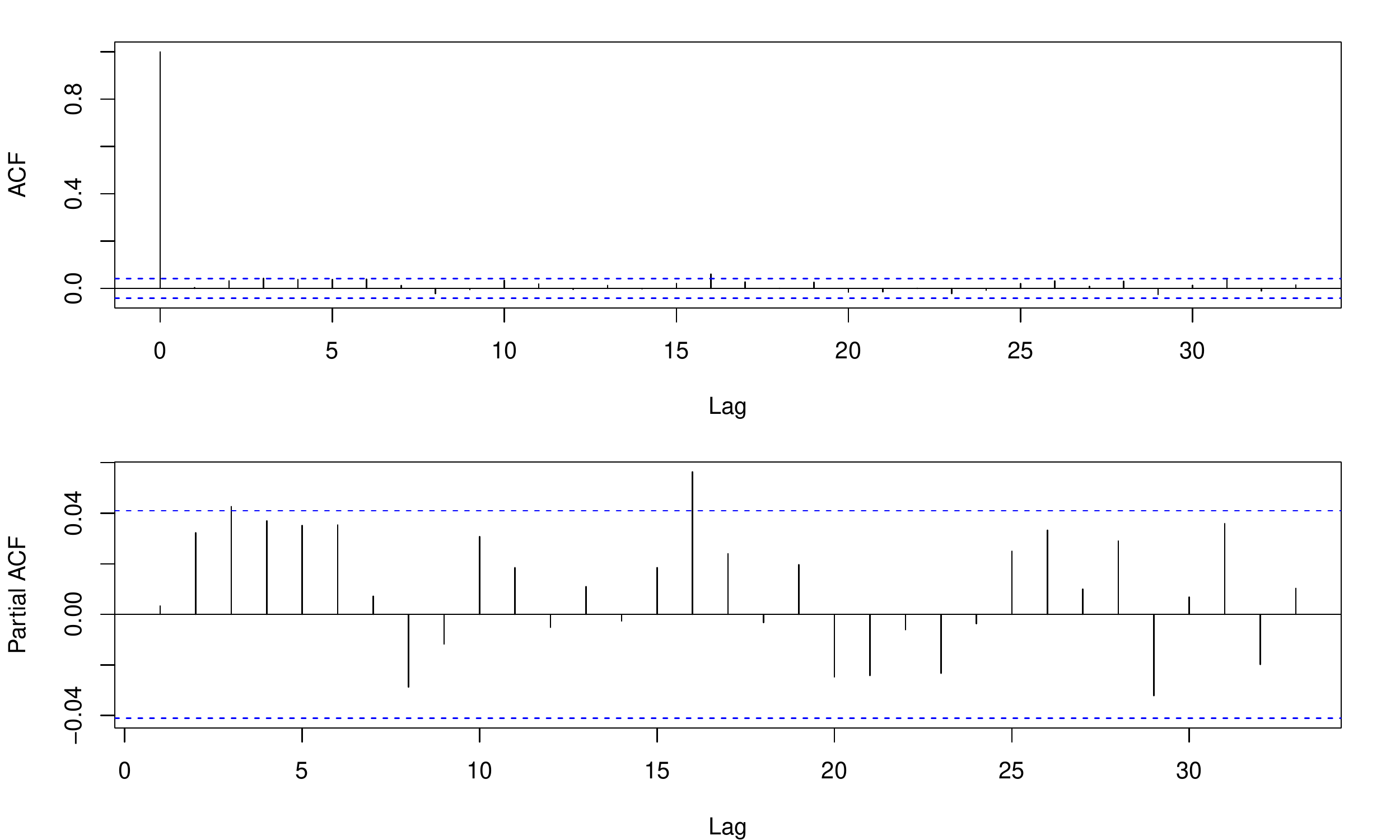}
    \caption{ACF/PACF plots of standardized residuals for Sys:BTC.}
\end{figure}

\begin{figure}[H]
    \centering
    \includegraphics[width = \textwidth]{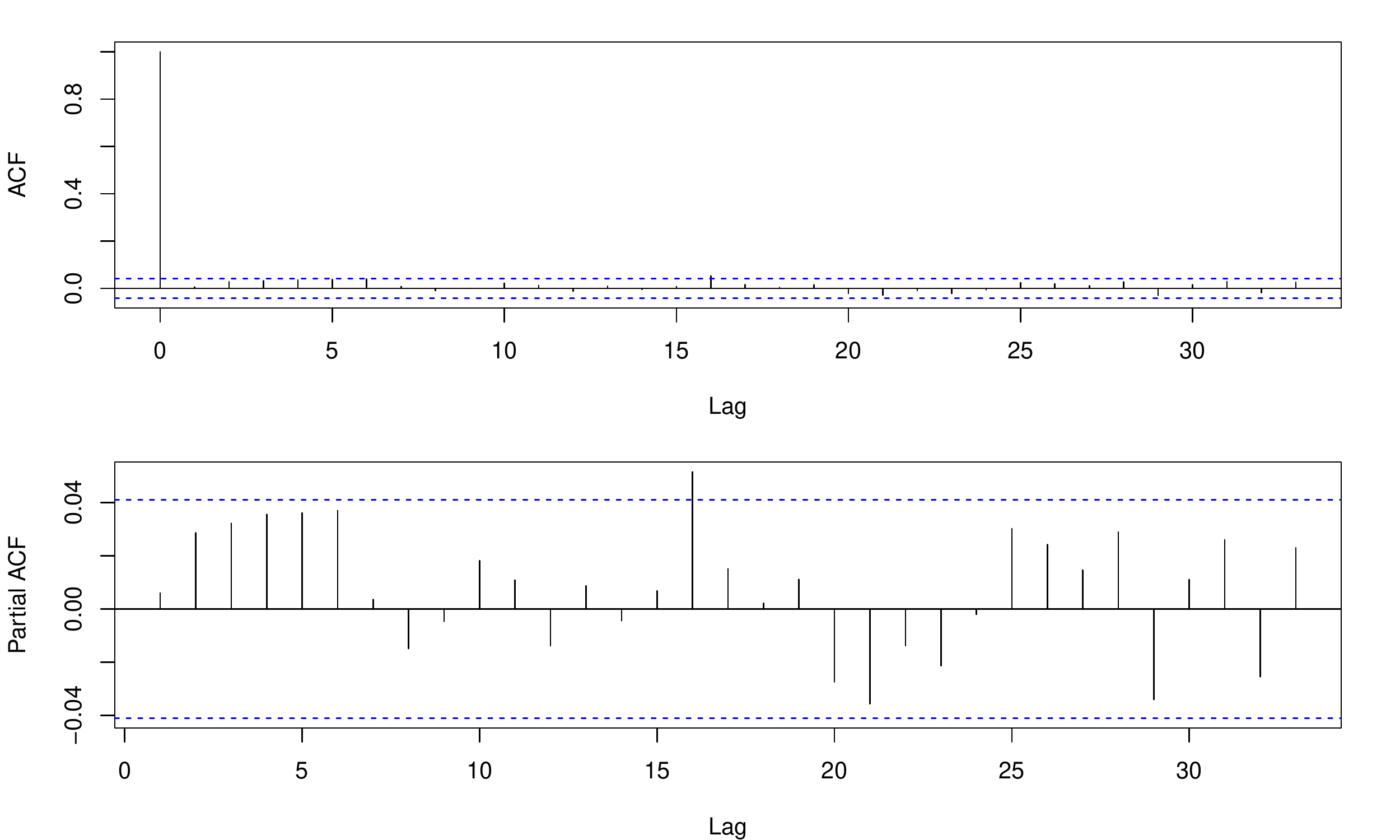}
    \caption{ACF/PACF plots of standardized residuals for Sys:ETH.}
\end{figure}

\begin{figure}[H]
    \centering
    \includegraphics[width = \textwidth]{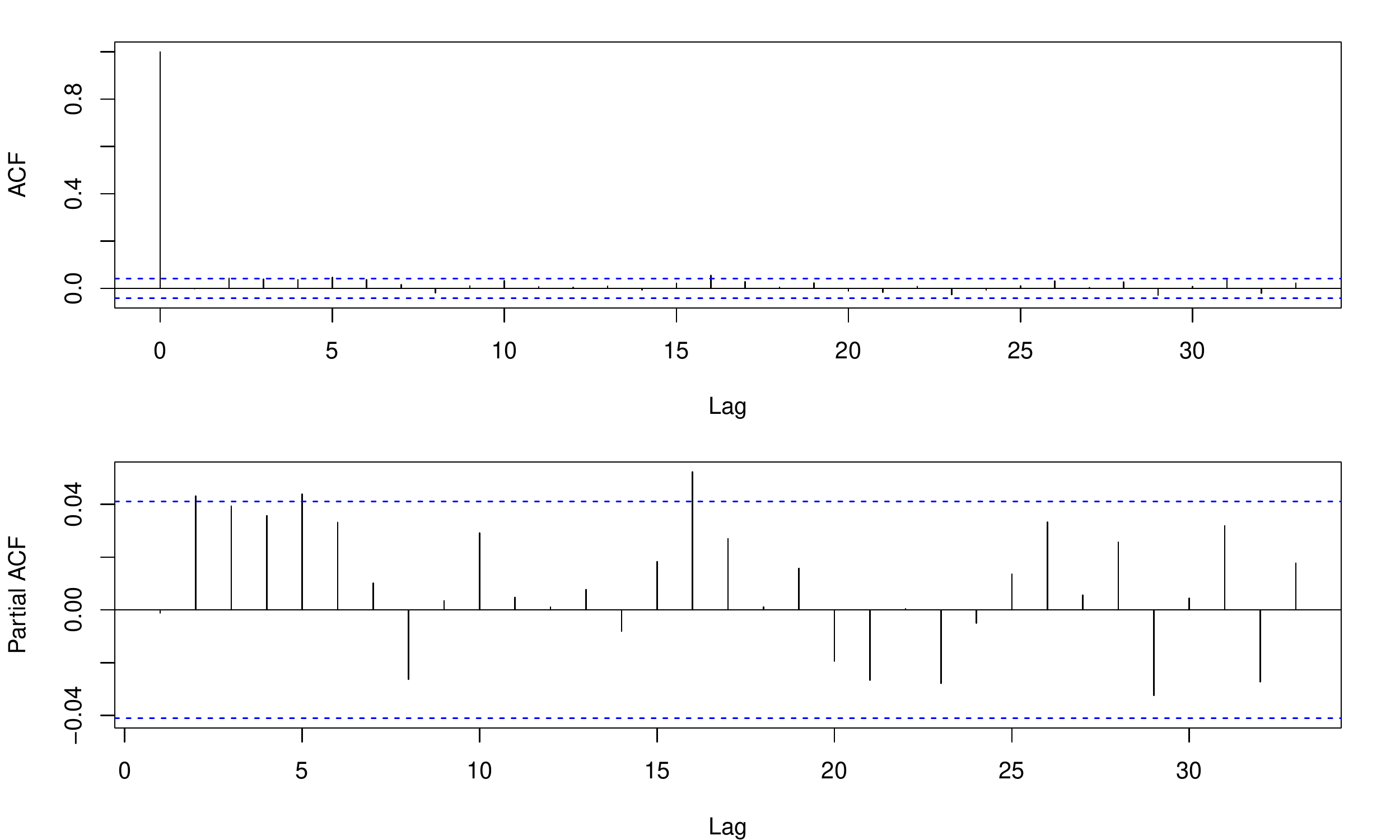}
    \caption{ACF/PACF plots of standardized residuals for Sys:LTC.}
\end{figure}

\begin{figure}[H]
    \centering
    \includegraphics[width = \textwidth]{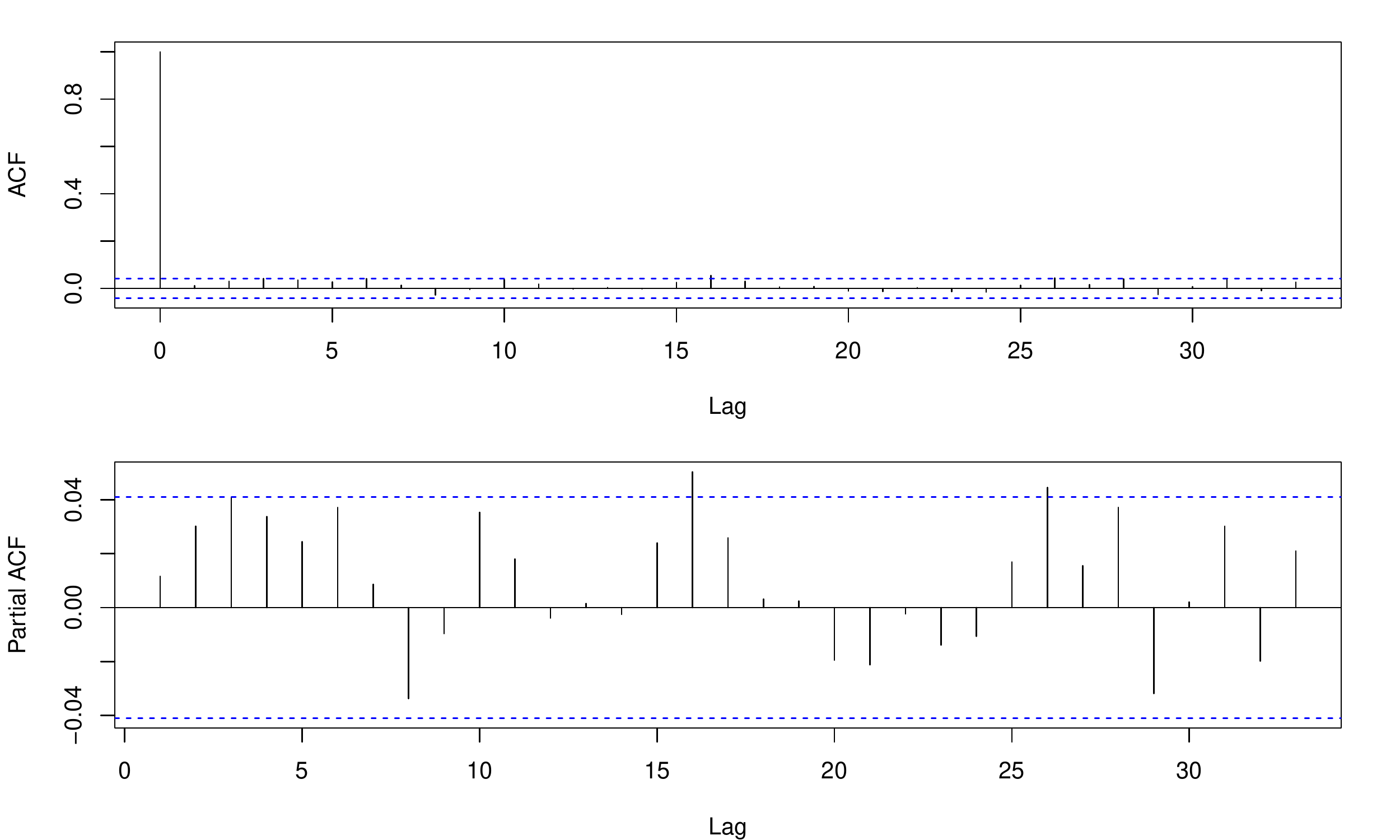}
    \caption{ACF/PACF plots of standardized residuals for Sys:XMR.}
\end{figure}

\begin{figure}[H]
    \centering
    \includegraphics[width = \textwidth]{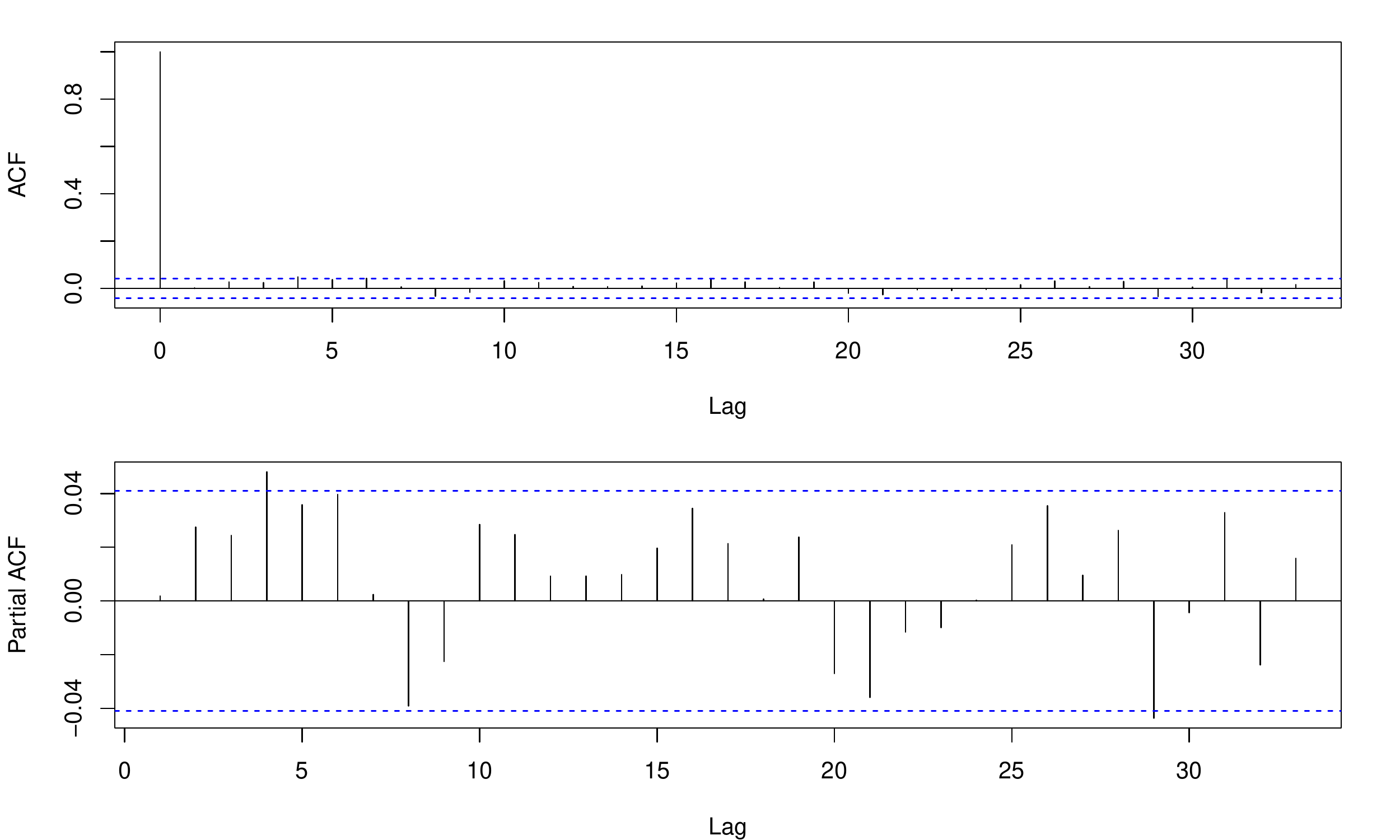}
    \caption{ACF/PACF plots of standardized residuals for Sys:XRP.}
\end{figure}

\subsection{Copula Models and Parameter Estimates}
This section provides details on the time-variant copula models used in the application. A crucial contribution in the field of dynamic copulae is \cite{patton2006modelling}, in which the copula parameter itself is assumed to follow an autoregressive scheme. Building on this idea, our work incorporates a time-variant $t$-copula. The number of degrees of freedom $\nu$ of the copula is assumed to be constant, while the copula parameter $\theta_t$ has the following dynamics:
\begin{align*}
    \theta_{t} = \Lambda\{\omega_{\theta} + \beta_{\theta} \theta_{t-1} + c_{\theta} \frac{1}{10} \sum_{l=1}^{10} t^{-1}_{\nu}(u_{i_1,t-l}) t^{-1}_{\nu}(u_{i_2,t-l})\},
\end{align*}
for $i_1, i_2 \in \{1, 2, 3, 4, 5\}, i_1 \neq i_2$, with $\Lambda(x) = \tanh(\frac{x}{2})$ to maintain $\theta_t \in [-1,1]$. $\omega_{\theta}$, $\beta_{\theta}$ and $c_{\theta}$ denote the parameters to be estimated, and $t_{\nu}^{-1}$ is the inverse univariate $t$-distribution with $\nu$ degrees of freedom. Following \cite{manner2012survey}, the model is limited to the bivariate case and will therefore be applied only to the bivariate CoVaR and the SCoVaR. The ML estimation routine depends on the initial values. For a second time-variant copula model, the DCC-copula approach proposed in \cite{jin2009large} is used. The model applies the idea of the DCC-model of \cite{engle2002dynamic} to the copula correlation matrix $R_t$ of the $t$-copula:
\begin{align*}
    &Q_t = (1 - a - b)\bar Q + a{z^{*}_{t - 1}}{{z^{*}_{t - 1}}}^{\top} + b{Q_{t - 1}},\\
    &R_t = diag{({Q_t})^{ - 1/2}}{Q_t}diag{({Q_t})^{ - 1/2}},
\end{align*} 
s.t. $a\geq 0$, $b\geq0$ and $a + b < 1$. $Q_t$ denotes an auxiliary process driving the dynamics of the copula correlation. If $R_t$ is five-dimensional, $z^{*}_t$ is the vector of transformed pseudo-observations $\{\tilde t_{\nu}^{-1}(u_{1,t}), \tilde t_{\nu}^{-1}(u_{2,t}), \tilde t_{\nu}^{-1}(u_{3,t}), \tilde t_{\nu}^{-1}(u_{4,t}),
\tilde t_{\nu}^{-1}(u_{5,t})\}^{\top}$ while ${\tilde t_{\nu}}$ is the standardized univariate $t$-distribution with unit variance and shape parameter $\nu$. $\bar Q$ denotes the unconditional matrix of $z^{*}_{t}$ and the degrees of freedom $\nu$ of the $t$-copula are assumed constant. The implementation of the model is realized using the \verb+R+-package \verb+rmgarch+ \citep{rmgarch}, which performs a two-stage ML estimation.

Tables \ref{table:cop_estimates_static_bivariate}, \ref{table:cop_estimates_dynamic_bivariate} shows the parameter estimates of the copulae for the bivariate case, while Table \ref{table:cop_estimates_multivariate} depicts the multivariate estimation results. $\theta$ denotes the copula correlation for the Gaussian and $t$-copulae in the bivariate case, which expands to the correlation matrix in the five-dimensional scenario. The tables additionally include the AIC values for the estimates based on the full log-likelihood. The information criterion states that the best copula is generally the $t$-copula, which further improves when considering the dynamic Patton and DCC models. This indicates that the dependencies of the CCs were not constant and changed over the years, as the dynamic dependence models provide a better fit. The Gumbel copula is considered the least appropriate one, while the Gaussian copula ranks between the $t$ and the Clayton one.

\input{tables/copula}

\subsection{Plots of Systemic Risk Measures}
Further illustrations of the estimated systemic risk measures are displayed below.

% BTC
\begin{figure}[H]
    \centering
    \includegraphics[width=\textwidth]{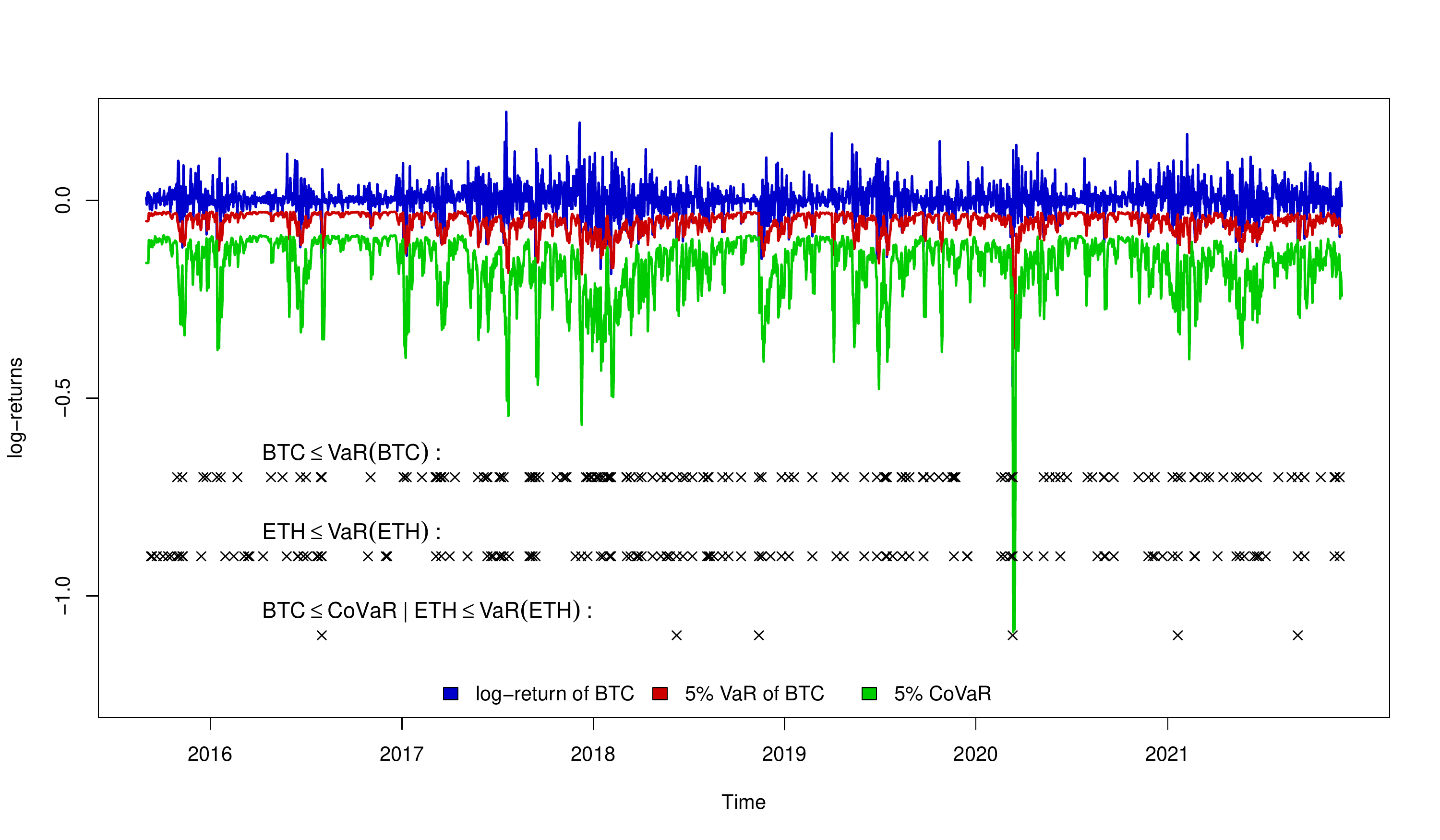}
    \caption{CoVaR of BTC-ETH using a time-invariant $t$-copula.}
\end{figure}

\begin{figure}[H]
    \centering
    \includegraphics[width=\textwidth]{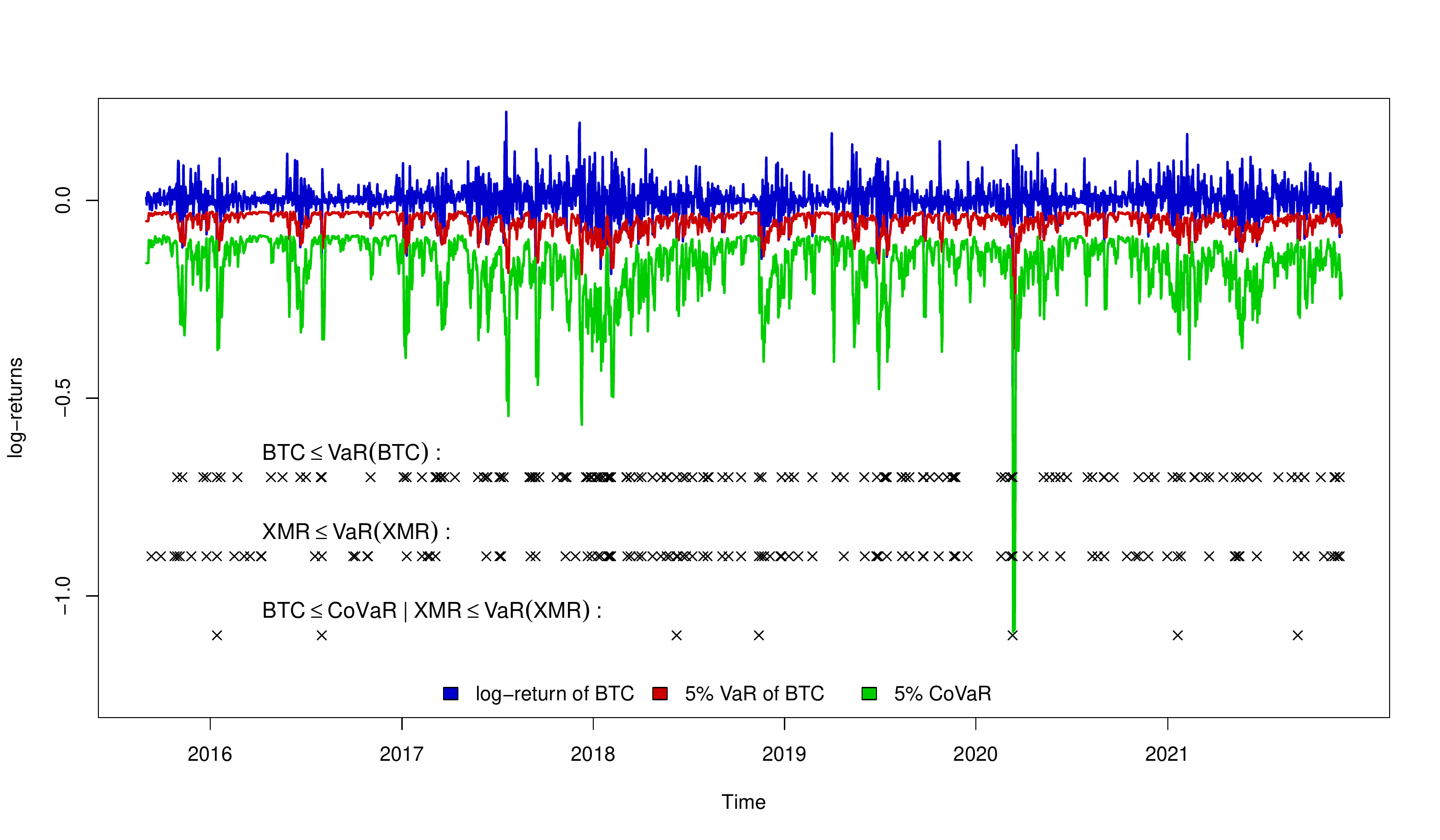}
    \caption{CoVaR of BTC-XMR using a time-invariant $t$-copula.}
\end{figure}

\begin{figure}[H]
    \centering
    \includegraphics[width=\textwidth]{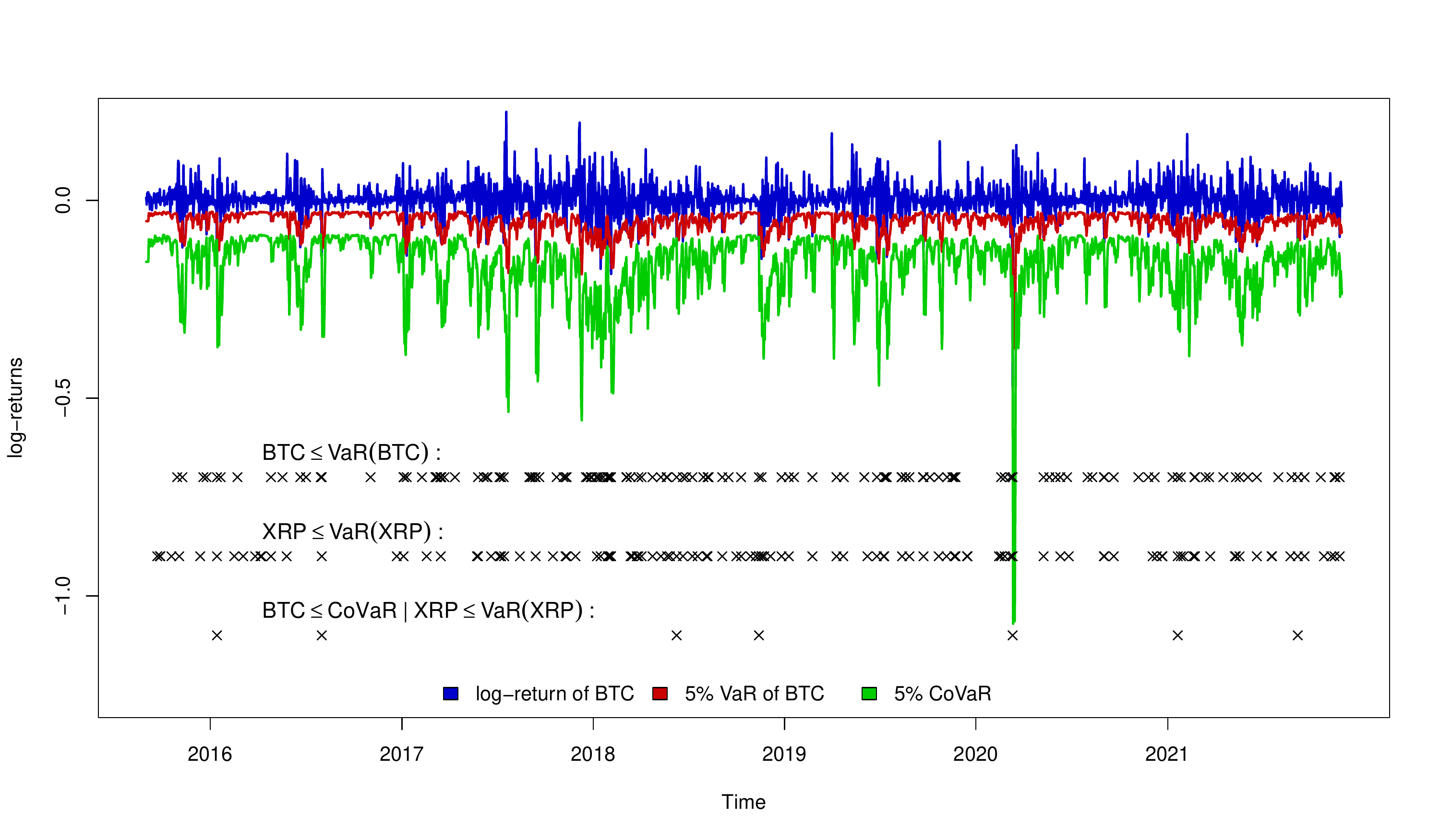}
    \caption{CoVaR of BTC-XRP using a time-invariant $t$-copula.}
\end{figure}

% ETH
\begin{figure}[H]
    \centering
    \includegraphics[width=\textwidth]{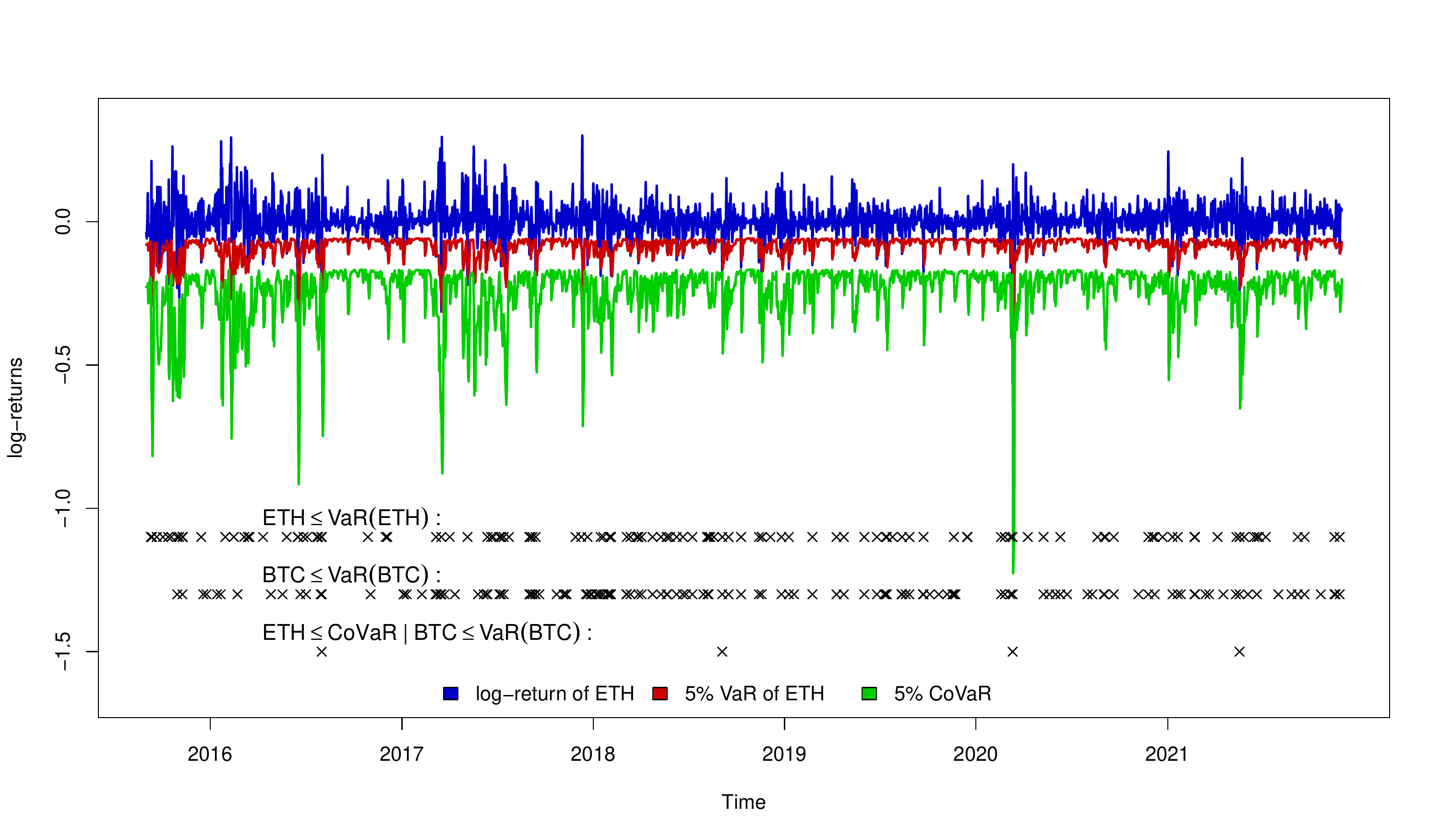}
    \caption{CoVaR of ETH-BTC using a time-invariant $t$-copula.}
\end{figure}

\begin{figure}[H]
    \centering
    \includegraphics[width=\textwidth]{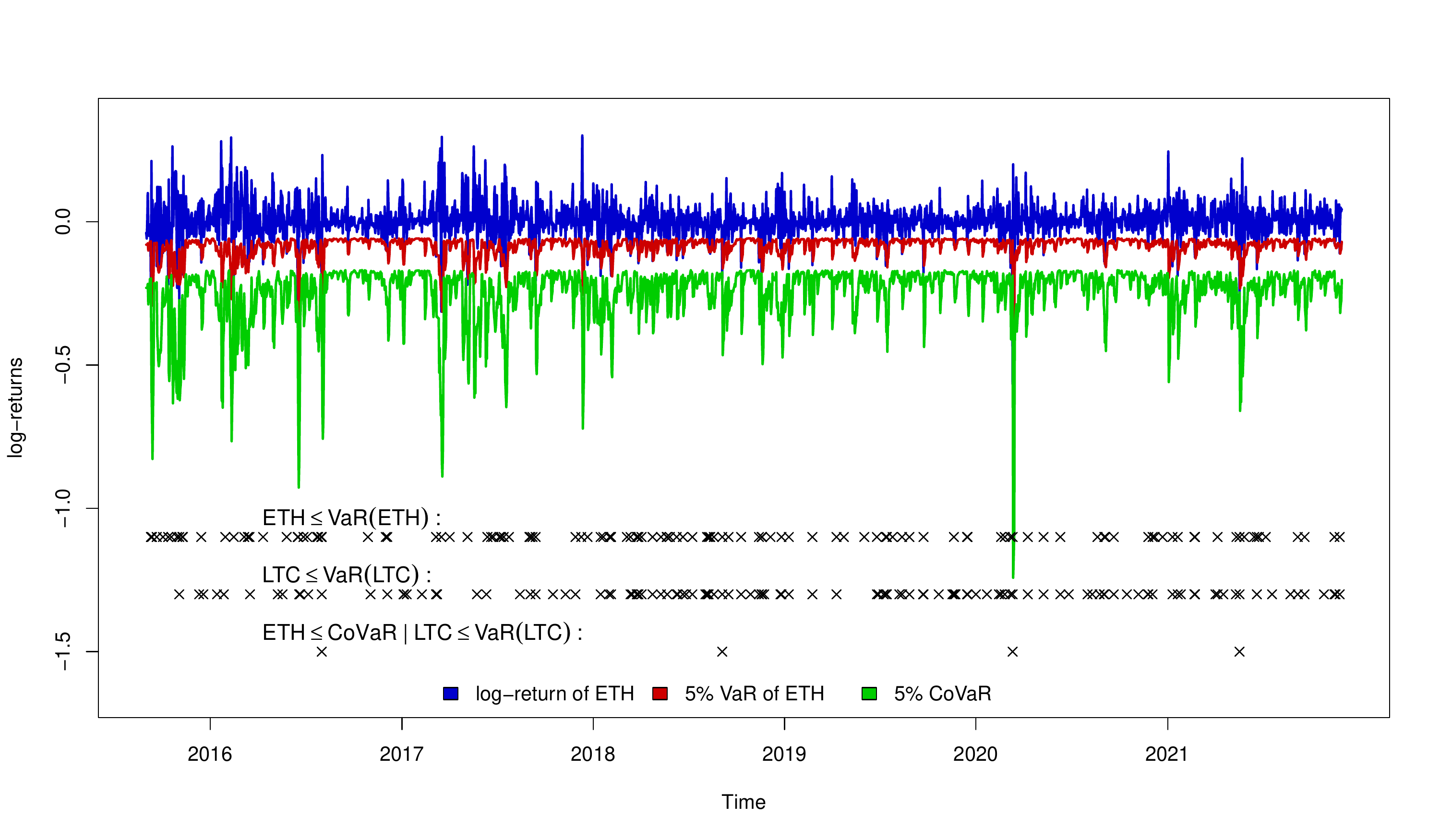}
    \caption{CoVaR of ETH-LTC using a time-invariant $t$-copula.}
\end{figure}

\begin{figure}[H]
    \centering
    \includegraphics[width=\textwidth]{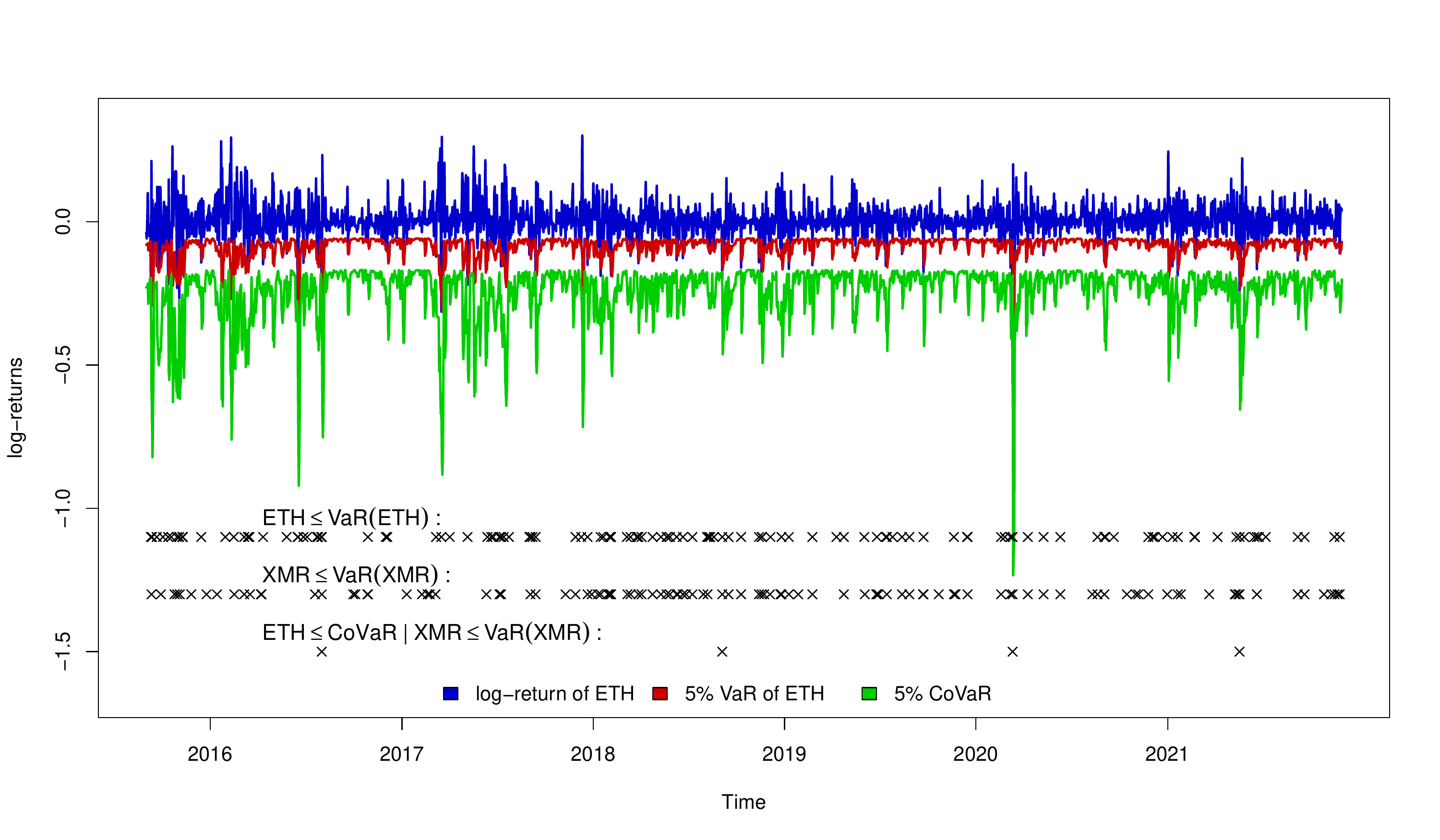}
    \caption{CoVaR of ETH-XMR using a time-invariant $t$-copula.}
\end{figure}

\begin{figure}[H]
    \centering
    \includegraphics[width=\textwidth]{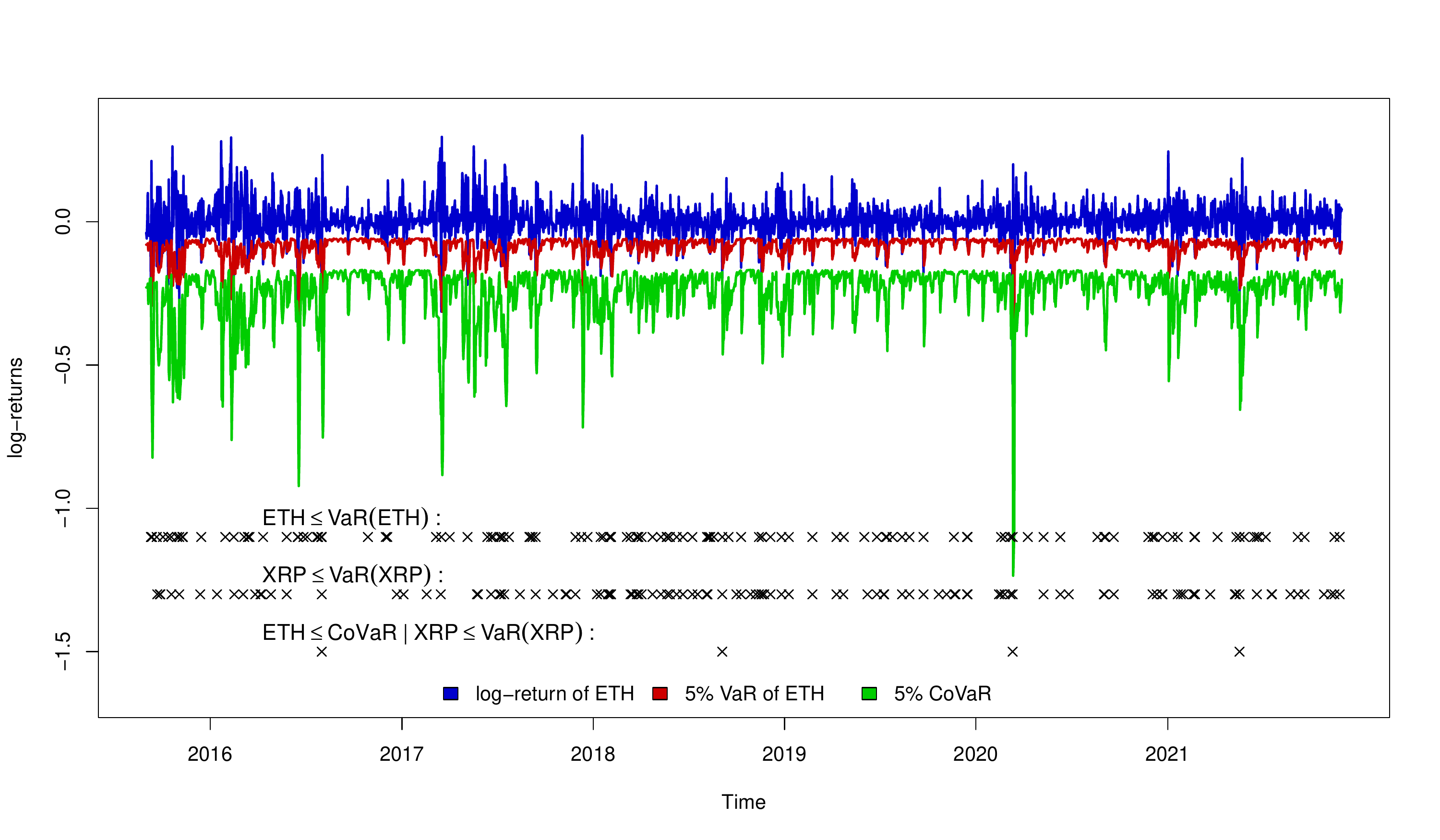}
    \caption{CoVaR of ETH-XRP using a time-invariant $t$-copula.}
\end{figure}

\begin{figure}[H]
    \centering
    \includegraphics[width=\textwidth]{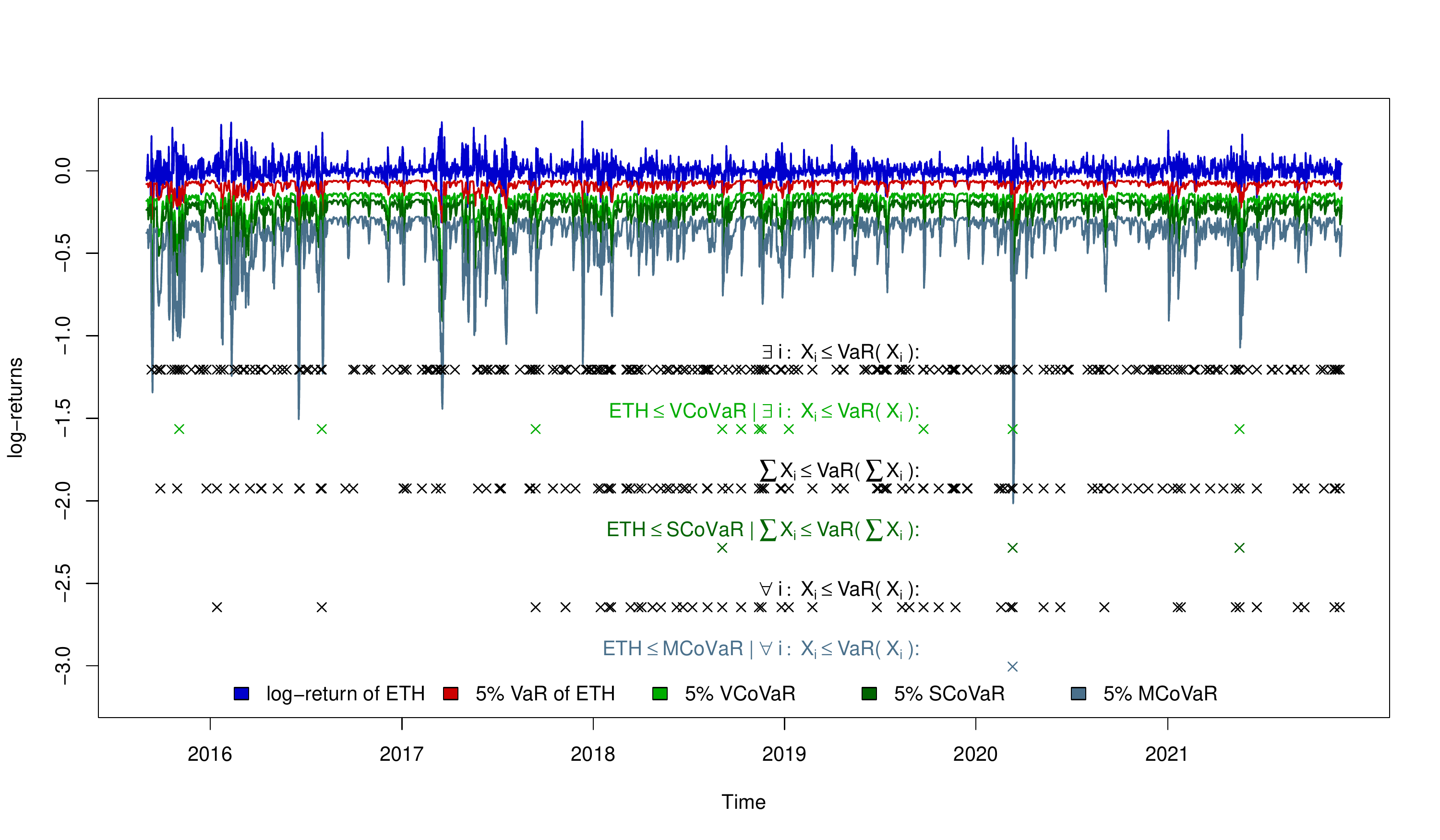}
    \caption{SCoVaR, MCoVaR, and VCoVaR of ETH using a time-invariant $t$-copula.}
\end{figure}

% LTC
\begin{figure}[H]
    \centering
    \includegraphics[width=\textwidth]{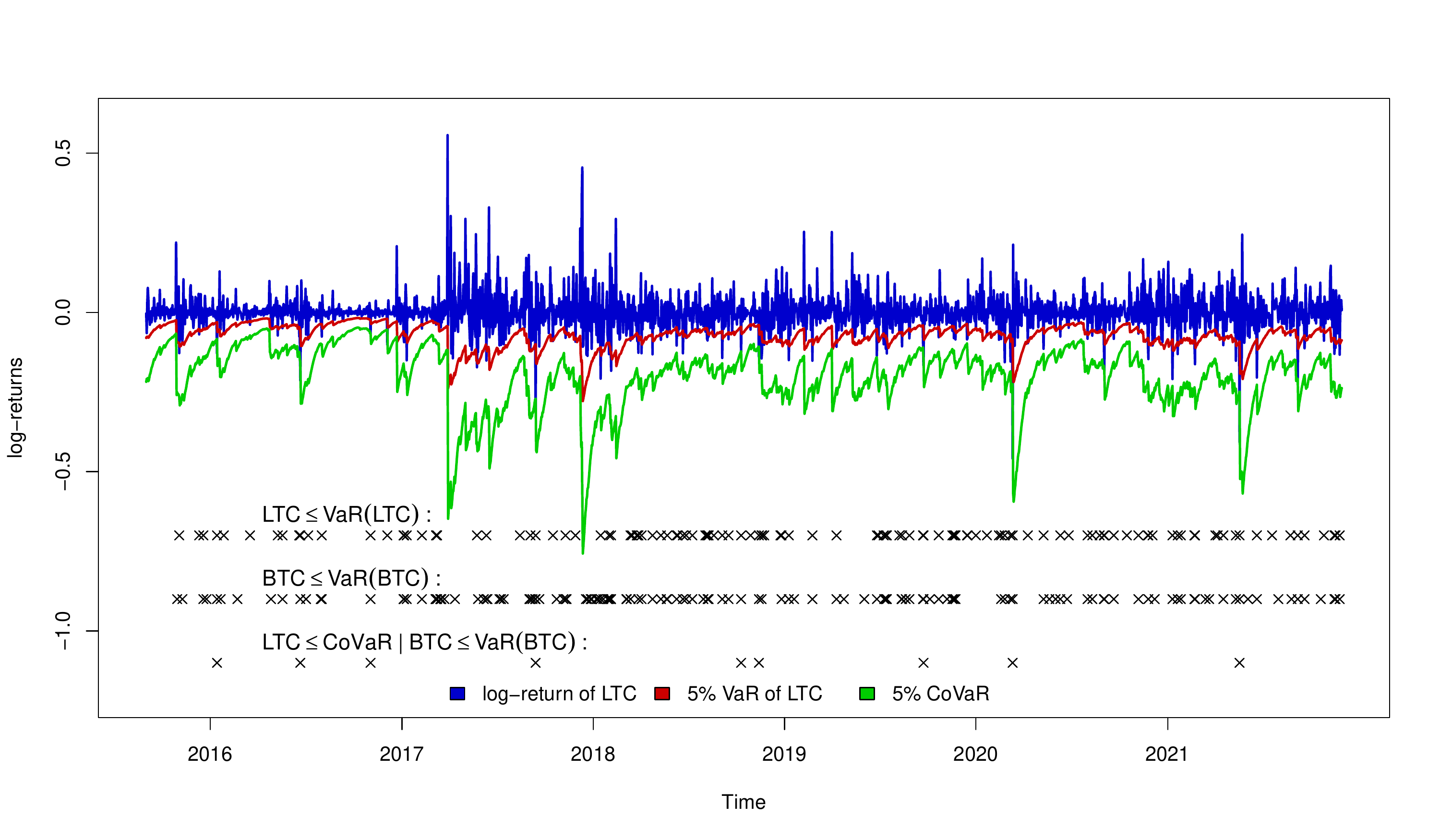}
    \caption{CoVaR of LTC-BTC using a time-invariant $t$-copula.}
\end{figure}

\begin{figure}[H]
    \centering
    \includegraphics[width=\textwidth]{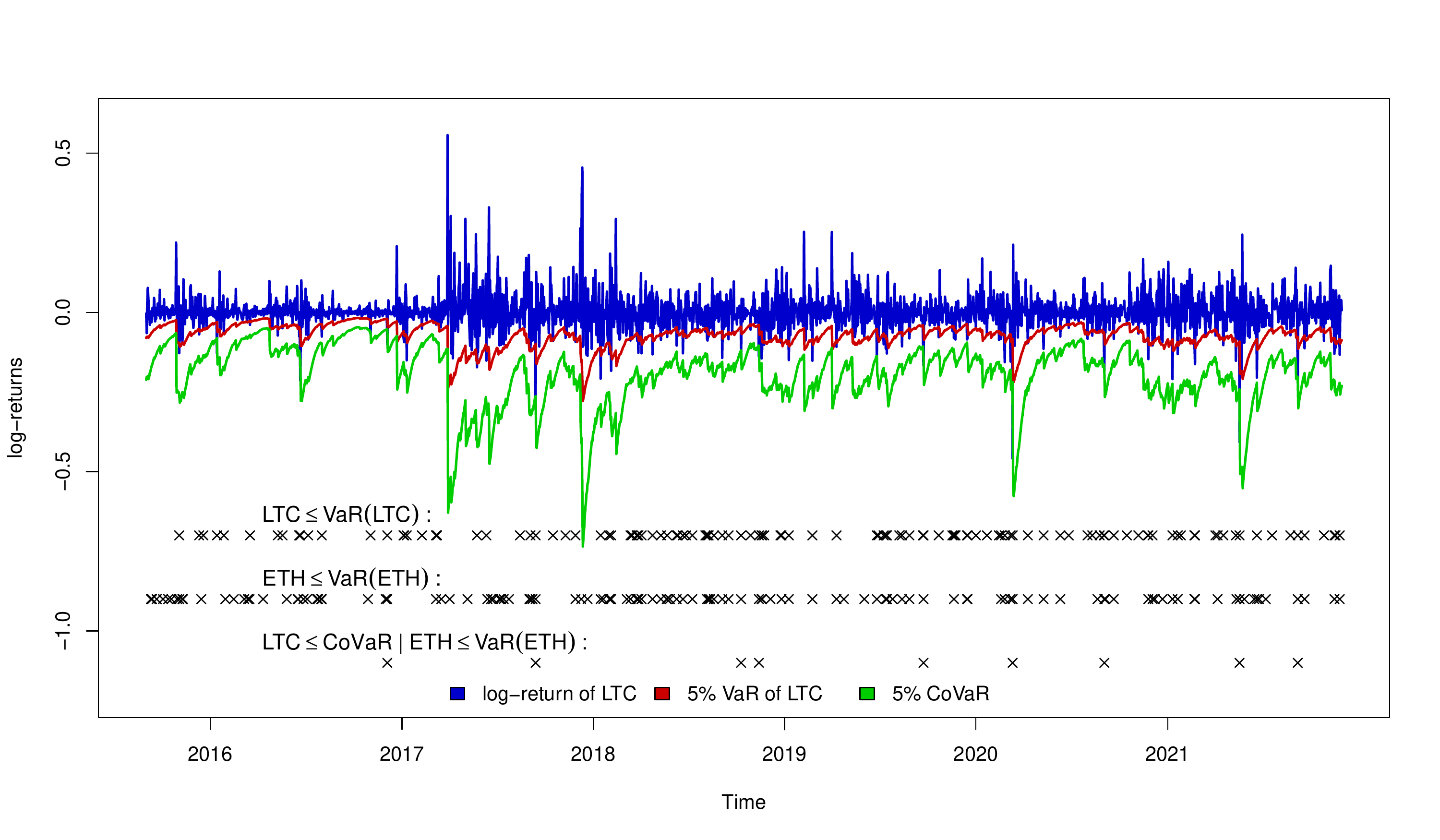}
    \caption{CoVaR of LTC-ETH using a time-invariant $t$-copula.}
\end{figure}

\begin{figure}[H]
    \centering
    \includegraphics[width=\textwidth]{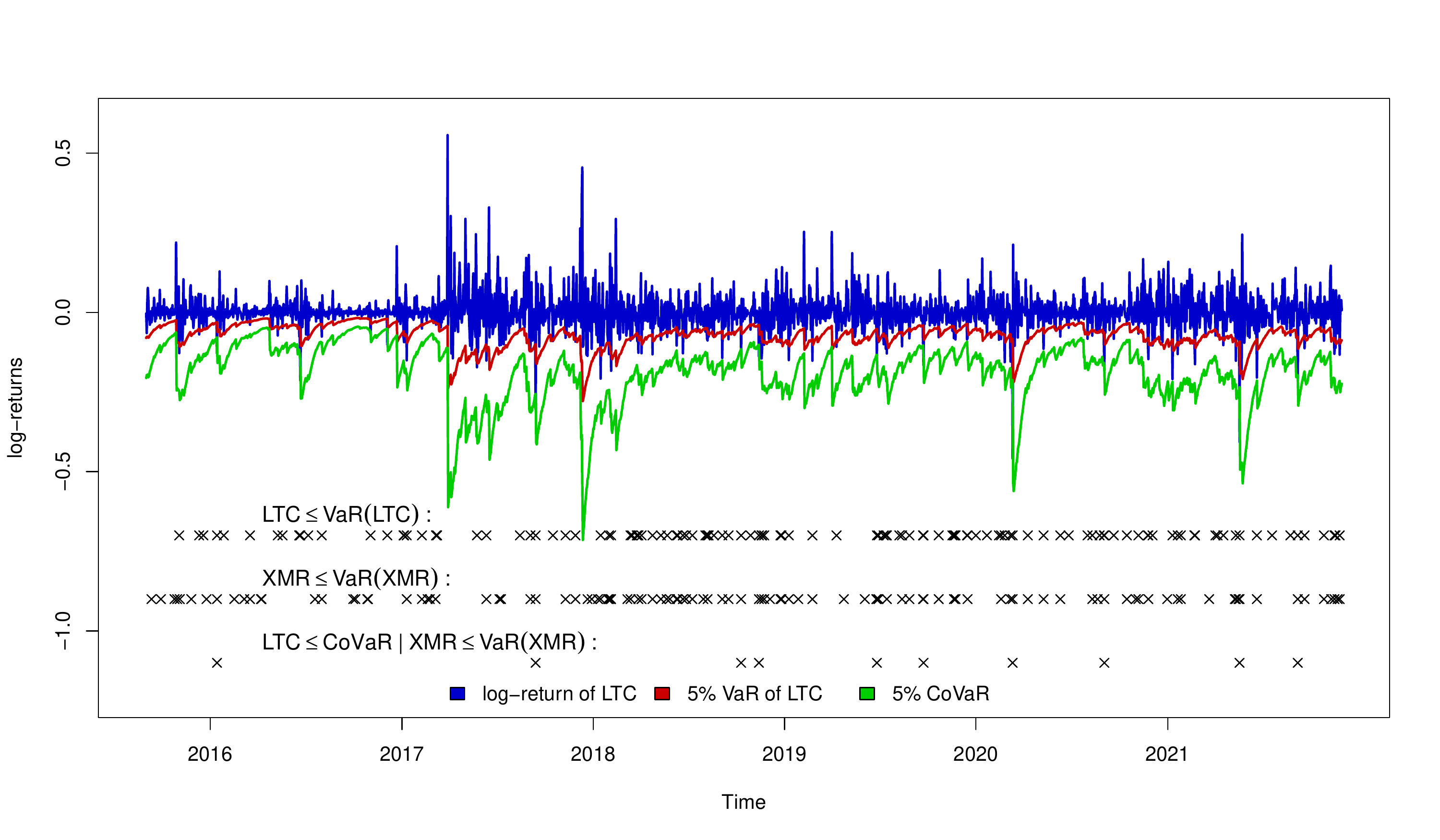}
    \caption{CoVaR of LTC-XMR using a time-invariant $t$-copula.}
\end{figure}

\begin{figure}[H]
    \centering
    \includegraphics[width=\textwidth]{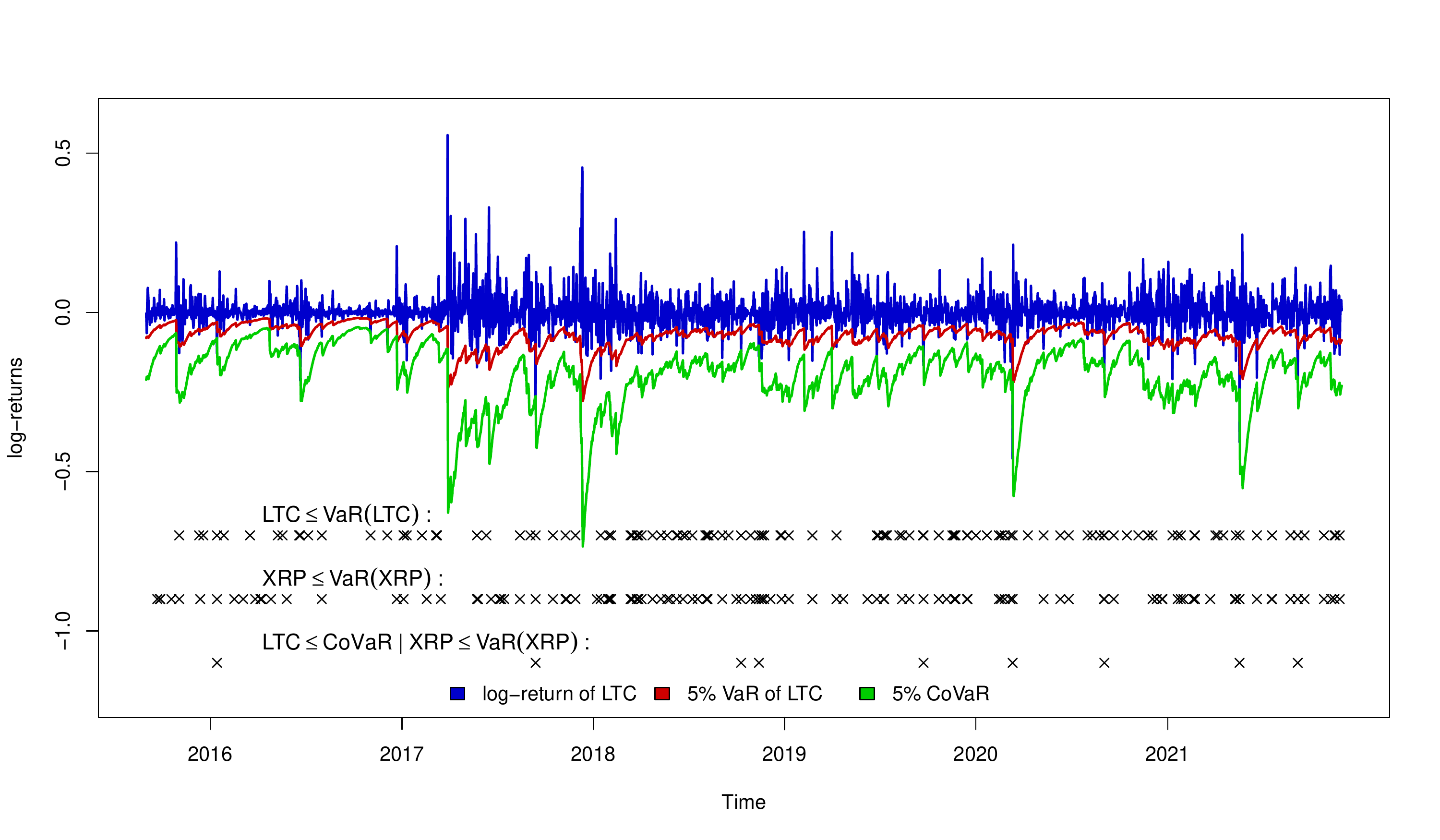}
    \caption{CoVaR of LTC-XRP using a time-invariant $t$-copula.}
\end{figure}

\begin{figure}[H]
    \centering
    \includegraphics[width=\textwidth]{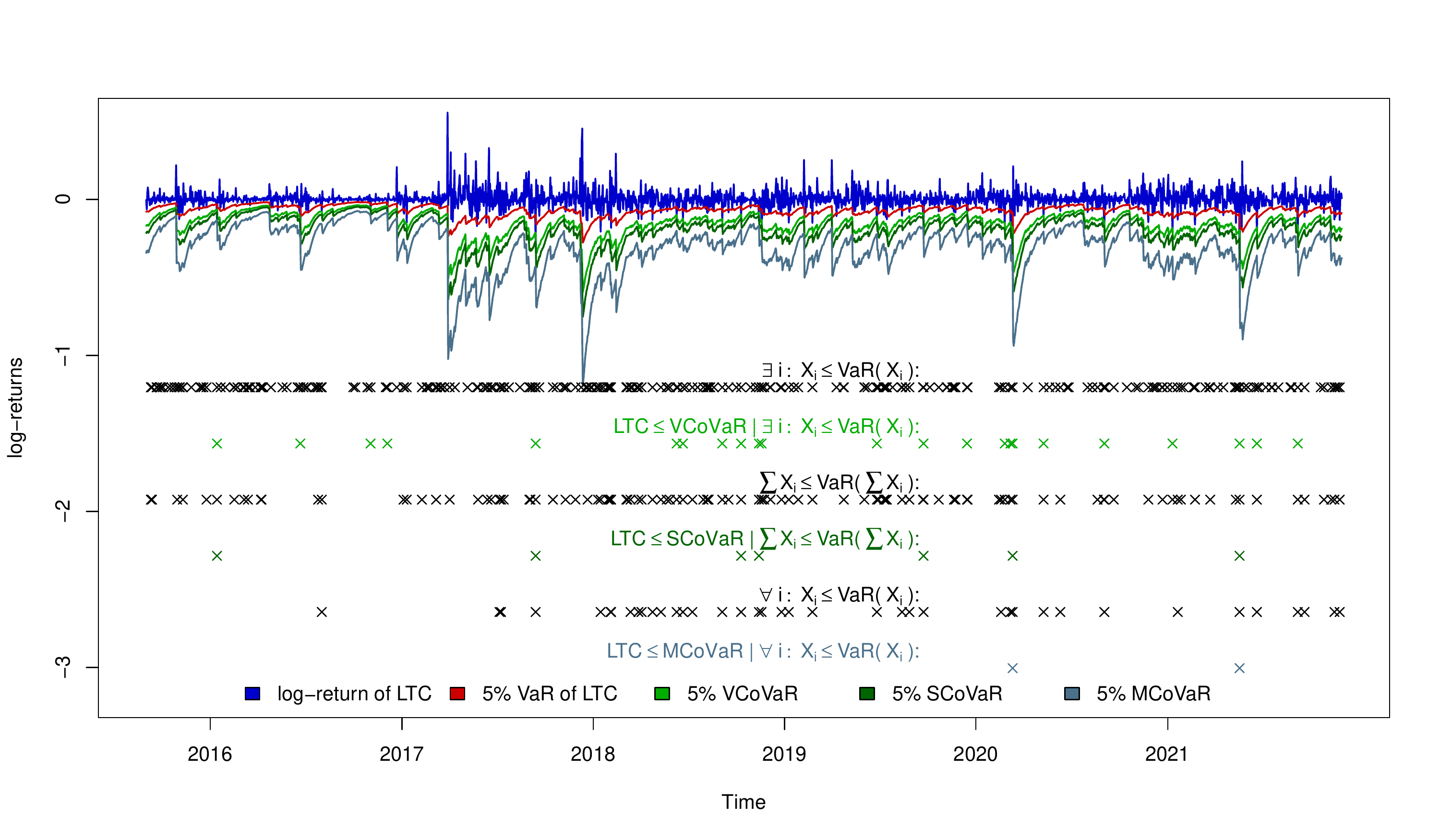}
    \caption{SCoVaR, MCoVaR, and VCoVaR of LTC using a time-invariant $t$-copula.}
\end{figure}

% XMR
\begin{figure}[H]
    \centering
    \includegraphics[width=\textwidth]{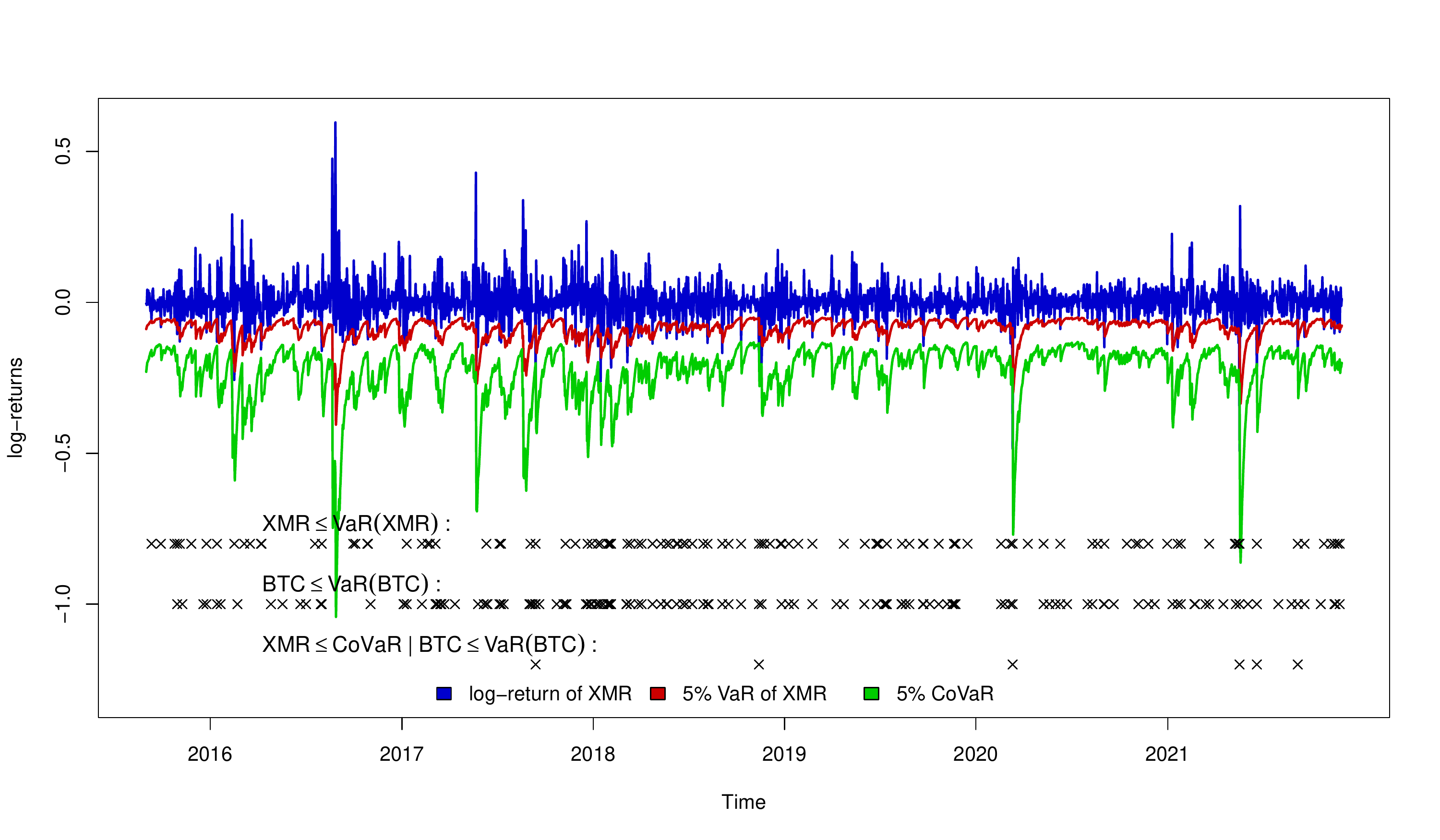}
    \caption{CoVaR of XMR-BTC using a time-invariant $t$-copula.}
\end{figure}

\begin{figure}[H]
    \centering
    \includegraphics[width=\textwidth]{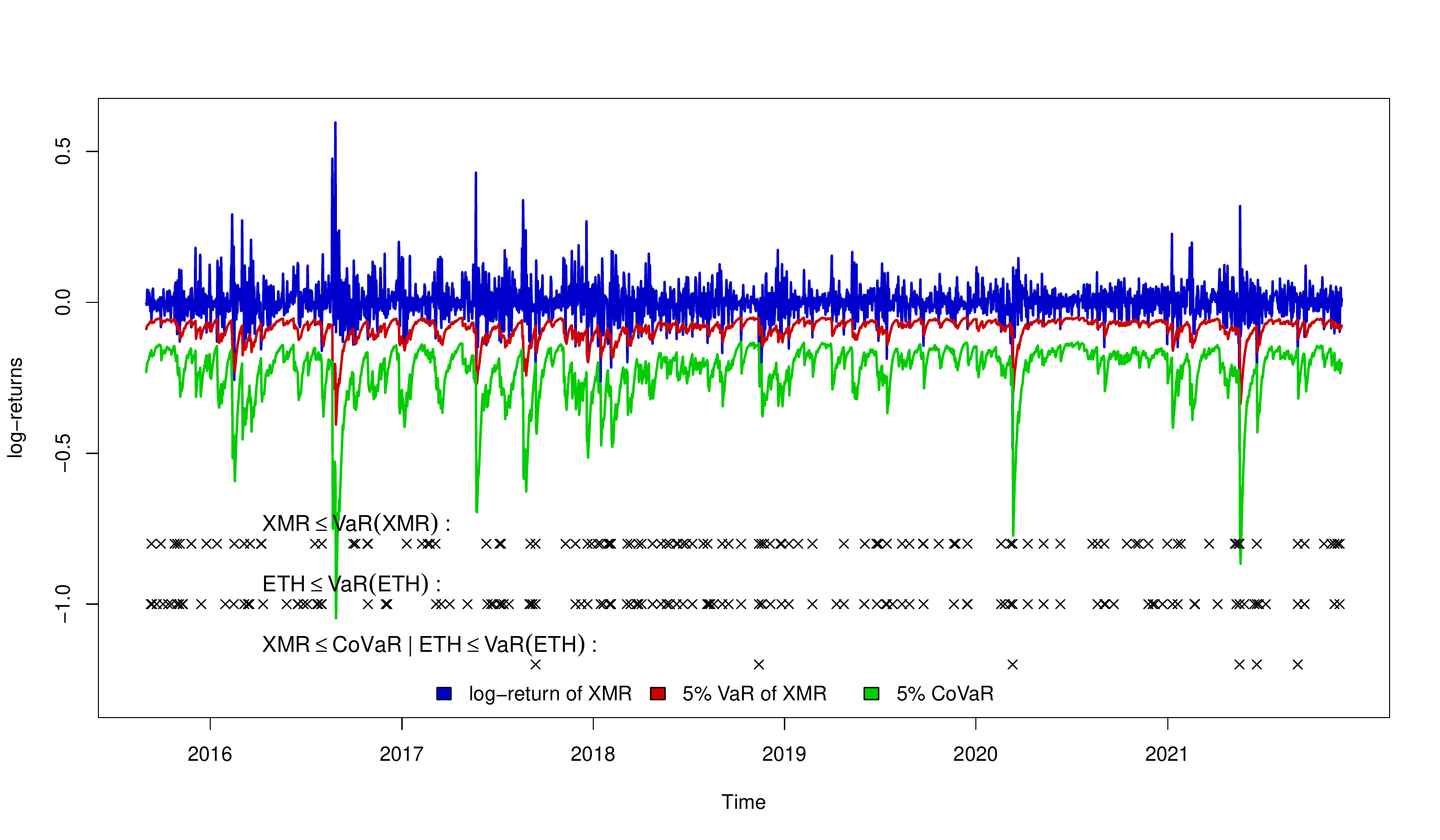}
    \caption{CoVaR of XMR-ETH using a time-invariant $t$-copula.}
\end{figure}

\begin{figure}[H]
    \centering
    \includegraphics[width=\textwidth]{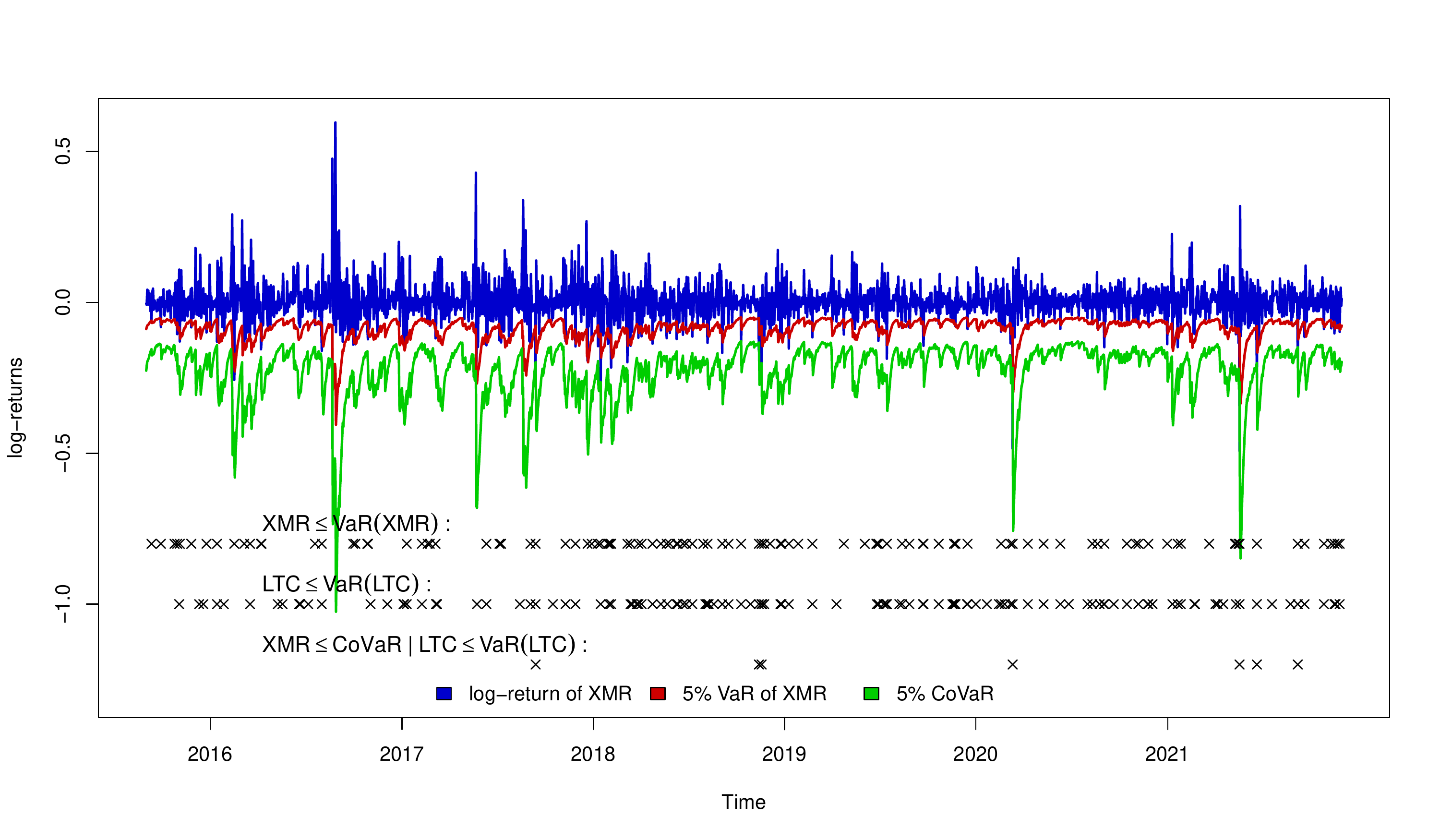}
    \caption{CoVaR of XMR-LTC using a time-invariant $t$-copula.}
\end{figure}

\begin{figure}[H]
    \centering
    \includegraphics[width=\textwidth]{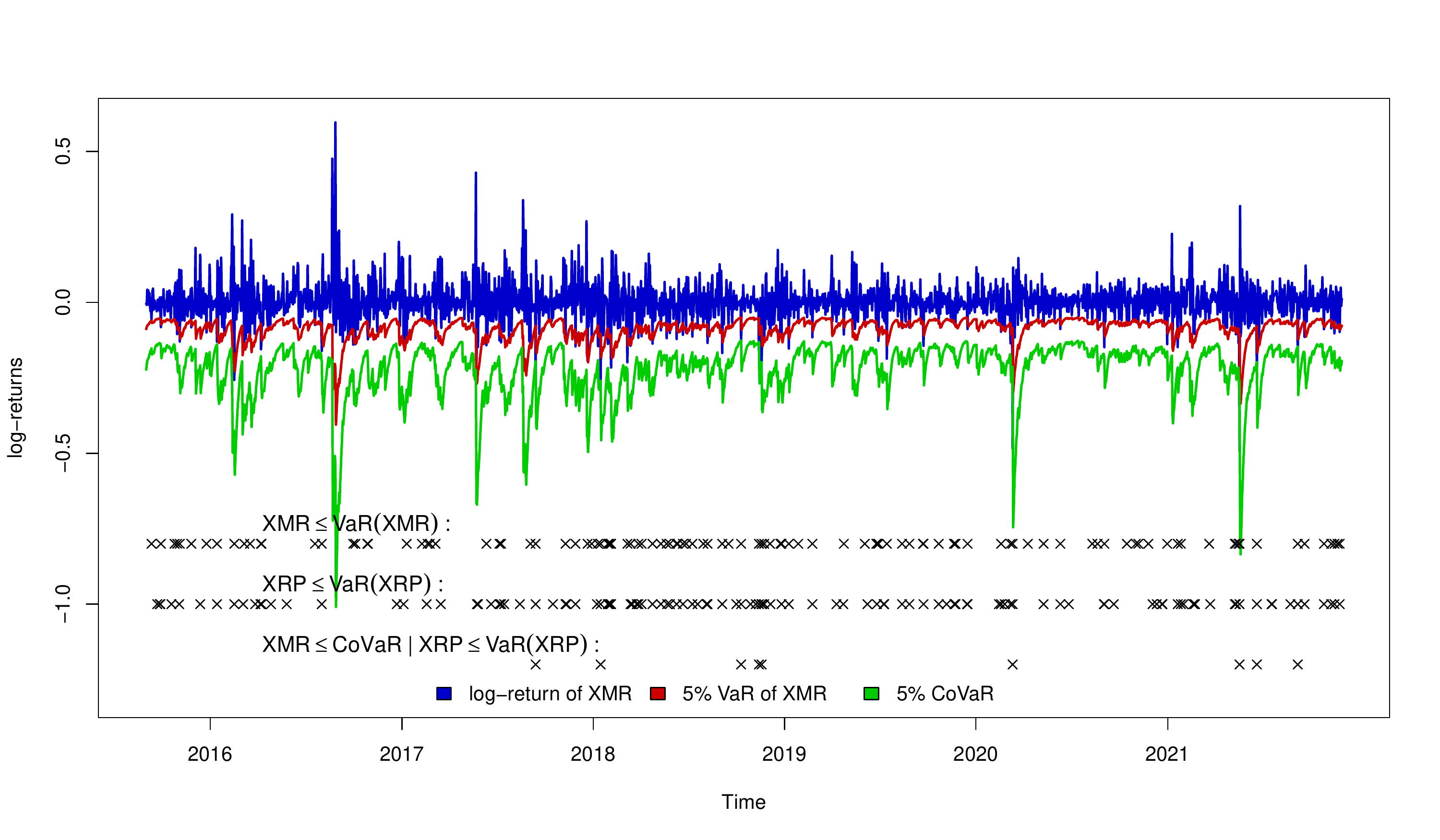}
    \caption{CoVaR of XMR-XRP using a time-invariant $t$-copula.}
\end{figure}

\begin{figure}[H]
    \centering
    \includegraphics[width=\textwidth]{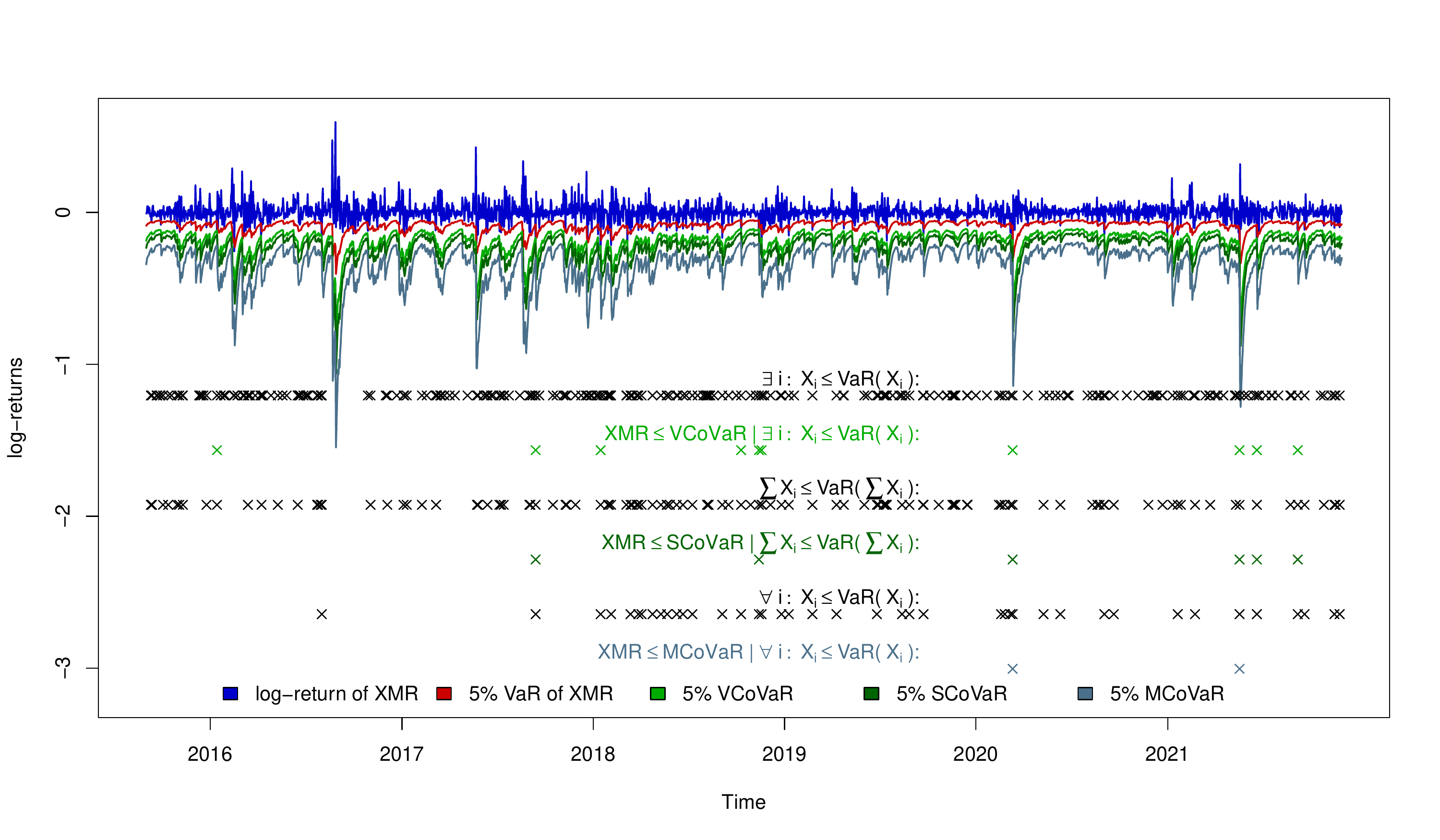}
    \caption{SCoVaR, MCoVaR, and VCoVaR of XMR using a time-invariant $t$-copula.}
\end{figure}

% XRP
\begin{figure}[H]
    \centering
    \includegraphics[width=\textwidth]{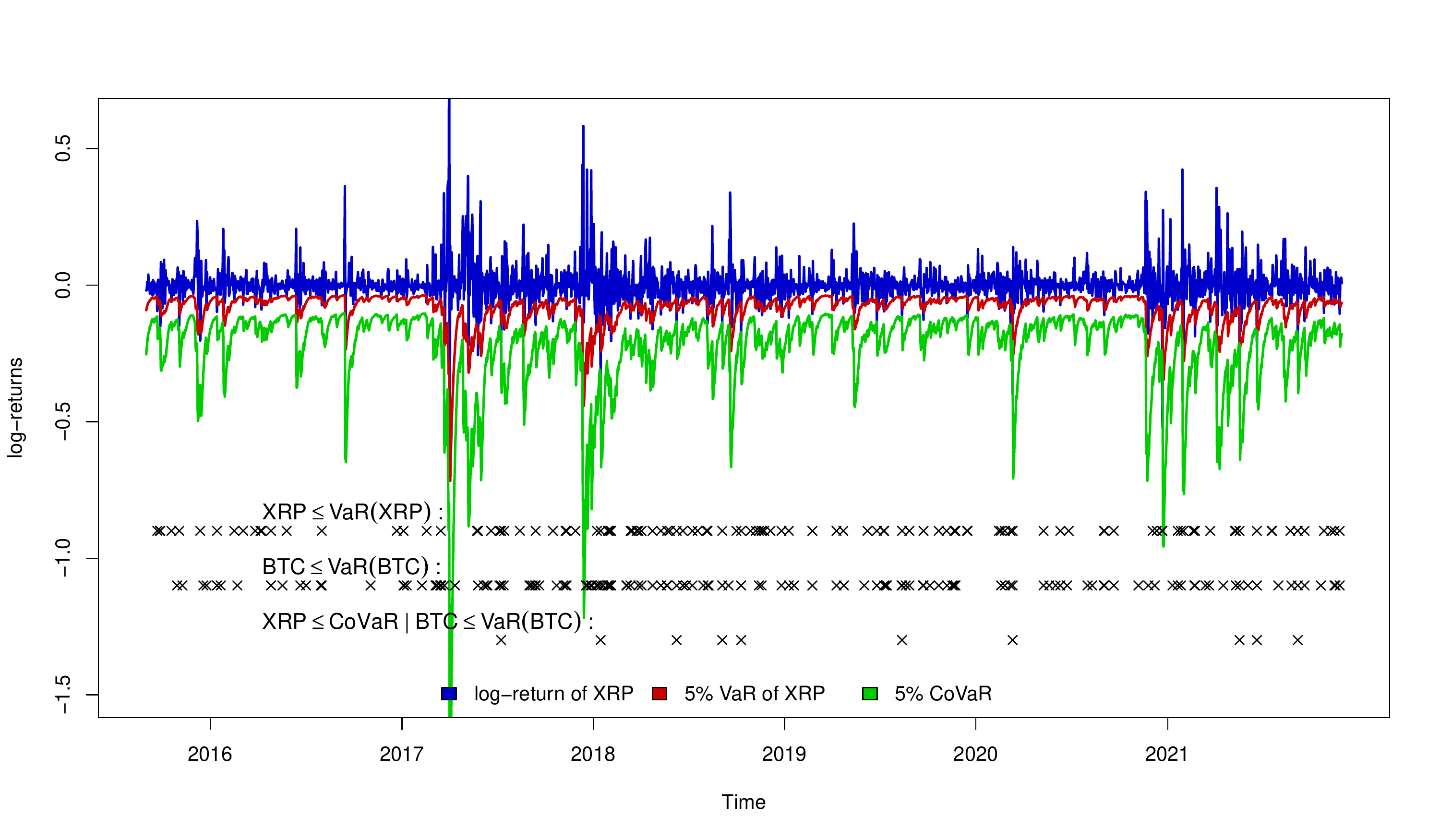}
    \caption{CoVaR of XRP-BTC using a time-invariant $t$-copula.}
\end{figure}

\begin{figure}[H]
    \centering
    \includegraphics[width=\textwidth]{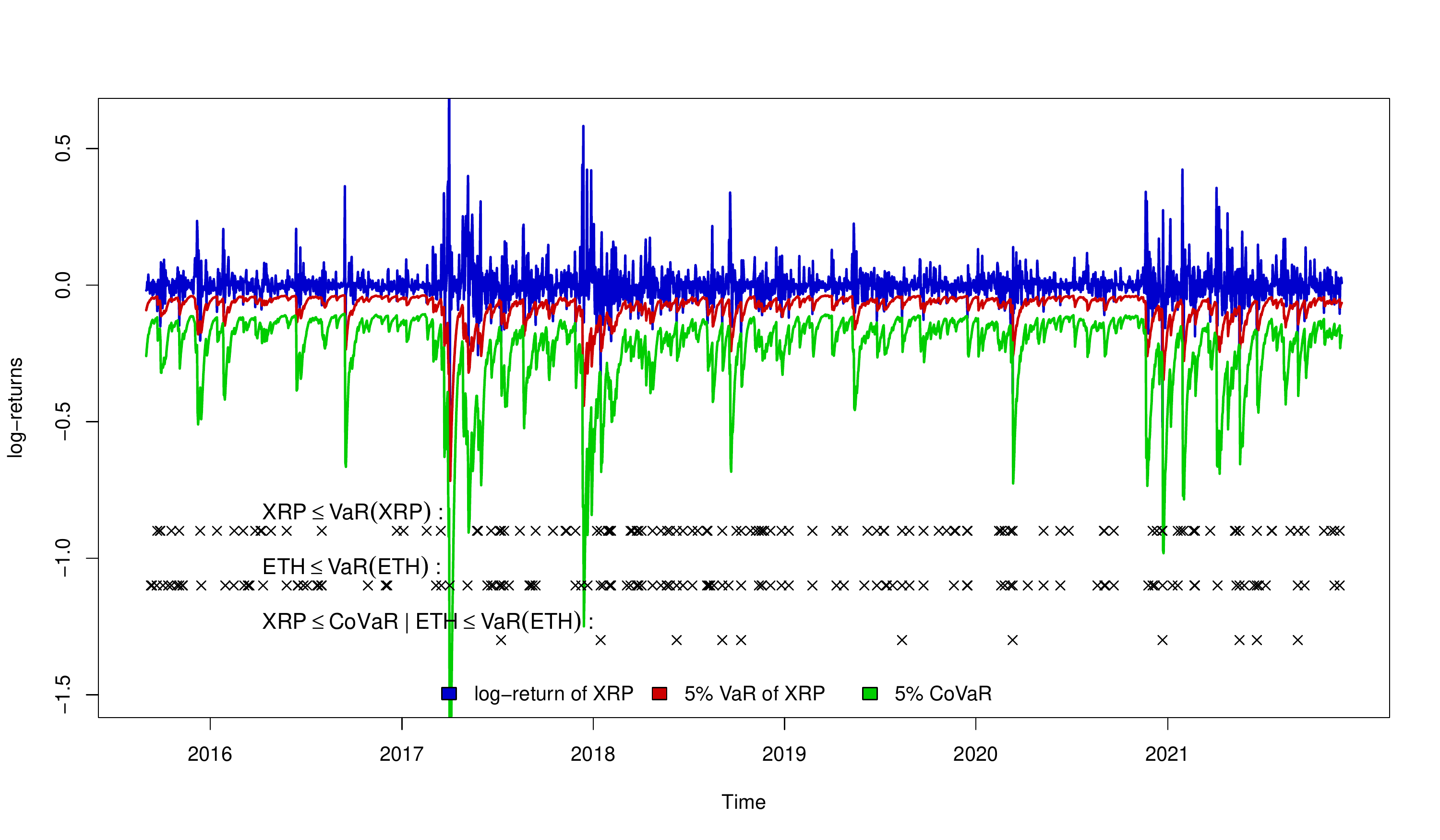}
    \caption{CoVaR of XRP-ETH using a time-invariant $t$-copula.}
\end{figure}

\begin{figure}[H]
    \centering
    \includegraphics[width=\textwidth]{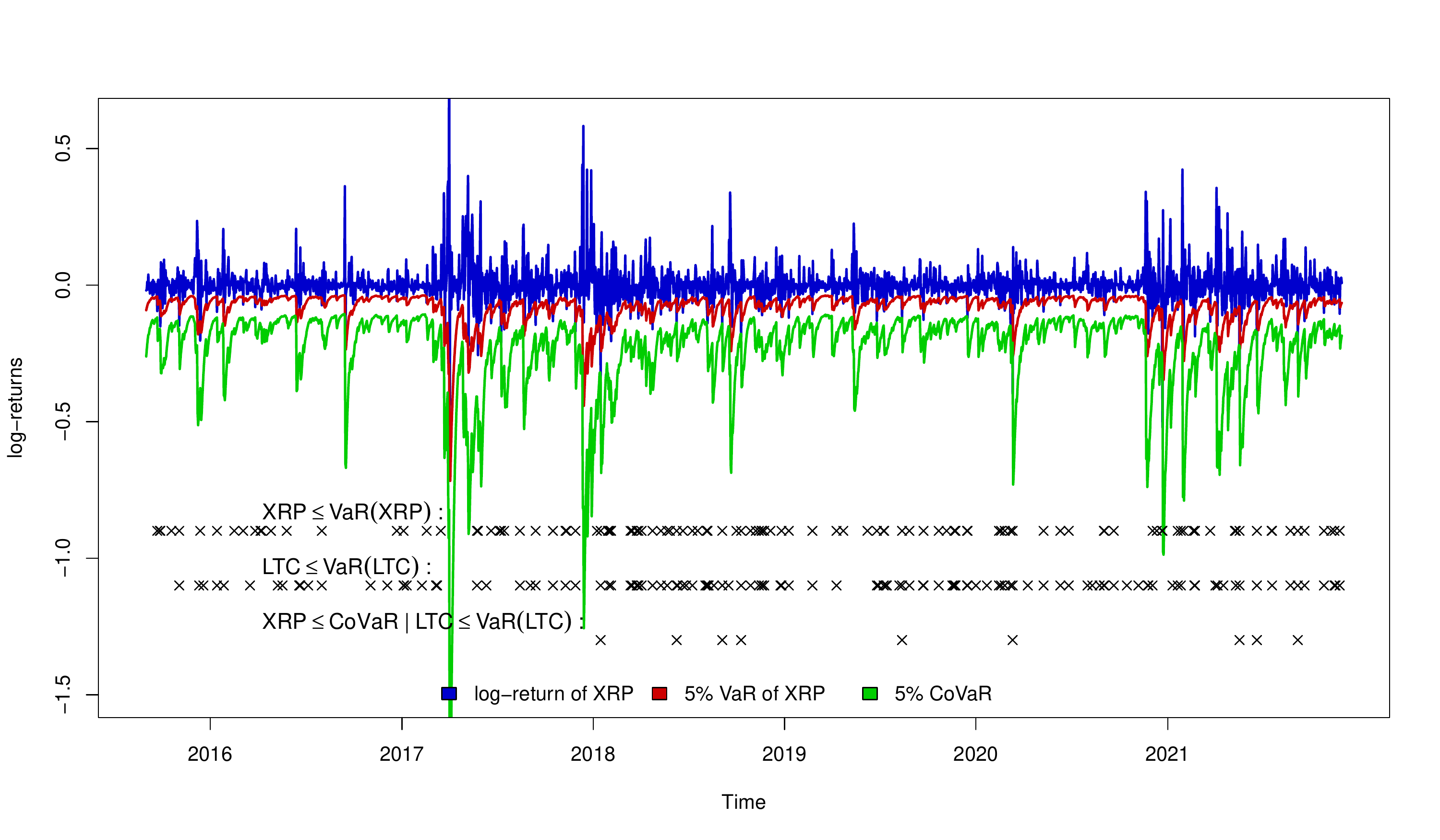}
    \caption{CoVaR of XRP-LTC using a time-invariant $t$-copula.}
\end{figure}

\begin{figure}[H]
    \centering
    \includegraphics[width=\textwidth]{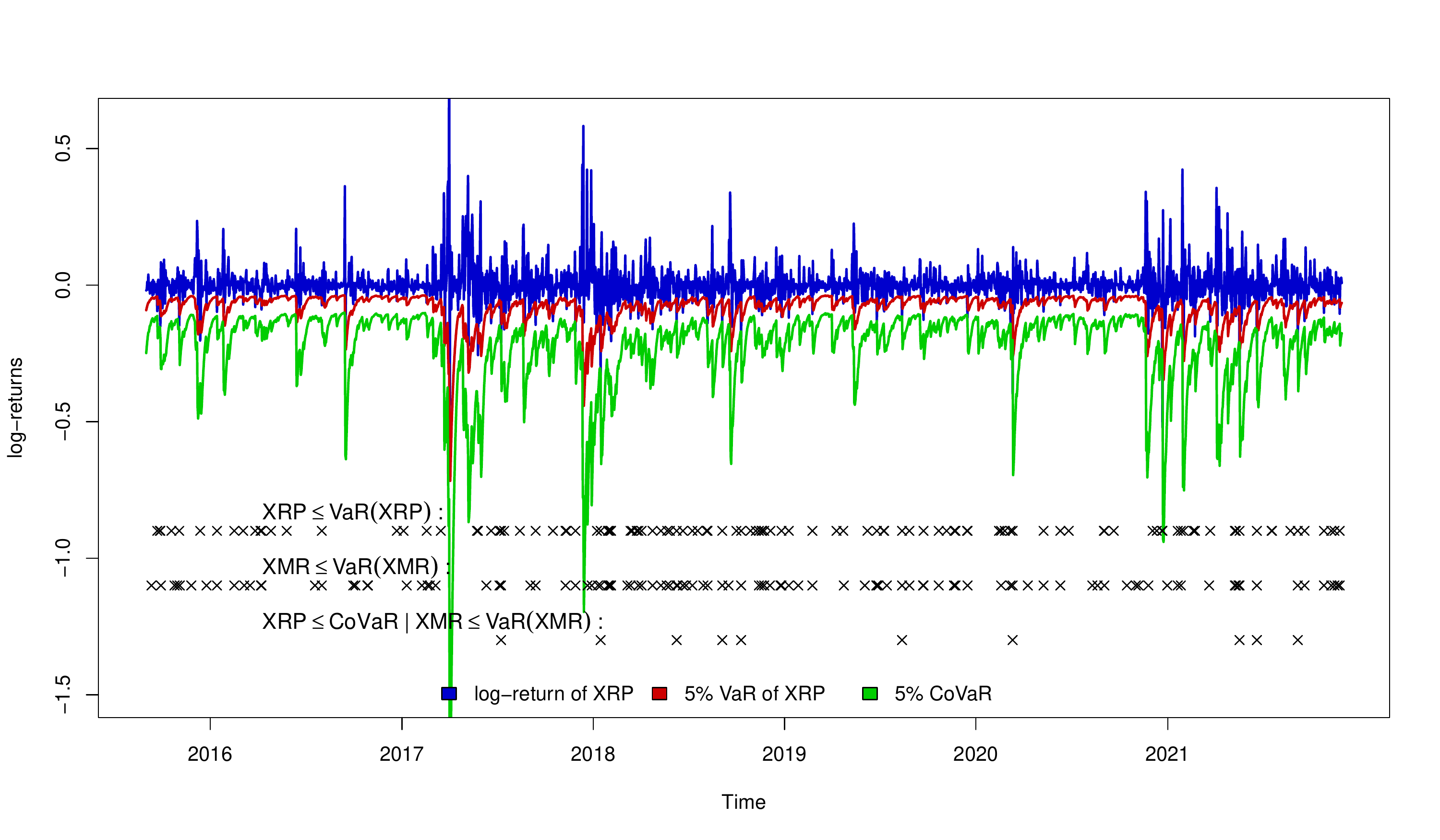}
    \caption{CoVaR of XRP-XMR using a time-invariant $t$-copula.}
\end{figure}

\begin{figure}[H]
    \centering
    \includegraphics[width=\textwidth]{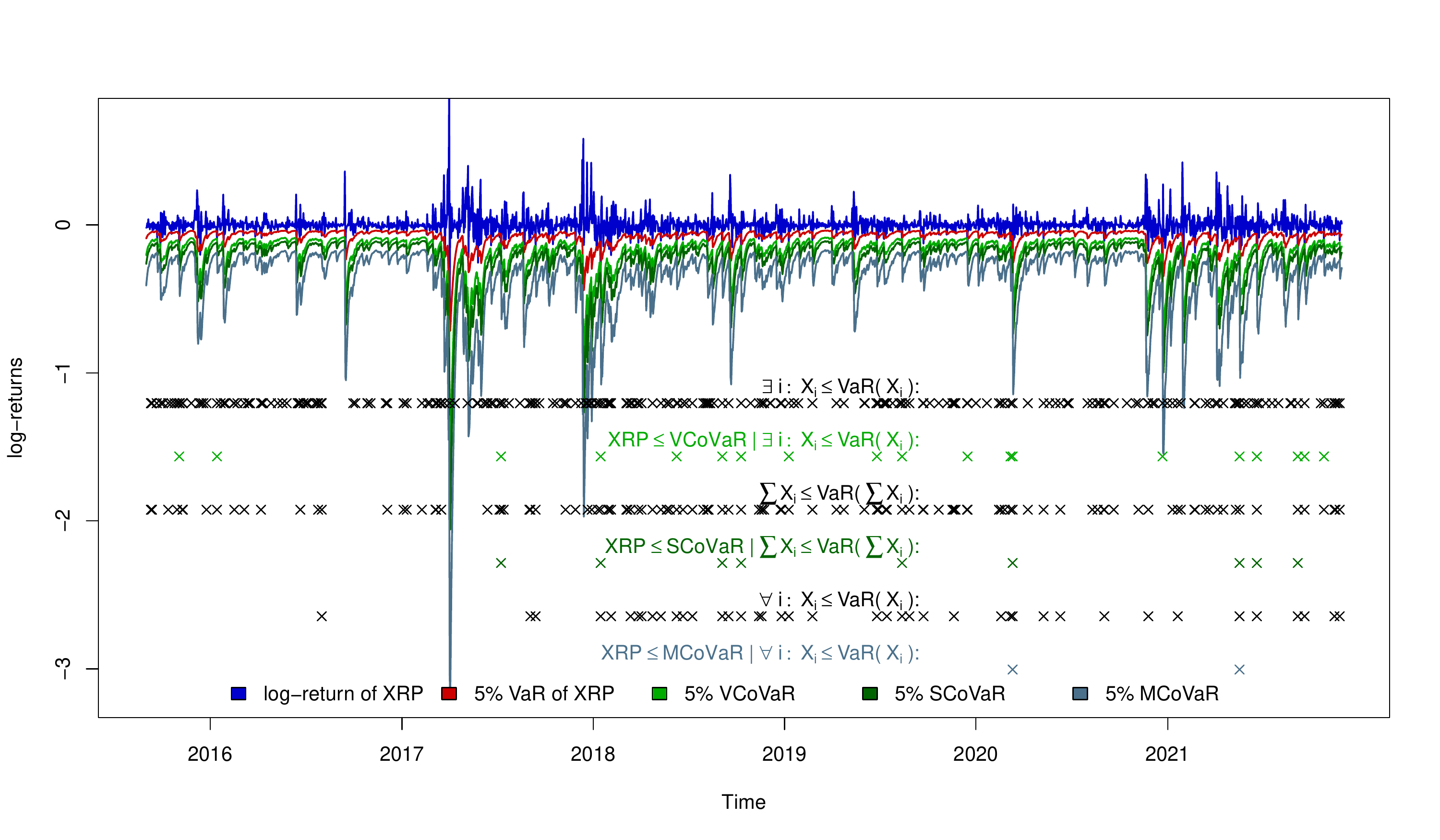}
    \caption{SCoVaR, MCoVaR, and VCoVaR of XRP using a time-invariant $t$-copula.}
\end{figure}

%% file: tables/stats_tests_logreturns.tex
\begin{table}[H]
\centering
 \resizebox{0.95\textwidth}{!}{
\begin{threeparttable}
\caption{Descriptive statistics and tests of the log-returns of the CC.}
\begin{tabular}{lrrrrrr}
\toprule[0.05em]
\midrule
& BTC & ETH & LTC & XMR & XRP \\ 
\midrule
Min & -0.4706 & -0.5656 & -0.4588 & -0.4922 & -0.6365 \\
 Mean & 0.0024 & 0.0036 & 0.0019 & 0.0027 & 0.0021 \\ 
  Median & 0.0024 & 0.0013 & -0.0005 & 0.0018 & -0.0015 \\ 
  Max & 0.2241 & 0.3006 & 0.5568 & 0.5963 & 1.0087 \\ 
  Sd & 0.0399 & 0.0615 & 0.0567 & 0.0620 & 0.0705 \\ 
  Kurtosis & 11.2569 & 6.7074 & 12.4750 & 11.2825 & 30.2736 \\ 
  Skewness & -0.7495 & -0.3117 & 0.7160 & 0.5130 & 1.9136 \\ 
  \midrule
  $p$-value: Jarque-Bera & 0.0000* & 0.0000* & 0.0000* & 0.0000* & 0.0000* \\ 
  $p$-value: Ljung-Box (8 lags) & 0.0950\phantom{*} & 0.0046* & 0.0004* & 0.0000* & 0.0001* \\ 
  \midrule
  Statistics: ADF & -15.7728* & -15.7839* & -16.0902* & -14.7867* & -13.2823* \\ 
  Statistics: PP & -49.6547* & -49.9640* & -48.7192* & -51.2537* & -49.6844* \\ 
  Statistics: KPSS & 0.1178 & 0.2188 & 0.1194 & 0.3638 & 0.1115 \\ 
  \bottomrule[0.05em]
\end{tabular}
\label{table:des_stat_tests}
\end{threeparttable}
}
\end{table}

\begin{table}[H]
\centering
 \resizebox{0.95\textwidth}{!}{
\begin{threeparttable}
\caption{Descriptive statistics and tests of the log-returns of the systems.}
\begin{tabular}{lrrrrrr}
\toprule[0.05em]
\midrule
& Sys:BTC & Sys:ETH & Sys:LTC & Sys:XMR & Sys:XRP \\ 
  \hline
Min & -1.9056 & -1.8106 & -1.9174 & -1.8918 & -1.9793 \\ Mean & 0.0103 & 0.0092 & 0.0109 & 0.0100 & 0.0106 \\ 
  Median & 0.0122 & 0.0090 & 0.0133 & 0.0111 & 0.0141 \\ 
  Max & 1.3151 & 1.0483 & 0.9200 & 1.2294 & 0.9068 \\ 
  Sd & 0.1950 & 0.1816 & 0.1812 & 0.1818 & 0.1807 \\
  \midrule
  Kurtosis & 8.5834 & 8.8517 & 8.8186 & 9.2215 & 9.7159 \\ 
  Skewness & -0.6068 & -0.6277 & -0.7920 & -0.5603 & -0.9170 \\ 
  \midrule
  $p$-value: Jarque-Bera & 0.0000* & 0.0000* & 0.0000* & 0.0000* & 0.0000* \\ 
  $p$-value: Ljung-Box (8 lags) & 0.0000* & 0.0000* & 0.0000* & 0.0001* & 0.0000* \\ 
  \midrule
  Statistics: ADF & -14.8125* & -14.8316* & -14.6531* & -15.0416* & -15.6975* \\ 
  Statistics: PP & -50.5398* & -50.8404* & -51.1797* & -50.3953* & -50.6902* \\ 
  Statistics: KPSS & 0.2609 & 0.2261 & 0.2738 & 0.1905 & 0.2673 \\ 
 \bottomrule[0.05em]
\end{tabular}
\label{table:des_stat_tests_systems}
\end{threeparttable}
}
\end{table}

%% file: tables/tau.tex
\begin{table}[H]
\centering
\caption{Estimates of Kendall's $\tau$ between log-returns.}
\resizebox{\textwidth}{!}{
\begin{tabular}{lrrrrrrrrrr}
\toprule[0.05em]
\midrule
 & BTC & ETH & LTC & XMR & XRP & Sys:BTC & Sys:ETH & Sys:LTC & Sys:XMR & Sys:XRP \\ 
\midrule
BTC & 1.0000 & 0.4218 & 0.5558 & 0.4278 & 0.3805 & 0.5070 & 0.5978 & 0.5570 & 0.5892 & 0.6049 \\ 
 ETH & & 1.0000 & 0.4558 & 0.4206 & 0.4166 & 0.6428 & 0.5013 & 0.6469 & 0.6617 & 0.6560 \\ 
 LTC & & & 1.0000 & 0.4228 & 0.4597 & 0.6262 & 0.6641 & 0.5467 & 0.6698 & 0.6543 \\ 
 XMR & & & & 1.0000 & 0.3737 & 0.6241 & 0.6416 & 0.6391 & 0.4794 & 0.6460 \\ 
 XRP & & & & & 1.0000 & 0.6120 & 0.6170 & 0.6061 & 0.6273 & 0.4655 \\ 
 Sys:BTC & & & & & & 1.0000 & 0.8298 & 0.8858 & 0.8283 & 0.8286 \\ 
 Sys:ETH & & & & & & & 1.0000 & 0.8073 & 0.7718 & 0.7788 \\ 
 Sys:LTC & & & & & & & & 1.0000 & 0.7997 & 0.8157 \\ 
 Sys:XMR & & & & & & & & & 1.0000 & 0.7704 \\ 
 Sys:XRP & & & & & & & & & & 1.0000 \\ 
  \bottomrule[0.05em]
\end{tabular}}
\label{table:tau_log_ret}
\end{table}

%% file: tables/margins.tex
\begin{table}[htp]
\centering
\captionsetup{font=large}
\resizebox{1.0\textwidth}{!}{
\begin{threeparttable}
\caption{Parameter estimates, AIC, and tests of the univariate models.}
\begin{tabular}{lrrrrrr}
\toprule[0.05em]
\midrule
 & BTC & ETH & LTC & XMR & XRP\\
\midrule
 Selected Model & \makecell{ARMA(2,2)-\\GJR-ARCH(5)} & ARCH(3) & GARCH(1,1) & GARCH(1,1) & GARCH(1,1)\\
 \midrule 
  $\hat\mu$ & 0.0024* (0.0012) & 0.0028* (0.0011) &  & 0.0019 (0.0010) &  \\ 
  $\hat\phi_1$  & -0.0043\phantom{*} (0.0036) &  &  &  &  \\ 
  $\hat\phi_2$ & 0.9846* (0.0035) &  &  &  &  \\ 
  $\hat\psi_1$  & 0.0162* (0.0003) &  &  &  &  \\ 
  $\hat\psi_2$  & -0.9759* (0.0000) &  &  &  &  \\ 
  \midrule
  $\hat\omega$  & 0.0005* (0.0001) & 0.0020* (0.0003) & 0.0000\phantom{*} (0.0000) & 0.0002* (0.0001) & 0.0002* (0.0000) \\ 
  $\hat\lambda_1$  & 0.1524* (0.0610) & 0.3246* (0.0720) & 0.0890* (0.0109) & 0.1669* (0.0302) & 0.2295* (0.0360) \\ 
  $\hat\lambda_2$  & 0.0771\phantom{*} (0.0452) & 0.2498* (0.0605) &  &  &  \\ 
  $\hat\lambda_3$  & 0.1236* (0.0501) & 0.2063* (0.0590) &  &  &  \\ 
  $\hat\lambda_4$  & 0.3485* (0.0835) &  &  &  &  \\ 
  $\hat\lambda_5$  & 0.2230* (0.0724) &  &  &  &  \\ 
  $\hat\gamma_1$ & 0.1687\phantom{*} (0.0935) &  &  &  &  \\ 
  $\hat\gamma_2$ & 0.0711\phantom{*} (0.0674) &  &  &  &  \\ 
  $\hat\gamma_3$ & 0.0442\phantom{*} (0.0747) &  &  &  &  \\ 
  $\hat\gamma_4$ & -0.0400\phantom{*} (0.1031) &  &  &  &  \\ 
  $\hat\gamma_5$ & -0.0935\phantom{*} (0.0815) &  &  &  &  \\ 
  $\hat\delta_1$&  &  & 0.9100* (0.0115) & 0.8045* (0.0291) & 0.7695* (0.0342) \\ 
  \midrule
  $\hat\zeta$ & 0.979* (0.0232) & 1.0426* (0.0268) & 1.0426* (0.0226) & 1.0141* (0.0294) & 1.0694* (0.0227) \\ 
  $\hat v$ & 3.0091* (0.1624) & 3.0941* (0.2385) & 3.4214* (0.1823) & 3.6145* (0.2881) & 2.9867* (0.1567) \\ 
  \midrule
  AIC & -9092.8599\phantom{*} & -6975.8293 & -7839.1305 & -7031.3708 & -7519.3082 \\ 
  $p$-val.: LB & 0.0524\phantom{*} & 0.0635 & 0.3947 & 0.0656 & 0.1635 \\ 
  $p$-val.: LB$^2$ & 0.0202* & 0.1608 & 0.9881 & 0.3693 & 0.9070 \\ 
  $p$-val.: WLM & 0.5244\phantom{*} & 0.2693 & 0.9923 & 0.4216 & 0.9117 \\ 
  $p$-val.: SB & 0.5536\phantom{*} & 0.2434 & 0.7885 & 0.7513 & 0.0916 \\ 
  $p$-val.: Neg. SB & 0.4161\phantom{*} & 0.4363 & 0.5272 & 0.1465 & 0.9266 \\ 
  $p$-val.: Pos. SB & 0.1552\phantom{*} & 0.8325 & 0.4973 & 0.6267 & 0.5489 \\ 
  $p$-val.: Joint SB & 0.4202\phantom{*} & 0.5247 & 0.8075 & 0.4924 & 0.1243 \\ 
  \bottomrule[0.05em]
\end{tabular}
\label{table:margin_model_res}
\end{threeparttable}}
\end{table}

\begin{table}[htp]
\centering
\captionsetup{font=large}
\resizebox{1.0\textwidth}{!}{
\begin{threeparttable}
\caption{Parameter estimates, AIC, and tests of the univariate models for the systems.}
\begin{tabular}{lrrrrrr}
\toprule[0.05em]
\midrule
 & Sys:BTC & Sys:ETH & Sys:LTC & Sys:XMR & Sys:XRP\\
\midrule
 Selected Model & GARCH(1,1) & GARCH(1,1) & GARCH(1,1) & GARCH(1,1) & GARCH(1,1)\\
 \midrule
$\hat\mu$ & 0.0052\phantom{*} (0.0034) & 0.0052\phantom{*} (0.0030) & 0.0071* (0.0032) & 0.0055\phantom{*} (0.0028) & 0.0088* (0.0033) \\ 
  $\hat\omega$ & 0.0018* (0.0005) & 0.0010* (0.0003) & 0.0018* (0.0005) & 0.0008* (0.0003) & 0.0015* (0.0004) \\ 
  $\hat\lambda_1$ & 0.1649* (0.0322) & 0.1452* (0.0266) & 0.1683* (0.0337) & 0.1521* (0.0247) & 0.1367* (0.0264) \\ 
  $\hat\delta_1$ & 0.8223* (0.0299) & 0.8538* (0.0257) & 0.8152* (0.0318) & 0.8469* (0.0248) & 0.8435* (0.0260) \\ 
   \midrule
  $\hat\zeta$ & 0.9446* (0.0267) & 0.9538* (0.0265) & 0.9404* (0.0260) & 0.9517* (0.0255) & 0.9363* (0.0265) \\ 
  $\hat v$ & 3.5653* (0.2926) & 3.3996* (0.2448) & 3.4711* (0.2838) & 3.5796* (0.2523) & 3.6289* (0.3025) \\ 
  \midrule
  AIC & -1667.4989\phantom{*} & -2109.0873 & -1972.6166\phantom{*} & -2114.0524\phantom{*} & -1955.2041\phantom{*} \\ 
  $p$-val.: LB & 0.0197* & 0.0625 & 0.0085* & 0.0290* & 0.0184* \\ 
  $p$-val.: LB$^2$ & 0.4058\phantom{*} & 0.6909 & 0.4429\phantom{*} & 0.5588\phantom{*} & 0.3737\phantom{*} \\ 
  $p$-val.: WLM & 0.5206\phantom{*} & 0.8245 & 0.4270\phantom{*} & 0.6103\phantom{*} & 0.3781\phantom{*} \\ 
  $p$-val.: SB & 0.7703\phantom{*} & 0.7930 & 0.1400\phantom{*} & 0.9673\phantom{*} & 0.7721\phantom{*} \\ 
  $p$-val.: Neg. SB & 0.5120\phantom{*} & 0.4868 & 0.2593\phantom{*} & 0.8352\phantom{*} & 0.7134\phantom{*} \\ 
  $p$-val.: Pos. SB & 0.4394\phantom{*} & 0.5198 & 0.6966\phantom{*} & 0.3526\phantom{*} & 0.1605\phantom{*} \\ 
  $p$-val.: Joint SB & 0.7278\phantom{*} & 0.7927 & 0.2627\phantom{*} & 0.7767\phantom{*} & 0.3126\phantom{*} \\ 
  \bottomrule[0.05em]
\end{tabular}
\label{table:margin_model_res_systems}
\end{threeparttable}}
\end{table}

%% file: tables/copula.tex
\begin{table}[htp]
\centering
\caption{Estimates of the time-invariant bivariate copulae.}
\resizebox{\textwidth}{!}{
\begin{tabular}{lccccccccc}
 \toprule[0.05em]
 \midrule
 \multirow{3}{*}{CC Pairs} & \multicolumn{9}{c}{Copula}\\
  \cmidrule[0.05em]{2-10}
  & \multicolumn{2}{c}{Gaussian} & \multicolumn{3}{c}{$t$} & \multicolumn{2}{c}{Clayton} & \multicolumn{2}{c}{Gumbel}\\
 \cmidrule[0.05em]{2-10}
& $\theta$ & AIC & $\theta$ & $\nu$ & AIC & $\theta$ & AIC & $\theta$ & AIC\\
 \midrule
 BTC-ETH & 0.559 & -16953.907 & 0.604 & 2.689 & -17264.039 & 1.174 & -17087.318 & 1.592 & -16914.514 \\ 
  BTC-LTC & 0.720 & -18716.545 & 0.747 & 3.827 & -18951.558 & 1.934 & -18872.348 & 1.930 & -18496.948 \\ 
  BTC-XMR & 0.581 & -17090.670 & 0.606 & 3.330 & -17306.663 & 1.251 & -17258.100 & 1.592 & -16976.563 \\ 
  BTC-XRP & 0.520 & -17379.916 & 0.557 & 3.271 & -17593.135 & 1.057 & -17553.168 & 1.491 & -17279.776 \\ 
  BTC-Sys:BTC & 0.667 & -12139.275 & 0.691 & 2.612 & -12450.922 & 1.639 & -12287.770 & 1.816 & -12052.933 \\ 
   \midrule
  ETH-BTC & 0.559 & -16953.907 & 0.604 & 2.689 & -17264.039 & 1.174 & -17087.318 & 1.592 & -16914.514 \\ 
  ETH-LTC & 0.582 & -15816.171 & 0.638 & 3.144 & -16108.938 & 1.288 & -15973.963 & 1.629 & -15743.510 \\ 
  ETH-XMR & 0.574 & -14917.515 & 0.618 & 3.023 & -15192.795 & 1.232 & -15081.750 & 1.623 & -14871.455 \\ 
  ETH-XRP & 0.564 & -15393.672 & 0.622 & 2.809 & -15696.008 & 1.221 & -15564.679 & 1.604 & -15337.560 \\ 
  ETH-Sys:ETH & 0.655 & -10370.179 & 0.707 & 2.447 & -10776.117 & 1.624 & -10549.724 & 1.832 & -10342.260 \\ 
   \midrule
  LTC-BTC & 0.720 & -18716.545 & 0.747 & 3.827 & -18951.558 & 1.934 & -18872.348 & 1.930 & -18496.948 \\ 
  LTC-ETH & 0.582 & -15816.171 & 0.638 & 3.144 & -16108.938 & 1.288 & -15973.963 & 1.629 & -15743.510 \\ 
  LTC-XMR & 0.555 & -15761.594 & 0.588 & 4.035 & -15923.228 & 1.211 & -15963.043 & 1.526 & -15602.025 \\ 
  LTC-XRP & 0.588 & -16410.185 & 0.637 & 3.321 & -16683.239 & 1.342 & -16606.216 & 1.619 & -16293.759 \\ 
  LTC-Sys:LTC & 0.690 & -11355.003 & 0.728 & 2.655 & -11705.841 & 1.850 & -11564.733 & 1.884 & -11241.507 \\ 
   \midrule
  XMR-BTC & 0.581 & -17090.670 & 0.606 & 3.330 & -17306.663 & 1.251 & -17258.100 & 1.592 & -16976.563 \\ 
  XMR-ETH & 0.574 & -14917.515 & 0.618 & 3.023 & -15192.795 & 1.232 & -15081.750 & 1.623 & -14871.455 \\ 
  XMR-LTC & 0.555 & -15761.594 & 0.588 & 4.035 & -15923.228 & 1.211 & -15963.043 & 1.526 & -15602.025 \\ 
  XMR-XRP & 0.516 & -15277.939 & 0.552 & 3.751 & -15450.299 & 1.071 & -15475.696 & 1.477 & -15158.206 \\ 
  XMR-Sys:XMR & 0.637 & -10350.501 & 0.671 & 3.045 & -10628.356 & 1.538 & -10556.765 & 1.732 & -10246.908 \\ 
   \midrule
  XRP-BTC & 0.520 & -17379.916 & 0.557 & 3.271 & -17593.135 & 1.057 & -17553.168 & 1.491 & -17279.776 \\ 
  XRP-ETH & 0.564 & -15393.672 & 0.622 & 2.809 & -15696.008 & 1.221 & -15564.679 & 1.604 & -15337.560 \\ 
  XRP-LTC & 0.588 & -16410.185 & 0.637 & 3.321 & -16683.239 & 1.342 & -16606.216 & 1.619 & -16293.759 \\ 
  XRP-XMR & 0.516 & -15277.939 & 0.552 & 3.751 & -15450.299 & 1.071 & -15475.696 & 1.477 & -15158.206 \\ 
  XRP-Sys:XRP & 0.623 & -10621.955 & 0.666 & 2.627 & -10946.718 & 1.471 & -10811.682 & 1.719 & -10548.650 \\ 
\bottomrule[0.05em]
\end{tabular}}
\label{table:cop_estimates_static_bivariate}
\end{table}

\begin{table}[htp]
\centering
\caption{Estimates of the time-variant bivariate copulae.}
\resizebox{\textwidth}{!}{
\begin{tabular}{lccccccccc}
 \toprule[0.05em]
 \midrule
 \multirow{3}{*}{CC Pairs} & \multicolumn{9}{c}{Copula}\\
  \cmidrule[0.05em]{2-10}
  & \multicolumn{5}{c}{Patton-$t$} & \multicolumn{4}{c}{DCC-$t$}\\
 \cmidrule[0.05em]{2-10}
& $\omega_\theta$ & $\beta_\theta$ & $c_\theta$ & $\nu$ & AIC & a & b & $\nu$ & AIC\\
 \midrule
BTC-ETH & 0.208 & 1.443 & 0.525 & 5.882 & -17582.578 & 0.085 & 0.910 & 5.612 & -17887.668 \\ 
  BTC-LTC & 1.090 & 0.622 & 0.344 & 5.767 & -19008.341 & 0.080 & 0.883 & 6.665 & -19134.243 \\ 
  BTC-XMR & 0.545 & 0.745 & 0.572 & 6.220 & -17468.442 & 0.071 & 0.919 & 5.201 & -17663.209 \\ 
  BTC-XRP & 0.345 & 0.980 & 0.527 & 5.678 & -17804.362 & 0.076 & 0.916 & 5.443 & -17972.693 \\ 
  BTC-Sys:BTC & 0.699 & 0.709 & 0.608 & 5.572 & -12656.446 & 0.088 & 0.902 & 5.578 & -12940.102 \\ 
 \midrule
  ETH-BTC & 0.208 & 1.443 & 0.525 & 5.882 & -17582.578 & 0.085 & 0.910 & 5.612 & -17887.668 \\ 
  ETH-LTC & 0.468 & 0.979 & 0.613 & 7.518 & -16358.821 & 0.037 & 0.959 & 7.637 & -16666.221 \\ 
  ETH-XMR & 0.683 & 0.755 & 0.427 & 5.598 & -15271.656 & 0.039 & 0.958 & 4.448 & -15473.335 \\ 
  ETH-XRP & 0.242 & 1.477 & 0.477 & 6.108 & -15919.341 & 0.051 & 0.948 & 5.785 & -16247.803 \\ 
  ETH-Sys:ETH & 0.748 & 0.798 & 0.615 & 6.218 & -10940.029 & 0.060 & 0.937 & 5.454 & -11405.215 \\ 
   \midrule
  LTC-BTC & 1.075 & 0.647 & 0.341 & 5.767 & -19008.544 & 0.080 & 0.883 & 6.665 & -19134.243 \\ 
  LTC-ETH & 0.468 & 0.979 & 0.613 & 7.518 & -16358.825 & 0.037 & 0.959 & 7.637 & -16666.221 \\ 
  LTC-XMR & 0.462 & 0.905 & 0.484 & 6.394 & -16084.549 & 0.062 & 0.929 & 6.501 & -16243.662 \\ 
  LTC-XRP & 0.468 & 1.102 & 0.409 & 6.287 & -16796.757 & 0.051 & 0.944 & 5.186 & -17045.160 \\ 
  LTC-Sys:LTC & 0.818 & 0.770 & 0.506 & 5.911 & -11847.605 & 0.070 & 0.922 & 5.347 & -12170.773 \\ 
   \midrule
  XMR-BTC & 0.545 & 0.745 & 0.572 & 6.220 & -17468.442 & 0.071 & 0.919 & 5.201 & -17663.209 \\ 
  XMR-ETH & 0.683 & 0.755 & 0.427 & 5.598 & -15271.656 & 0.039 & 0.958 & 4.448 & -15473.335 \\ 
  XMR-LTC & 0.462 & 0.905 & 0.484 & 6.394 & -16084.549 & 0.062 & 0.929 & 6.501 & -16243.662 \\ 
  XMR-XRP & 0.475 & 0.785 & 0.491 & 5.757 & -15580.298 & 0.044 & 0.952 & 5.779 & -15747.537 \\ 
  XMR-Sys:XMR & 0.825 & 0.662 & 0.435 & 5.508 & -10713.380 & 0.053 & 0.938 & 4.537 & -10897.825 \\ 
\midrule
  XRP-BTC & 0.345 & 0.980 & 0.527 & 5.678 & -17804.362 & 0.076 & 0.916 & 5.443 & -17972.693 \\ 
  XRP-ETH & 0.309 & 1.336 & 0.502 & 6.013 & -15914.526 & 0.051 & 0.948 & 5.785 & -16247.803 \\ 
  XRP-LTC & 0.468 & 1.102 & 0.409 & 6.287 & -16796.757 & 0.051 & 0.944 & 5.186 & -17045.160 \\ 
  XRP-XMR & 0.475 & 0.785 & 0.491 & 5.757 & -15580.298 & 0.044 & 0.952 & 5.779 & -15747.537 \\ 
  XRP-Sys:XRP & 0.392 & 1.253 & 0.478 & 5.036 & -11150.509 & 0.063 & 0.934 & 4.466 & -11445.793 \\ 
\bottomrule[0.05em]
\end{tabular}}
\label{table:cop_estimates_dynamic_bivariate}
\end{table}

\begin{table}[htp]
\centering
\caption{Estimates of the multivariate copulae for BTC-ETH-LTC-XMR-XRP.}
\begin{tabular}{lcc}
 \toprule[0.05em]
 \midrule
Copula & Characteristic & Estimate/Value \\
 \midrule
\multirow{2}{*}{Gaussian} & $\theta$ & $\left[ \begin{array}{ccccc}
1.000 & 0.554 & 0.721 & 0.577 & 0.517\\
 & 1.000 & 0.582& 0.568 & 0.559\\
 & & 1.000 & 0.555 & 0.590\\
 & & & 1.000 & 0.511\\
 & & & & 1.000\\
\end{array} \right]$\\
 & AIC & -44087.260 \\
  \midrule
\multirow{3}{*}{$t$} & $\theta$ & $\left[ \begin{array}{ccccc}
1.000 & 0.639 & 0.746 & 0.633 & 0.593\\
 & 1.000 & 0.659 & 0.621 & 0.629\\
 & & 1.000 & 0.609 & 0.655\\
 & & & 1.000 & 0.569\\
 & & & & 1.000\\
\end{array} \right]$\\
& $\nu$ & 4.804 \\
& AIC & -45201.470 \\
\midrule
\multirow{2}{*}{Clayton} & $\theta$ & 1.077\\
& AIC & -44174.270 \\
\midrule
\multirow{2}{*}{Gumbel} & $\theta$ & 1.616\\
& AIC & -43531.790 \\
\midrule
\multirow{4}{*}{DCC-$t$} & a & 0.036\\
& b & 0.959 \\
& $\nu$ & 8.029 \\
& AIC & -46720.380 \\
\bottomrule[0.05em]
\end{tabular}
\label{table:cop_estimates_multivariate}
\end{table}